\documentclass[fleqn,usenatbib]{mnras}

\usepackage{newtxtext,newtxmath}

\usepackage[T1]{fontenc}

\DeclareRobustCommand{\VAN}[3]{#2}
\let\VANthebibliography\thebibliography
\def\thebibliography{\DeclareRobustCommand{\VAN}[3]{##3}\VANthebibliography}

\usepackage{graphicx}	
\usepackage{amsmath}	

\title[VMC: Classifying extragalactic sources with a PRF]{The VMC Survey -- LI. Classifying extragalactic sources using a probabilistic random forest supervised machine learning algorithm\thanks{While we here refer to the `Magellanic' Clouds as their common names, we acknowledge and disapprove of the associated colonial heritage of which the inequities and disparities still endure today.}}

\author[C. M. Pennock et al.]{
Clara M. Pennock$^{1,2}$\thanks{E-mail: Clara.Pennock@ed.ac.uk},
Jacco Th. van Loon$^{1}$,
Maria-Rosa L. Cioni$^{3}$,
Chandreyee Maitra$^{4}$,
\newauthor Joana M. Oliveira$^{1}$,
Jessica E. M. Craig$^{1}$,
Valentin D. Ivanov$^{5}$,
James Aird$^{2}$,
Joy O. Anih$^{1}$,
\newauthor Nicholas J. G. Cross$^{2}$,
Francesca Dresbach$^{1}$,
Richard de Grijs$^{6, 7, 8}$,
Martin ~A.~T.~Groenewegen$^{9}$
\\
$^{1}$Lennard-Jones Laboratories, Keele University, ST5 5BG, UK\\
$^{2}$Institute for Astronomy, University of Edinburgh, Royal Observatory, Blackford Hill, Edinburgh, EH9 3HJ, UK\\
$^{3}$Leibniz-Institut f\"ur Astrophysik Potsdam, An der Sternwarte 16, D-14482 Potsdam, Germany\\
$^{4}$Max-Planck-Institut f\"ur extraterrestrische Physik, Gießenbachstraße, D-85748 Garching, Germany\\
$^{5}$ European Southern Observatory, Karl-Schwarzschild-Strasse 2, D-85748 Garching bei M\"unchen, Germany\\
$^{6}$School of Mathematical and Physical Sciences, Macquarie University, Balaclava Road, Sydney, NSW 2109, Australia\\
$^{7}$Astrophysics and Space Technologies Research Centre, Macquarie University, Balaclava Road, Sydney, NSW 2109, Australia\\
$^{8}$International Space Science Institute-Beijing, 1 Nanertiao, Zhongguancun, Beijing 100190, China\\
$^{9}$Koninklijke Sterrenwacht van Belgi\"e, Ringlaan 3, B--1180 Brussels, Belgium\\
}

\date{Accepted XXX. Received YYY; in original form ZZZ}

\pubyear{2024}

\begin{document}
\label{firstpage}
\pagerange{\pageref{firstpage}--\pageref{lastpage}}
\maketitle

\begin{abstract}
We used a supervised machine learning algorithm (probabilistic random forest) to classify $\sim$130 million sources in the VISTA Survey of the Magellanic Clouds (VMC). We used multi-wavelength photometry from optical to far-infrared as features to be trained on, and spectra of Active Galactic Nuclei (AGN), galaxies and a range of stellar classes including from new observations with the Southern African Large Telescope (SALT) and SAAO 1.9m telescope. We also retain a label for sources that remain unknown. This yielded average classifier accuracies of $\sim$79\% (SMC) and $\sim$87\% (LMC).
Restricting to the 56,696,719 sources with class probabilities (P$_{\rm class}$) $>$ 80\% yields accuracies of $\sim$90\% (SMC) and $\sim$98\% (LMC).  
After removing sources classed as `Unknown', we classify a total of 707,939 (SMC) and 397,899 (LMC) sources, including $>$77,600
extragalactic sources 
behind the Magellanic Clouds. The extragalactic sources are distributed evenly across the field, whereas the Magellanic sources concentrate at the centres of the Clouds, and both concentrate in optical/IR colour--colour/magnitude diagrams as expected. We also test these classifications using independent datasets, finding that, as expected, the majority of X-ray sources are classified as AGN (554/883) and the majority of radio sources are classed as AGN (1756/2694) or galaxies (659/2694), where the relative AGN--galaxy proportions vary substantially with radio flux density. 
We have found: $>$49,500 hitherto unknown AGN candidates, likely including more AGN dust dominated sources which are in a critical phase of their evolution; $>$26,500 new galaxy candidates and $>$2800 new Young Stellar Object (YSO) candidates.
\end{abstract}

\begin{keywords}
Magellanic Clouds -- methods: data analysis -- galaxies: active -- galaxies: photometry
\end{keywords}



\section{Introduction}


As the depth and field of view of survey telescopes continues to improve it becomes increasingly more unfeasible for astronomers to manually verify every individual source. Separating the extragalactic from the non-extragalactic is important for population studies, and for the study of individual systems. Identifying galaxies, and whether they are hosting an active galactic nucleus (AGN) or not, allows us to study how galaxies evolve over cosmic time and what role AGN have to play in this process. Galaxies, especially those that host an AGN that have the potential to produce emission across the entire electromagnetic spectrum \citep[e.g.][]{padovani2017}, are more easily identified from combinations of multi-wavelength photometric survey data. 

Different wavelength regimes probe different parts of the AGN's structure \citep[e.g.][]{2015Netzer, padovani2017, 2018Hickox}. The ultraviolet/optical are sensitive to the accretion disk and the infrared (IR) is sensitive to the dusty obscuring material surrounding the accretion disk. The emission of a corona is observed in X-rays, whilst possible non-thermal radiation (which often take the form of jets/lobes) is picked up in the radio.  One of the most reliable methods of identifying an AGN is with optical spectroscopy, which can reveal the broad and/or narrow emission lines of an AGN. However, spectroscopically observing every source would be a time consuming process, which is why we often turn to photometric surveys, which can observe large areas of the sky much more quickly. 

Spectral energy distributions can be produced from photometric surveys, which for an AGN would exhibit a “big blue bump\textquotedblright\, due to the accretion disk and another bump in the mid-infrared (mid-IR) due to re-processed emission from dust heated by accretion from the central supermassive black hole, which is a feature that has been well used to select large samples of AGN \citep[e.g.][]{2004Lacy, stern2005,2015Secrest,2018Assef}. However, factors such as obscuration from dust, emission from the host galaxy and the abundance of stars in our field of view can make AGN harder to select in one wavelength band, hence the need to make use of multiple wavelength regimes to ascertain a source's true nature.

The Magellanic Clouds are an often over-looked but - while challenging - eminently suitable and worthwhile location to search for the extragalactic sources behind them. The Clouds span $\sim$ 100 sq.\ degrees on the sky that have been covered as part of all-sky multi-waveband surveys such as the optical \textit{Gaia} \citep{Gaia2021}, the near-infrared (near-IR) 2MASS \citep{2MASS} and the mid-IR AllWISE \citep{Cutri2013} surveys. Furthermore, there have been Magellanic Cloud specific surveys, where depth and angular resolution are improved compared to all-sky surveys. They include ultraviolet (UV) with the Galaxy Evolution Explorer (\textit{GALEX}) and the UltraViolet Imaging Telescope (\textit{UVIT}) \citep[e.g.][]{2014AAS...22335511T, UVIT} and the optical Survey of the Magellanic Stellar History \citep[SMASH;][]{Nidever2017} and the MAGellanic Inter-Cloud Project \citep[MAGIC;][]{2013MAGIC, 2015MAGIC, 2017MAGIC}, as well as surveys observed with the \textit{Sptizer Space Telescope} in the mid-IR as part of the \textit{Sptizer} Agents of Galaxy Evolution (SAGE) survey of the Large Magellanic Cloud \citep[LMC;][]{Meixner2006} and Small Magellanic Cloud \citep[SMC;][]{gordon2011}, and the \textit{Herschel Space Observatory} in the far-IR as part of the \textit{HERschel} Inventory of The Agents of Galaxy Evolution \citep[HERITAGE;][]{Meixner2010}. Additionally, there have been many observations in the radio domain \citep[e.g. MOST, ATCA;][]{Mauch2003,Murphy2010} and X-ray \citep[XMM--Newton;][]{2013XMM}. These galaxies are also located away from the Galactic Plane and Galactic Centre, reducing source confusion in the radio band and extinction at UV/optical/near-IR wavelengths. The main caveat for searching in the field of the Magellanic Clouds is the increased stellar confusion. The new and deeper surveys towards the Magellanic Clouds, such as the near-IR VISTA Survey of the Magellanic Clouds \citep[VMC;][]{2011A&A...527A.116C} and radio Evolutionary Map of the Universe all-sky \citep[EMU;][]{Joseph2019,2021MNRAS.506.3540P} survey greatly enhance such attempts. 

The VMC ESO public survey showcases a great improvement in depth and angular resolution compared to previous near-IR surveys, and has detected stars encompassing most phases of evolution such as main sequence stars, sub-giants, upper and lower red giant branch (RGB) stars, red clump stars, RR Lyr\ae\ and Cepheid variables, asymptotic giant branch (AGB) stars, post-AGB stars, young stellar objects (YSOs), planetary nebul\ae\ (PNe) and supernova remnants (SNRs) populations \citep[e.g.][]{2012AGB,2015Cepheid,2018YS,2019RGB,2020YS,2020AGB,2021RRL,2021RGB} that can be used to help assess the age, metallicity, 3D structure, etc. within the Magellanic systems. This survey has also had success in discovering background extragalactic sources \citep{2013Cioni, Ivanov2016, Bell2019, Bell2020, Bell2022, 2022MNRAS.515.6046P}.

Machine learning algorithms are a use of artificial intelligence to automate tasks, such as identification and classification, on large sets of data that would otherwise prove time consuming. They can also replace subjective approaches that depend on user choices by objective approaches that are data driven. Another advantage of machine learning is that they can combine information from multiple datasets, effectively classifying within a highly multi-dimensional parameter space. 
Machine learning algorithms are usually divided into two types, supervised and unsupervised. Unsupervised machine learning is predominantly used for clustering and dimensionality reduction, where objects with similar properties are grouped together in 2D space to find patterns/trends in the data. 

Supervised machine learning algorithms predict classifications/values based on example data with features (e.g. photometry) and known classifications/values. They do this by analysing a known dataset, the training set, and producing a model from this dataset, which can then be used to make predictions of the output of an unseen dataset. A disadvantage of supervised learning is that it is only as good as the data it is trained upon, and is therefore not best suited to finding new or unusual objects. Furthermore, imbalanced training sets (when the amount of objects for one or more of the classes dominates the training set) can lead to poor performance of the classifier, though this can be mitigated by artificially balancing the training sets \citep[e.g.][]{Kinson2021, Kinson2022}.

The VMC survey consists of $\sim$ 130 million sources, the identities of the majority of which remain unknown. In \citet{2022MNRAS.515.6046P} unsupervised machine learning was used to cluster similar sources together in the radio-detected population of the VMC near-IR sources \citep{2021MNRAS.506.3540P}, showing that machine learning can be a valuable tool in separating objects into different classifications, especially for separating dusty/evolved stars (such as YSOs, PNe and post-AGB/RGB) that are often confused with AGN and vice versa in the optical and IR. This work is a continuation of these studies, where we use supervised machine learning with multi-wavelength data from UV to far-IR to classify all of the sources in the VMC survey. In a follow-on paper we will apply an unsupervised approach.

This paper is laid out as follows: Section \ref{Data} outlines the photometric surveys used in this work from which the features for the machine learning algorithm are selected, as well as the spectroscopic datasets from which the data with known labels is selected as a training set. Section \ref{PRF} describes the machine learning algorithm used in this work, the choice of parameters and how it was trained. In Section \ref{Results} we explore the spatial distributions of the newly classified sources across the VMC fields of the Magellanic Clouds. Then in Section \ref{Discussion} we test the classifier against sources with known classes that were not used in training in Sections \ref{dustyAGN} and \ref{Unseenclasses}, as well as exploring their distributions of the high-confidence classifications across colour--colour/magnitude diagrams in the optical, near-IR and mid-IR regimes in Section \ref{CCD}. Furthermore, in Sections \ref{Radio} and \ref{Xray} we use the radio and X-ray detected sources as an independent check to test the classifications, as the majority of X-ray and radio detected sources are expected to be extragalactic, and explore these populations. We explore the classifications of \textit{Gaia} low-resolution spectroscopically confirmed QSOs from \citet{2023Quaia} in Section \ref{Quaia} and the classifications of spectroscopically and photometrically selected YSOs in the LMC from \citet{2023Kokusho} in Section \ref{YSOs}. Then, in Section \ref{Unknownclass}, we explore the sources that have been confidently classed as Unknown by using the known photometric selection techniques. Lastly, we summarise our results in Section \ref{conclusions}.


\section{Data}\label{Data}
Machine learning algorithms require “features\textquotedblright\, for each object, such as photometric measurements in various wavebands. Supervised machine learning requires an additional training set of known objects with labels that can be used to learn from, in order to classify unseen data. Here the photometry and the training sets are outlined.

\subsection{Photometry}
\begin{figure*}
\centering
    \begin{tabular}{c}
	\includegraphics[width=\textwidth, trim=4mm 1mm 1mm 4mm, clip]{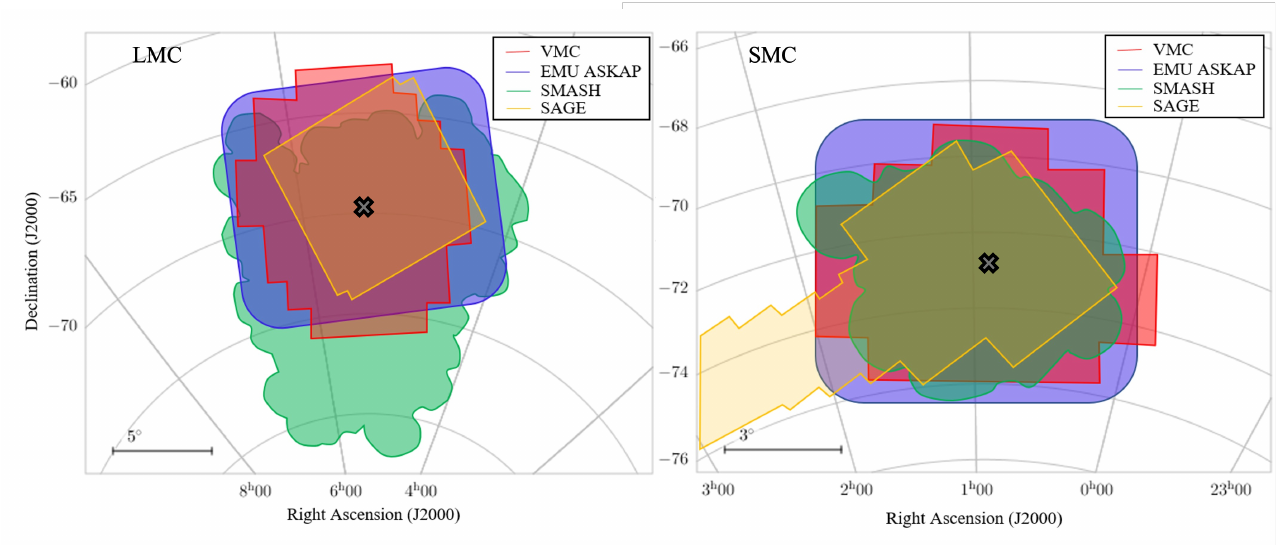}
	\end{tabular}
    \caption{Area of the sky covered by the VMC (near-IR, \textit{red}), EMU ASKAP (radio, \textit{blue}), SMASH (optical, \textit{green}), SAGE (mid-IR, \textit{yellow}) surveys of the LMC (left) and SMC (right). The approximate centres of the Clouds are marked with a black `X'.}
    \label{MCcoverage}
\end{figure*}

In this study we use photometry obtained from dedicated observations of the Magellanic Clouds, from the optical Survey of the Magellanic Stellar History \citep[SMASH;][]{Nidever2017}, VMC \citep{2011A&A...527A.116C}, \textit{Spitzer} SAGE \citep{Meixner2006, gordon2011}, \textit{Herschel} HERITAGE \citep{Meixner2010} and XMM-Newton \citep{2013XMM} X-ray and ASKAP radio \citep{Joseph2019,2021MNRAS.506.3540P} imaging surveys. Figure \ref{MCcoverage} shows the comparisons between the area covered by the VMC, EMU ASKAP, SMASH and SAGE surveys of the SMC and LMC. Below we describe each of the individual survey datasets used in this work. Details of how the UV/optical/IR surveys are combined to produce the multiwavelength catalogue for the machine learning process will be given in Section \ref{createTS}.

The X-ray and radio catalogues are not used in the machine learning, but as an independent check of the final source classifications, as radio/X-ray sources tend to be extragalactic in origin, as opposed to stars. Also, they are only a small fraction of the total sources so would introduce a lot of missing data if used as features.


\subsubsection{Optical SMASH survey}

The Dark Energy Camera \citep[DECam;][]{DECAM} on NOAO’s 4-m Blanco telescope was used as part of the Survey of the Magellanic Stellar History \citep[SMASH;][]{Nidever2017} to map 480 square degrees of sky to depths of $ugriz$ $\sim$ 23.9, 24.8, 24.5, 24.2, 23.5 mag (Vega) at median seeing of 1$\rlap{.}^{\prime\prime}$22, 1$\rlap{.}^{\prime\prime}$13, 1$\rlap{.}^{\prime\prime}$01, 0$\rlap{.}^{\prime\prime}$95, 0$\rlap{.}^{\prime\prime}$90, respectively. The main goal of this survey was to identify broadly distributed, low surface brightness stellar populations associated with the stellar halos and tidal debris of the Magellanic Clouds. The catalogue contains $\sim$ 360 million objects in 197 fields. Note that \citet{Nidever2017} adjusted the DECam $ugriz$ photometry to be comparable to SDSS and are therefore “pseudo-SDSS\textquotedblright\, $ugriz$ bands.

\subsubsection{Near-IR VISTA Magellanic Clouds survey}

The Visible and Infrared Survey Telescope for Astronomy \citep[VISTA;][]{2006vista} is a 4.1-m near-infrared optimised telescope, which is equipped with the VISTA InfraRed CAMera \citep[VIRCAM;][]{VIRCAM} which is composed of a large array of 16 detectors that fill about a 1.5 square degree field. 

The VMC survey \citep{Cioni2011} is a near-IR deep, multi-epoch and wide-field study of the Magellanic Clouds, covering an area of about 170 deg$^{2}$. VISTA observations for the VMC main survey started in November 2009 and ended in October 2018. It has a spatial resolution of 1$\rlap{.}^{\prime\prime}$0 -- 1$\rlap{.}^{\prime\prime}$1, 0$\rlap{.}^{\prime\prime}$9 -- 1$\rlap{.}^{\prime\prime}$0 and 0$\rlap{.}^{\prime\prime}$8 -- 0$\rlap{.}^{\prime\prime}$9 in the $YJK$\textsubscript{s} filters, respectively, where the two values specified for seeing indicate maximum allowed seeing for crowded and uncrowded regions, respectively. It also reaches a sensitivity at 5$\sigma$ level of about 22, 22 and 21.5 mag (Vega; in AB this is 22.5, 22.9 and 23.4 mag) in the $YJK$\textsubscript{s} bands, respectively. Its depth and coverage can be compared to the VISTA Deep Extragalactic Observations \citep[VIDEO; ][]{2013MNRAS.428.1281J} survey, which was specifically designed to study galaxy and cluster/structure evolution up to $z \sim 4$ in a 12 deg$^{2}$ area, reaching depths of about 24.5, 24.4 and 23.8 mag (AB) at 5$\sigma$ detection level in the $YJK$\textsubscript{s} bands, respectively. The VMC data provide an opportunity to expand on the effort of the VIDEO survey and cover more area to better overcome cosmic variance, and has already proven successful in discovering more AGN \citep[e.g.][]{Ivanov2016}. This however comes with the caveat of increased stellar confusion with the presence of the LMC and SMC.

The catalogues created from the VMC survey provide both aperture and PSF photometry, where PSF photometry reaches sources on average 0.3 magnitudes fainter than aperture photometry. The PSF catalogue is created as described in \cite{2015Rubele} and are publicly available as part of VMC DR6\footnote{\url{http://archive.eso.org}}\footnote{\url{http://vsa.roe.ac.uk}}. The magnitudes in each band are calculated from deep tile images, which are a combination of single exposure images from different epochs. Due to the PSF photometry's increased depth and ability to distinguish sources in crowded regions, it is the PSF photometry that is used in this work.

\subsubsection{Infrared SAGE and HERITAGE surveys}


The Magellanic Clouds were observed by \textit{Spitzer} as part of the SAGE survey of the LMC \citep{Meixner2006} and SMC \citep{gordon2011} which map 49 deg$^2$ and 30 deg$^2$ respectively. It produced a list of about 8.4 million sources taken with IRAC filters 3.6, 4.5, 5.8, 8.0 $\mu$m with an angular resolution of 2$^{\prime\prime}$. The faint limits for SAGE are 18.3, 17.7, 15.7 and 14.5 mag, respectively.

The \textit{Herschel} Space Observatory \citep{Herschel} was a 3.5-m infrared telescope that was active from 2009 to 2013 and was sensitive to the far infrared and submillimetre wavebands (55 –- 672 $\mu$m). \textit{HERschel} Inventory of The Agents of Galaxy Evolution \citep[HERITAGE;][]{Meixner2010} used the \textit{Herschel}’s Photodetector Array Camera and Spectrometer \citep[PACS, 100 and 160 $\mu$m;][]{PACS} and the Spectral and Photometric Imaging REceiver \citep[SPIRE, 250, 350, 500 $\mu$m;][]{SPIRE} bands to image the LMC, SMC and Magellanic Bridge. This survey is complementary to the SAGE survey.

\subsubsection{All-sky surveys}

Various all-sky surveys have also observed the Magellanic Clouds. However, this comes with the caveat of not reaching the same depths as the Magellanic specific surveys. All-sky surveys used in this work include optical \textit{Gaia} \citep{GaiaDR3} and mid-IR AllWISE and unWISE \citep{WISE,Cutri2013,UnWISE} surveys.

The \textit{Gaia} mission \citep{Gaiamission} was launched on 19 December 2013, with the aim of measuring the 3D spatial and velocity distribution of stars, as well as determine their astrophysical properties. The \textit{Gaia} on-board system is designed to detect point-like sources, but can detect extragalactic sources \citep{GaiaEx} if their central region is sufficiently bright and compact. The latest data release, DR3 \citep{GaiaDR3}, is based on 34 months of \textit{Gaia} operations. The catalogue provides celestial positions, proper motions, parallaxes, and broad band photometry in the wide G (centred on 650 nm), blue-enhanced G$_{BP}$ (centred on 360 nm), and red-enhanced G$_{RP}$ (centred on 750 nm) pass-bands. This data release also includes class probabilities (QSO, galaxy or stellar source) for 1.5 billion sources.

WISE \citep{WISE} is a telescope launched in 2009 to repeatedly map the entire sky in infrared. WISE mapped the whole sky in four bands W1, W2, W3, W4 centred at 3.4 $\mu$m, 4.6 $\mu$m, 12 $\mu$m, and 22 $\mu$m, respectively, using a 40-cm telescope feeding arrays with a total of four million pixels. The sensitivities of W1, W2, W3 and W4 correspond to Vega magnitudes of 16.5, 15.5, 11.2, and 7.9, respectively, in the all-sky WISE survey. The AllWISE \citep{Cutri2013} programme extended the work of the WISE survey mission by combining W1 and W2 data from the cryogenic and post-cryogenic survey phases to form the most comprehensive view of the mid-infrared sky currently available. W3 and W4 measurements remain unchanged from the All-Sky Release because no additional data were included in those bands. 

Further addition to the WISE mission, is the unWISE \citep{UnWISE} catalogue, which used the deep unWISE coadded images built from five years of publicly available WISE imaging, as well as improved modelling of crowded regions. This resulted in a catalogue of $\sim$ 2 billion unique objects detected in the W1 and/or W2 channels, reaching depths $\sim$ 0.7 mag fainter than those achieved by AllWISE. 

\subsubsection{X-ray XMM-Newton}

An SMC-survey point-source catalogue was created from archival XMM-Newton data with additional newer observations from the same facility \citep{2013XMM}, which covers 5.6 deg$^2$, including the bar and eastern wing of the SMC. The catalogue contains 3053 unique X-ray sources with a median position uncertainty of 1\rlap{.}$^{\prime\prime}$3 down to a flux limit of $\sim$ 10$^{-14}$ erg cm$^{-2}$ s$^{-1}$. The majority of the sources are expected to be AGN. One limitation of this survey is that it only covers the central part of the SMC, and therefore does not cover the same breadth as the VMC survey. 

There is no similar X-ray catalogue specifically for the LMC. There is, however, an XMM-Newton serendipitous source catalogue \citep{XMMS}, which is a collection of the sources detected in all the publicly available XMM-Newton observations. This catalogue includes observations in the direction of the LMC. 

\subsubsection{Radio ASKAP survey}



The Evolutionary Map of the Universe \citep[EMU;][]{norris2011} is a wide-field radio continuum survey which uses the Australian Square Kilometre Array Pathfinder \citep[ASKAP;][]{Johnston2008,Hotan2021} telescope. EMU's primary goal is to make a deep (RMS $\sim$ 10 $\mu$Jy/beam) radio continuum survey of the Southern sky, extending as far north as +30$^{\circ}$  declination, with a resolution of 10$^{\prime\prime}$. It is expected to catalogue about 70 million galaxies, including AGN up to the edge of the visible Universe.

Two radio continuum images from the ASKAP survey in the direction of the SMC were taken as part of the EMU Early Science Project (ESP) survey of the Magellanic Clouds \citep{Joseph2019}. The two source lists that were produced from these images by \cite{Joseph2019} contain radio continuum sources observed at 960 MHz (4489 sources) and 1320 MHz (5954 sources) with a bandwidth of 192 MHz and beam sizes of $30^{\prime\prime}\times30^{\prime\prime}$ and $16\rlap{.}^{\prime\prime}3\times15\rlap{.}^{\prime\prime}1$, respectively. The median RMS noise values were 186 $\mu$Jy beam$^{-1}$ (960 MHz) and 165 $\mu$Jy beam$^{-1}$ (1320 MHz). The observations of the SMC were made with only 33 per cent and 44 per cent (for 960 MHz and 1320 MHz respectively) of the full ASKAP antenna configuration and 66 per cent of the final bandwidth that was available in the final array, with which the LMC was observed, so the resolution and depth is not as good as for the LMC observation.

The LMC was observed at 888 MHz \citep{2021MNRAS.506.3540P} with a bandwidth of 288 MHz taken on 2019 April 20 using ASKAP's full array of 36 antennas (scheduling block 8532). The LMC was observed as part of the ASKAP commissioning and early science (ACES, project code AS033) verification \citep{DeBoer2009,Hotan2014,McConnell2016} in order to investigate issues that were found in higher-frequency higher-spectral-resolution Galactic-ASKAP \citep[GASKAP;][]{Dickey2013} survey observations, as well as to test the rapid processing with ASKAPsoft \citep{Whiting2020}. The observations cover a total ﬁeld of view of 120 deg$^{2}$, with a total exposure time of $\sim$12h40m. They were compiled by four pointings ($\sim$3h10m each) with three interleaves\footnote{interleaves are overlapping pointings where the telescope slews between them at a more rapid cadence.}, each to result in more uniform depth across the field -- effectively 12 pointings. The three interleaves overlap by $\sim$0.5$^{\circ}$ to improve the uniformity of sensitivity across the field. The largest angular scales that can be recovered in this survey are 25 -- 50$^{\prime}$ \citep{McConnell2020}. 

\subsection{Training sets for machine learning}
A set of known sources is required to train a supervised machine 
learning classifier. We focused on using sources with spectroscopic observations. The total number of sources for each class can be seen in Table \ref{classtable}. The training sets for the SMC and LMC are made available alongside this paper as online supplementary material.

We chose to focus on ten classes: AGN; galaxies; stars of O and B type (OB); red giant branch stars (RGB); asymptotic giant branch stars (AGB); red supergiants (RSG), post-AGB and post-RGB stars (pAGB/RGB); planetary nebul\ae\ (PNe), YSOs and compact H\,\textsc{ii} regions (H\,\textsc{ii}/YSOs) and Milky-Way high proper-motion stars (PM). These classes were chosen for having larger numbers of sources with classifications based on spectroscopy and/or due to their tendency to be mistaken for AGN (dusty and/or emission-line sources) and vice versa.

Using spectroscopically observed sources however introduces bias into the training sample, since the sources observed tend to be chosen based on colour cuts that similar previously observed objects conform to. Furthermore, there is also a bias in magnitude, as the faintest sources would be too faint for spectroscopy. This therefore leaves the rarer/unusual and fainter versions of each class to not be observed, which would make the machine learning algorithm less certain about these sources. 

A potential problem could be that classes that have little to no spectroscopic observations, and therefore not trained upon, could be misclassified as one of the classes trained upon if they are similar enough. A class that encompasses these sources that are not part of the known classes could be needed to prevent confusion.

\begin{table}
\caption{The number of sources for each class and the region of the Clouds they were spectroscopically observed in, as well as the references of the literature they originated from. (1) From own observations using SALT or SAAO's 1.9m telescope; (2) Spitzer-spec surveys \citep{Ruffle2015,Jones2017}; (3) From a Simbad \citep{Simbad} search of`PM' stars in the VMC footprint that had listed spectral type and reference given. }
\begin{tabular}{|l|c|c|l|}
\hline\hline
\llap{C}lass & SM\rlap{C} & LM\rlap{C} & References\\
\hline
\llap{A}GN & 3\rlap{03} & \mbox{~~}6\rlap{39} & (1); (2); \cite{Flesch2019pap,Kozlowski2012}; \\
 & & & \cite{Kozlowski2013,Geha2003};\\
  & & & \cite{Esquej2013,Ivanov2016};\\
  & & & Ivanov et al. (\textit{in press.})  \\
\llap{G}alaxie\rlap{s}  & 1\rlap{24} & \mbox{~~}4\rlap{30} & (1); (2); \cite{Jones2009} \\
\llap{O}B    & 4\rlap{17} & 10\rlap{73} & (2); \cite{Walborn2014,Evans2015b,Evans2015a};\\
& & & \cite{Lamb2016, Grin2017};\\
& & &  \cite{RomDuv2019};\\
& & & \cite{Jones2020}\\
\llap{R}GB  & 5\rlap{19} & \mbox{~~}4\rlap{89} & (2); \cite{Cole2005,Neugent2020}\\
& & & \cite{Parisi2009,Parisi2010,Parisi2022};\\
& & & \cite{debortoli2022} \\
\llap{H}\,\textsc{ii}/YS\rlap{Os} & \mbox{~~}\rlap{86} & \mbox{~~}4\rlap{59} & (2);\cite{Seale2009,Oliveira2011,Oliveira2013};\\
& & & \cite{Oliveira2019,vanGelder2020} \\
\llap{P}Ne  & \mbox{~~}\rlap{53} & \mbox{~~~~}\rlap{50} & (2); \cite{Shaw2001}\\
\llap{A}GB   & 1\rlap{65} & \mbox{~~}2\rlap{21} & (2); \cite{vanLoon1998};\\
 & & & \cite{Groenewegen1998};\\
 &&& \cite{vanLoon1999a,vanLoon1999b,vanLoon2005,vanLoon2006,vanLoon2008};\\
 &&& \cite{Kamath2014}\\
\llap{R}SG   & \mbox{~~}\rlap{44} & \mbox{~~~~}\rlap{70} & (2); \cite{Neugent2020} \\
\llap{p}AGB/\rlap{RGB}   & \mbox{~~}\rlap{46} & \mbox{~~~~}\rlap{33} & (2); \cite{vanLoon2008,Kamath2014} \\ 
\llap{P}M  & \mbox{~~}\rlap{78} & \mbox{~~~}3\rlap{03} & (3)\\
\hline
\end{tabular}
\label{classtable}
\end{table}

\subsubsection{SAAO 1.9m}

We observed 174 new optical spectra (see Appendix Section \ref{specsec} for full list) at the South African Astronomical Observatory (SAAO) 1.9-m telescope with SpUpNIC \citep[Spectrograph Upgrade: Newly Improved Cassegrain;][]{2019JATIS...5b4007C} during observing runs in 2019 and 2021. Grating 7 (grating angle of $16^\circ$) and the order blocking ‘BG38’ filter were used, delivering a resolving power $R = \frac{\lambda}{\Delta\lambda}\sim$ 500 over a wavelength range of 3800 \AA\ -- 9000 \AA. Dome-flats and bias images were taken at the beginning of each night. The CuAr lamp was used for wavelength calibration. Three 600 s (300 s for sources brighter than $\sim$ 16 mag) exposures were obtained for each source. The standard stars \citep[EG 21, Feige 110 or LTT 1020;][]{1994PASP..106..566H} were observed on the same night under the same conditions for 30 s. 

The data was processed using the standard IRAF\footnote{IRAF is distributed by the National Optical Astronomy Observatory, which is operated by the Association of Universities for Research in Astronomy, Inc., under cooperative agreement with the National Science Foundation.} tools \citep{1986SPIE..627..733T,1993ASPC...52..173T}. 

The sources that we observed with the 1.9m telescope, and have been classified based on their optical spectroscopy, were added to the training set. This added 26 sources (18 AGN, 7 galaxies, 1 H\textsc{ii}/YSO).

\subsubsection{SALT}

We also observed 40 sources with the Southern African Large Telescope (SALT) \citep{2006SPIE.6267E..0ZB}, located in Sutherland, South Africa that has an effective diameter of 7 -- 9 m. SALT was used to observe AGN candidates that had the potential to be similar to SAGE0536AGN (\citealt{2022MNRAS.515.6046P}, Pennock et al. \textit{in prep.}). The Robert Stobie Spectrograph \citep[RSS;][]{2003SPIE.4841.1463B, 2003SPIE.4841.1634K} was used, a combination of three CCD detectors with total 3172 $\times$ 2052 pixels and spatial resolution of $0\rlap{.}^{\prime\prime}1267$ per pixel. We used the long-slit with width $1\rlap{.}^{\prime\prime}5$ or $1\rlap{.}^{\prime\prime}25$, grating PG0300 or PG0900 and an Argon arc lamp. 
Initial processing (basic CCD data reductions) was done automatically by the SALT pipeline \citep{SALTpipeline}. We processed these data by performing cosmic ray removal, wavelength calibration and source extraction also using also using the standard IRAF tools \citep{1986SPIE..627..733T,1993ASPC...52..173T}.

The sources we observed with SALT that have been classified were added to the training set. This numbered 22 sources: 1 AGB, 1 H\textsc{ii}/YSO, 1 galaxy and 19 AGN.

\subsubsection{SAGE-spec}

The Infrared Spectrograph onboard the \textit{Spitzer} Space Telescope was used to observe the LMC and SMC in low and high resolution modes for the wavelength range of 5 -- 38 $\mu$m. The resolving power varies between  60 -- 130 for low resolution mode whereas high resolution mode has resolving power of $\sim$ 600.

All the spectra obtained by \textit{Spitzer} within the SAGE footprint were looked at as part of the SAGE-Spec project. In the SMC \citep{Ruffle2015}, this survey found 58 AGB stars, 51 YSOs, 4 post-AGB objects, 22 RSGs, 27  undefined stars (of which 23 are dusty OB stars), 24 PNe, 10 Wolf-Rayet (WR) stars, 3 H\,\textsc{ii} regions, 3 R Coron\ae\ Borealis (R CrB) stars, 1 blue supergiant and six other objects.

In the LMC \citep{Jones2017}, this survey observed $\sim$ 800 sources, the majority of which are YSO and H\,\textsc{ii} regions and (post-)AGB stars, PNe and massive stars. Also observed were two SNRs, a nova and several background galaxies. 

\subsubsection{Extragalactic classes}

There are 657 spectroscopically observed AGN from the Milliquas catalogue of \citet{Flesch2019cat,Flesch2019pap} in the field of the VMC footprint of the LMC. The largest contributions are from \cite{Kozlowski2012,Kozlowski2013}, \cite{Geha2003}, \cite{Esquej2013} and \cite{Ivanov2016}, contributing 547, 24, 23 and 10 objects, respectively.

There are 240 spectroscopically observed AGN from the Milliquas catalogue in the field of the VMC footprint of the SMC. The largest contributions are from \cite{Kozlowski2011,Kozlowski2013} and \cite{Ivanov2016}, contributing 194 and 10 objects, respectively.

Galaxies (with no signs that they are hosting AGN) were taken from the 6dFGS survey \citep{Jones2009}. The observations for this survey were carried out using the Six Degree Field (6dF) fibre-fed multi-object spectrograph at the UK Schmidt Telescope (UKST; Siding Spring Observatory, Australia) over 2001 May to 2006 January \citep{6dfgs_v1}. Target fields covered $\sim$ 17 000 deg$^2$ of the southern sky more than 10$^{\circ}$ from the Galactic Plane. This survey data however comes with the caveat that it is limited to the brightest (it is complete to total extrapolated 2MASS magnitude limits of 13.75, 12.95, 12.65 mag for $J$, $H$ and $K$, respectively)  and closest (median redshift of whole survey is $z\sim 0.053$) of galaxies. 

To further augment the extragalactic sample, galaxies and AGN were added from the Galaxy and Mass Assembly \citep[GAMA][]{2011GAMA,2013GAMAspec} survey, specifically the GAMA09 region, which was also observed with VISTA as part of the VIKING survey \citep{2017VIKING}\footnote{Based on observations made with ESO Telescopes at the La Silla Paranal Observatory under programme ID 179.A-2004.}. The SDSS DR16 survey \citep{2017SDSS, 2020SDSS} covered this region, and spectroscopically observed extragalactic sources and further separated them into galaxies and AGN, adding 8504 and 2337 sources, respectively. For these sources \textit{Sptizer} IRAC band and \textit{Herschel} 250 $um$ were left as missing data.

\subsubsection{Galactic and Magellanic classes}

For the training sample, sources that are often mistaken for AGN were needed, whilst also including other sources that are more distinct from AGN that are prevalent throughout the Magellanic Clouds. There have been many spectroscopic surveys of the LMC and SMC, generally looking at specific types of stellar objects that can be found within the Magellanic Clouds. This led to the accumulation of 1490 OB stars, 1008 RGB stars, 545 YSO or compact H\,\textsc{ii} regions, 386 AGB stars, 114 RSGs, 103 PNe, 79 pAGB/pRGB and 382 high proper-motion/foreground stars (see Table \ref{classtable} for details). A large part of these sources were from the SAGEspec surveys \citep{Ruffle2015,Jones2017}, which observed 209 and 862 sources in the direction of the SMC and LMC, respectively.

The more common stars (e.g. main-sequence M-type stars 2700--3800K) however tend not to be spectroscopically observed as they are an already well observed/studied class of objects (in the Milky Way), whereas spectroscopic studies preferentially target (and confirm the identities of) the rarer less well-studied stellar classes, therefore leading to a lack of main-sequence stars for training. They are however distinct from AGN, with a lack of emission in the IR, so should not be mistaken for extragalactic sources, and be preferentially associated with the stellar sources based on proper motions and colours.

The sources with high-proper motion and that are in the foreground were identified using a Simbad \citep{Simbad} search for`PM' stars in the VMC footprint that had a listed spectral type and reference given. This yielded 78 and 303 sources for the SMC and LMC, respectively, with a cross-match with a source in the VMC catalogue with a 1$^{\prime\prime}$ search radius.

\subsubsection{Unknown class}

Not all classes can be accounted for, as some classes have too few spectroscopically observed sources to create a robust training sample from, and others are simply unknown. However, all the sources must be classified as one of the classes it has been trained upon, which would inevitably lead to contamination within each class. Therefore, to ensure that these sources are not classified incorrectly, an Unknown class is created for the LMC and SMC. This was done by randomly selecting a number of sources (same amount as the largest training set, galaxies, at 9118) from the VMC catalogues for both the LMC and SMC. This creates a sample of sources that have no structure in feature space as it includes a mix of everything. This should allow the clearly defined classes in feature space (collection of features, in this case photometry, that are used to characterise the different classes) to be classed correctly whilst setting the sources that lay away from the known classes to be set as Unknown. Note that none of the spectroscopically observed sources are in this class. 

We found that creating this class was necessary to ensure that faint and/or difficult to classify sources were not (erroneously) allocated to one of the other classes by the machine learning algorithm (i.e. that this class is needed to fully capture our remaining uncertainty/ignorance).

\section{Probabilistic Random Forest}\label{PRF}

A random forest \citep{breiman2001} is a supervised machine learning algorithm that can be used for both classification and regression problems. The algorithm builds several decision trees (a decision tree consists of a series of nodes where at each node a condition is given that is either true or false) independently and then averages the predictions of these to obtain the final prediction, as well as the probability the prediction is correct from the fraction of trees that agree with the final prediction. This reduces variance over using a single estimator and creates an overall more stable model. It is called a random forest because randomness is injected into the training process of each individual tree via a method called ‘bagging'. This method splits up the training set into randomly selected subsets, and each decision tree is then trained on one of those subsets. Furthermore, at each node of the decision tree, only a randomly selected subset of the features is considered.

The Probabilistic Random Forest\footnote{Python code can be found here: \url{https://github.com/ireis/PRF}} \citep[PRF;][]{reisprob} is a random forest algorithm that can handle and take into account measurement uncertainties and missing data (where data is missing that is required for a condition at a node, the PRF will propagate down both true and false paths with equal weight given to the probability that either path is correct). Compared to an ordinary random forest it has been proven to provide an up to $10\%$ increase in classification accuracy with noisy features and proven to be more accurate than the original random forest when up to $45\%$ objects in the training set are misclassified. 

This algorithm was created in Python and requires the Python module \textsc{Scikit-learn} \citep{scikit-learn} to run.

\subsection{Creating the multi-wavelength dataset}\label{createTS}

The base of the multi-wavelength data set is the near-IR VMC PSF survey catalogue. All coordinate matchings were made to the VMC coordinates. We matched the VMC catalogue with SMASH \citep{Nidever2017}, \textit{Gaia} DR3 \citep{GaiaDR3}, SAGE \citep{Meixner2006,gordon2011}, unWISE \citep{UnWISE} and AllWISE \citep{Cutri2013} using \textsc{Topcat} \citep{Taylor2005}. The parameters of different surveys in the dataset can be seen in Table \ref{table:PRFparam}. The cross-matchings between all catalogues were done with a 1$^{\prime\prime}$ search radius. Note that the parameters from X-ray and radio surveys have not been used as features, and will be used as an independent check (e.g., they should favour extragalactic sources) for the classifications. 

\begin{table}
\caption[Parameters taken from various surveys to act as features in the PRF algorithm.]{Parameters taken from various surveys to act as features in the PRF algorithm. For each parameter the error on the values is also taken from the corresponding surveys. New features were created by subtracting each feature from all the other features to create colours. This did not include the sharpness (a measure of the difference between the observed width of the object and the width of the PSF model, where stars should have a sharpness value of $\sim$0 and resolved objects values of $>$0. Sharpness values $<$ 0 indicate artefacts such as bad pixels or cosmic ray impacts) and proper motions in RA and DEC (pmRA and pmDEC, respectively).}
\vspace{5pt}
\begin{tabular}{|l|l|l|}
        \hline\hline
        Parameter & Units & Survey \\
        \hline
        $Y$ PSF & mags (Vega) & VMC \\
        $J$ PSF & mags (Vega) & VMC \\
        $K$\textsubscript{s} PSF & mags (Vega) & VMC \\
        $Y$ sharp PSF & -- & VMC \\
        $J$ sharp PSF & -- & VMC \\
        $K$\textsubscript{s} sharp PSF & -- & VMC \\
        $u$ & mags (Vega) & SMASH \\
        $g$ & mags (Vega) & SMASH \\
        $r$ & mags (Vega) & SMASH \\
        $i$ & mags (Vega) & SMASH \\
        $z$ & mags (Vega) & SMASH \\
        sharp & -- & SMASH \\
        pmRA & mas/yr & Gaia DR3 \\
        pmDEC & mas/yr & Gaia DR3 \\
        $G$ & mags (Vega) & Gaia DR3 \\
        $G$\textsubscript{BP} & mags (Vega) & Gaia DR3 \\
        $G$\textsubscript{RP} & mags (Vega) & Gaia DR3 \\
        IRAC 3.6 $\mu$m & mags (Vega) & SAGE \\
        IRAC 4.5 $\mu$m & mags (Vega) & SAGE \\
        IRAC 5.8 $\mu$m & mags (Vega) & SAGE \\
        IRAC 8.0 $\mu$m & mags (Vega) & SAGE \\
        unW1 & mags (Vega) & unWISE \\
        unW2 & mags (Vega) & unWISE \\
        W1 & mags (Vega) & AllWISE \\
        W2 & mags (Vega) & AllWISE \\
        W3 & mags (Vega) & AllWISE \\
        W4 & mags (Vega) & AllWISE \\
        SPIRE PSW 250 $\mu$m & MJy & HERITAGE \\
        \hline
\end{tabular}
\label{table:PRFparam}
\end{table}

Some of the parameters required calculation. For instance, the unWISE catalogue only provided fluxes rather than Vega magnitudes like the rest of the catalogues. For consistency the fluxes were converted using the method recommended in the notes of the table on CDS\footnote{\url{https://cds.unistra.fr}} (Centre de Données astronomiques). The fluxes in Vega nanomaggies \citep[nMgy; ][]{2004AJ....128.2577F} were converted to Vega magnitudes using $m = 22.5 - 2.5\log(flux)$. These fluxes showed slight discrepancies with the AllWISE values and a correction was applied \citep[see][]{UnWISE} of subtracting 0.004 mag and 0.032 mag from unWISE W1 and unWISE W2, respectively. The differences in the values of W1 and W2 bands between the AllWISE and unWISE catalogues is, as expected, centred around 0. Differences beyond this could be explained by variability. Where there are no differences, this should not affect the classifications as no new information is being provided, so using the feature again, once from AllWISE and then once from unWISE, would make no difference in splitting up the data. 

Other parameters that had to be calculated were colours between all photometry bands and their corresponding errors, which were calculated with standard propagation of errors. Note that if, for example, $Y - J$ was calculated, the reverse, $J - Y$, would not be calculated and added as a feature. This led to a total of 237 features.

The far-IR measurement is taken from the SPIRE PSW 250 $\mu$m images of the SMC and LMC that were taken with \textit{Herschel}, that are first smoothed with a 2D box kernel across ten pixels. After smoothing the image, the flux values were taken from the image at each co-ordinate, where the median within a 5$^{\prime\prime}$ radius was taken. This provides a measurement of the background flux instead of the individual source flux, where the flux is expected to be higher when looking through the Magellanic Clouds.

Parameters which were not observed for a source in a given survey were assigned null values within their respective survey databases, which differs between surveys. For example, the VMC applies the large negative value of $-$9.99999 x 10$^8$, whereas SMASH represents a null value as 99. It is also commonplace to leave null parameters unassigned or to assign a value of 0. Therefore, these data must be homogenised prior to the implementation of the PRF. To achieve this, we assigned the standard null value, “$NaN$", to all null values across our input data, regardless of their origin.

\subsection{Extinction}\label{extinction}
The effect of extinction from the Magellanic Clouds \citep[e.g.,][]{Bell2019, Bell2020, Bell2022} is a non-explicit factor that is included in the photometry/colours that help to separate the stellar from the extragalactic. Galactic extinction from the Milky Way \citep[e.g.,][]{1998Schlegel, 2011Schlafly}, though minor in this case compared to the Clouds, is also a factor that needs to be taken into consideration. 

However, it is not straightforward to correct for LMC/SMC extinction, because it is unknown in advance which objects are extragalactic and which belong to the LMC or SMC. The idea of this classifier is that it can take the raw photometric data, with examples of sources from across the Clouds affected by different amounts of extinction, and learn from this. This is why it is important to have spectroscopically classified extragalactic sources right across the Clouds, so that the PRF is appropriately trained to recognise such populations even in cases of substantial foreground reddening.

As shown in Section \ref{FI} and Figure \ref{featureimport}, the far-IR HERITAGE SPIRE PSW 250 $\mu$m average flux density at each source position is shown to be the most important feature for the classifier. This band traces the emission from cold dust across the Clouds (responsible for the extinction of other bands) and provides our classifier with information related to the reddening that is likely being used to aid the classification of different source classes, including extragalactic sources that lie behind the central regions of the Clouds. Furthermore, a higher average flux density would tend to be found in areas of high star-formation (found in the centre of the Magellanic Clouds), therefore the far-IR feature would most likely bias the classifier positively towards young stellar populations, and negatively bias against background galaxies and AGN.


\subsection{Training}
In this section we discuss the various aspects of the training of the classifiers, including the configuration of the input training set and the PRF parameter choices.

\subsubsection{Inputs}
The data were arranged in three configurations, individual ‘LMC' and ‘SMC' datasets, as well as a further ‘MC' (LMC and SMC) data set. Each configuration had two versions, one with no colour features and another with colour features. Individual classifiers are trained for the SMC and LMC due the different stellar populations, population histories, extinction distributions and metallicities \citep[e.g.,][]{Rubele2012, 2014Rezaeikh, Rubele2018, Bell2019, Bell2020, Bell2022} between the two Clouds. Therefore, the Magellanic classes were specifically trained for their respective galaxy, but, the same extragalactic and foreground stars training sets were used for training both classifiers.

For training the classifiers, datasets were split into features, $X$, errors on features, $dX$, and class, $y$. 

To ascertain the accuracies of the trained classifiers each dataset was split into training and testing sets, where 75$\%$ of the data were trained on and 25$\%$ were retained to test the classifier on. For each of the training runs the data split was randomised. Note that when testing a machine learning model a training dataset is often split into training, validation and test set. A validation set is used to tune the parameters of the model, whilst a test set is used to test the final model. Both these datasets are not trained upon. In the interest of not splitting the different sets into too small groupings, and therefore not providing a good overview of how well the trained classifier works, we combined the validation and test sets together into an overall ‘test' set. The configuration of the training and test sets were then randomised over multiple runs of tuning and testing the classifier, so that the classifier is not overfitting to one training set.

\subsubsection{Probability threshold parameter}

The probabilistic random forest classifier has parameters that can be varied. Most parameters are set to default. 


The probability threshold parameter, $p_{\rm th}$, determines the probability threshold at which to stop propagating along a branch. In an ideal PRF, $p_{\rm th} = 0$, where all objects propagate along all branches to all terminal nodes, unlike $p_{\rm th} = 1$  which denotes a classical RF, where each object propagates to only one terminal node. The former requires a higher amount of computation time, and for any given object there may be nodes with small propagation probability. Stopping the propagation at these nodes reduces the run time without decreasing overall performance. \cite{reisprob} found that reducing the probability threshold below a value of $p_{\rm th} = 0.05$ does not significantly improve the prediction accuracy, and only increases computation time, so we used this value in our work.

\subsubsection{Number of trees parameter}\label{PRFtreetest}

To determine the optimum number of trees for the PRF, for each of the datasets, the dataset was split into training and testing and then the classifier was trained on this at $n_{\rm trees} = 1, 5, 10, 25, 50, 100, 200, 500$, and then the score (fraction of correct classifications when the classifier is used on the test set) was calculated. It should be noted that since the extragalactic sources dominate the test set, the score therefore is dominated by the extragalactic accuracy, and the score is not representative of all classes. Next the whole dataset was split randomly again and then the classifier was trained again on the different number of trees. This is done for five iterations for each $n_{\rm trees}$ and then the score is averaged for each value of $n_{\rm trees}$.


From this it was seen that for all dataset configurations after $n_{\rm trees} > 100$ the score for each classifier plateaus. Therefore the value of the number of trees was set to $n_{\rm trees} = 100$. This was done for training sets with and without colour information and it was found that the overall accuracy/score is greater when colour features are included. It was also found that the classifier trained and tested on the `MC' training dataset was found to have a lower accuracy than the classifiers trained and tested on the individual `LMC' and `SMC' training datasets.

\subsubsection{Balanced vs imbalanced datasets}
A balanced training set would have all classes roughly equal in size. An imbalanced training set would have large differences between class sample sizes. This imbalance can cause a poor predictive performance for the minority classes \citep[e.g.][]{2007Khosh, 2017MoreRana} as most machine learning algorithms operate under the assumption of an equal sample size for each class.

Ensemble methods such as random forests can mitigate the effects of imbalanced datasets by training each tree on an independently randomly selected subset of the training set and then combining the results of all the trees together.
However, for extremely imbalanced datasets, when randomly selecting a subset to train a tree on, if a minority class is too small then only a few or even none at all of the minority class may be selected for a particular tree, meaning there will be trees that have not seen the minority class at all, so will not know how to classify them. This effect can be mitigated by balancing the dataset.

Balancing the dataset can either be done by downsampling, which reduces all the class sizes to the smallest class size, or upsampling, which increases/augments the minority class with synthetic data so that all the class sizes are the same as the largest class size. Downsampling works well if spread in parameter space is preserved \citep[such as in][]{Kinson2021,Kinson2022}, however, comes with the caveat of potentially losing important information if the training sets are heavily imbalanced and therefore will not be used here. Upsampling maintains the same amount of information (though with the possible caveat of overfitting due to replication of non-relevant features) so this strategy was adopted. See Appendix Section \ref{sampling} for the comparison between not balancing, downsampling and upsampling the training set effects the classifier's precision and recall.

Upsampling can be done in one of two ways. Either by using machine learning on the minority class to generate synthetic data points based on the real data of the minority class sample; or by randomly copying objects from the minority class sample to increase the sample size. The latter method was used as it maintains that only real data is used whilst balancing the dataset so that each tree will randomly sample sources from each class. To do this the `resample' function of Python's \textsc{Scikit-learn} module was used to upsample all the class samples to the same size as the majority class, so that all classes have an equally sized training set. 

The SMC and LMC training samples were upsampled to the size of the galaxy/Unknown class. This was only done after the training sample was split into training and test sets, and only on the training sets. This process was so that the same objects did not end up in both the training and test set. This was trained and tested three times and the results averaged. From this we found that the addition of the upsampling overall increased the number of confident (P$_{\rm class}$ $>$ 80\%) correct classifications, especially for those with the smaller training sets. 

Note that an imbalance in the dataset can also be caused by bias within the classes themselves. In this case, it would be selection bias of the sources being bright enough in the optical to be spectroscopically observed. We do not address this issue in this paper as it is beyond the scope of our work and is due to limiting to the spectroscopic data that are already in hand. It is however, somewhat mitigated by the inclusion of the Unknown class, which was found to decrease the number of confident wrong predictions. One way this could be improved upon is by finding suitable, brighter targets in the Milky Way for some classes, for example, RGB stars of SMC metallicity in the globular cluster 47 Tucanae. Another is by spectroscopically observing more sources that are not necessarily bright enough in the optical, but perhaps instead in the IR, with a telescope such as the recent \textit{James Webb Space Telescope} \citep[JWST;][]{2006JWST}.

\subsubsection{Final data configuration}

The overall accuracy does not tell one how the classifier performs on individual classes, this can be shown through the use of confusion matrices. A confusion matrix shows the comparison between the true labels (y-axis) vs the predicted labels (x-axis). A perfect classifier would show a value of 1 (100\%) in a diagonal line from top left to bottom right of the confusion matrix, which represents the recall (the ratio of  $\frac{tp}{tp + fn}$, where $tp$ is the number of true positives and $fn$ is the number of false negatives) of each class, whilst all the other values would be 0 (0\%).  This would show that all classes have been predicted correctly. The values for an entire row should sum to 1, showing the distribution of class predictions for each class. 

\begin{figure*}
\centering
\includegraphics[width=0.95\textwidth, trim=0 0 4mm 2mm, clip]{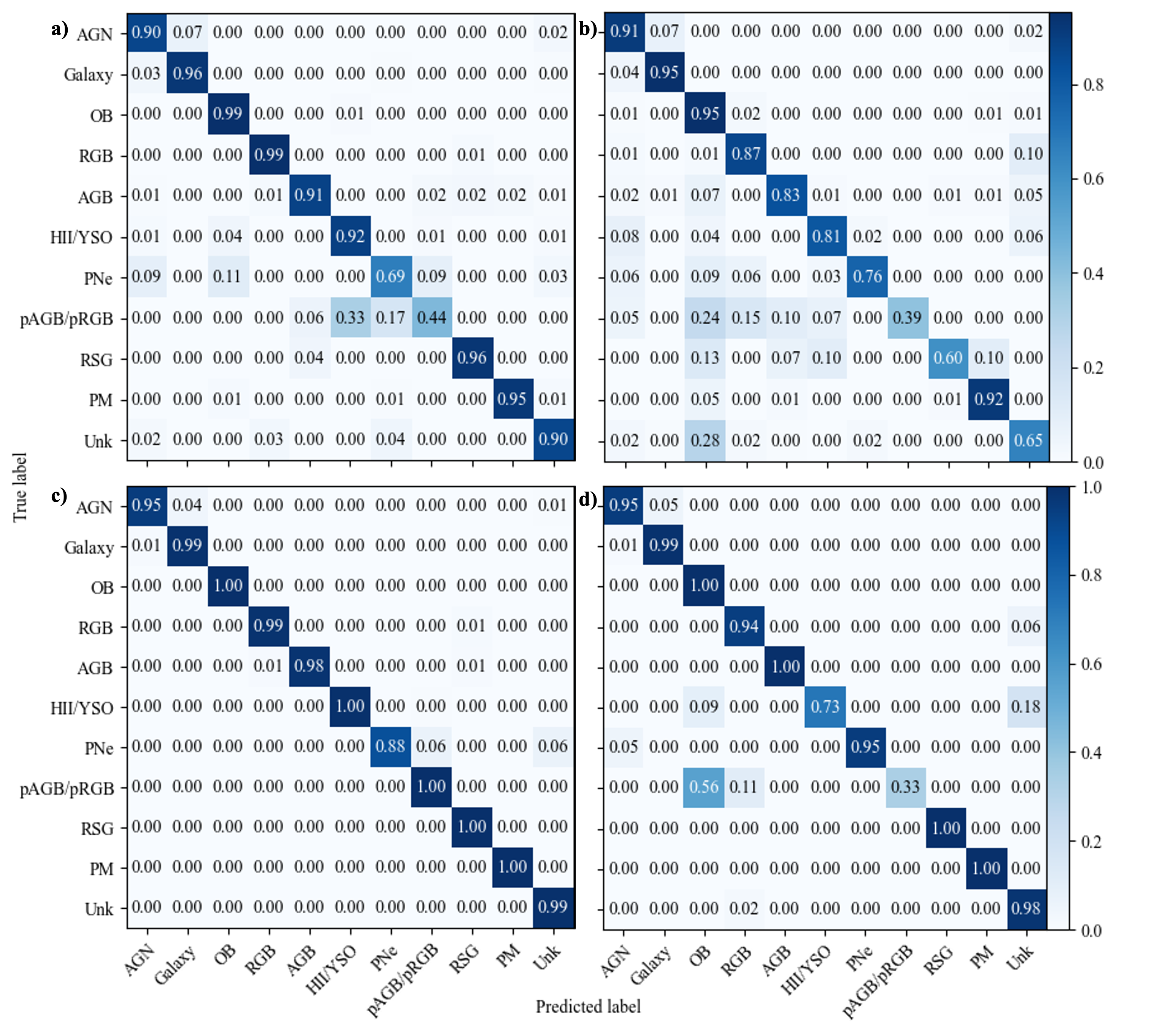}
\caption{Confusion matrix of final classifier trained and tested on the LMC (left) and SMC (right) datasets. The top panels, a) and b), show all sources, whilst the bottom panels, c) and d), are restricted to sources that were classified with probabilities $>$ 80\%. This dataset configuration includes extragalactic and foreground sources from both the LMC and SMC, but Magellanic stellar sources only from the respective Clouds. }
\label{MCfinalCM}
\end{figure*}

The final dataset configuration to be trained upon was for separate classifiers for the LMC and SMC, where both datasets will share extragalactic sources from both regions, whilst keeping stellar sources specific to the Cloud they are from. The PRF has 100 trees. As the PRF randomly selects a subset of sources from the training set to train from for each decision tree, each run of the classifier creates different trees. How it randomises the selection can be locked in by setting a `random seed state'. The classifier was trained and tested on the dataset ten times, using ten different random seed states, to create ten confusion matrices. The values were then averaged to create an ‘average' confusion matrix. The confusion matrix for the SMC-trained classifier tested on SMC data can be found in Figure \ref{MCfinalCM} (right panels), and the confusion matrix for the LMC-trained classifier tested on LMC data can be found in Figure \ref{MCfinalCM} (left panels).

The overall accuracy (score) of the SMC classifier is found to be 0.79 $\pm$ 0.01, and the overall accuracy of the LMC classifier is found to be 0.87 $\pm$ 0.01. For both classifiers the AGN class has one of the highest recall, $\sim$ 90\% of all AGN in the test set are classified correctly for the SMC and LMC. For the AGN misclassified as other sources, they are most often misclassified as galaxies, which is not unexpected.  
The precision (the ratio of  $\frac{tp}{tp + fp}$, where $fp$ is the number of false positives) of the AGN class is not as great, as other sources are misclassified as AGN. For the LMC sightlines $\sim$ 3\% of galaxies and $\sim$ 9\% of PNe are misclassified as AGN. However, for the SMC sightlines, $\sim$ 6\% of PNe, $\sim$ 8\% of H\textsc{ii}/YSOs, $\sim$ 5\% of pAGB/pRGB, $\sim$ 4\% of galaxies,$\sim$ 2\% of AGB and $\sim$ 1\% of RGB and OB are misclassified as AGN. Overall, most AGN will be classified correctly, with some expected confusion between AGN and galaxies, which is not unexpected as AGN are hosted in galaxies with varying levels of obscuration and luminosity of the AGN emission, making it hard to discern a heavily obscured or low luminosity AGN from a galaxy with no AGN. There will, however, be some stellar interlopers, most often PNe. Though, when limited to only the high confidence sources ($>$ 80\% probability of class being correct), as seen in the bottom panels of Figure \ref{MCfinalCM}, we see that the precision is improved as only $\sim$1\% of galaxies are misclassified as AGN for both SMC and LMC classifiers, and only $\sim$5\% of PNe are misclassified as AGN for the SMC classifier.

The recall of the post-AGB/RGB class is the worst (39\% and 44\% for the SMC and LMC, respectively), though when restricting to high confidence sources it is then PNe that have the worst recall ($\sim$88\%) for the LMC classifier. For the SMC and LMC, post-AGB/RGB stars are mostly misclassified as other stellar sources, with only $\sim$5\% ($<$1\% for high confidence sources) misclassified as AGN for the SMC, and $<$1\% for the LMC. It is not surprising that post-AGB/RGB have the lowest recall, since they are intrinsically one of the rarest source populations in our fields and thus our spectroscopic sample is also limited to a small number of examples (46 and 33 sources in the SMC and LMC respectively, with only 75\% of these used for training) and is likely a biased sample that does not  accurately probe the full range of source properties for this class. With the upsampling, it is possible that the classifier has been overfit to this class.

\begin{figure*}
\centering
\includegraphics[width=0.80\textwidth, trim=1mm 1mm 1mm 2mm, clip]{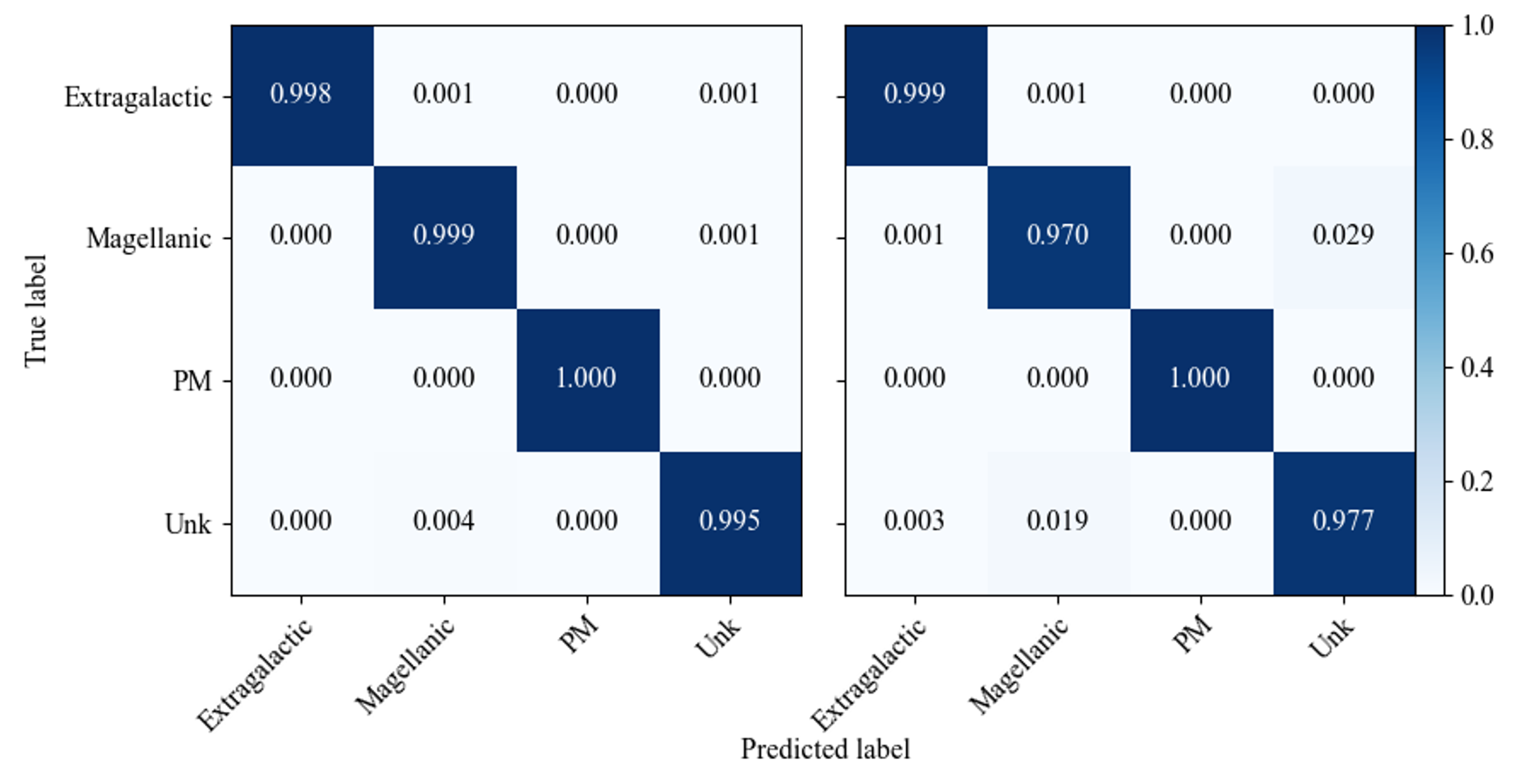}
\caption{Simplified confusion matrix of final classifier trained and tested on the LMC (left) and SMC (right) datasets, restricted to sources that were classified with probabilities $>$ 80\%. The AGN and galaxy classes have been combined into the `Extragalactic' label. The Magellanic stellar classifications have been combined into the `Magellanic' label.}
\label{MCfinalMiniCM}
\end{figure*}

Simplifying the confusion matrix by combining the classes into extragalactic, Magellanic, PM and Unknown, as seen in Figure \ref{MCfinalMiniCM}, shows that, when restricting to probabilities $>$ 80\%, the classifier is working well. 99.8\% and 99.9\% of all extragalactic sources are predicted as extragalactic, for the LMC and SMC fields, respectively. These confusion matrices show that most of the misclassification occurring is not between extragalactic and Magellanic classes, but within them, showing that in instances where the classifiers do not obtain the correct class, it will likely classify it as either within or outside the Clouds correctly.

The misclassification of stellar sources as AGN, and AGN as stellar sources, reflects what has been found anecdotally in the Magellanic Clouds. Classifications based on photometry have led to stars masquerading as AGN and vice versa in the Magellanic Clouds, such as SAGE0536AGN \citep{Hony2011, vanLoon2015} and SAGE0534AGN \citep[][]{2022MNRAS.515.6046P}, two AGN which were first thought to be evolved stars in the LMC, and Source 5 and Source 8 from the study of a small sample of AGN in \cite{2022MNRAS.515.6046P}, which were revealed to be stars in the SMC instead. These sources are within the datasets to be trained upon, which increases the likelihood the PRF would classify similar sources correctly, but it is possible their small number might not be enough. 

Overfitting of a machine learning model can generally be spotted by using the classifier on the training set, and if the performance is much better than on the test set, then the model is overfitting. For the SMC classifier used on the SMC training set, the average accuracy was 0.92 $\pm$ 0.01, and for the LMC classifier used on the LMC training set the average accuracy was 0.88 $\pm$ 0.01. Both classifiers only performed marginally better on the training sets compared to the test sets (0.90 $\pm$ 0.01 for the SMC and 0.87 $\pm$ 0.01 for the LMC), meaning the machine learning model is not overfitting.

\subsubsection{Feature importance}\label{FI}
The PRF algorithm can calculate overall feature importance for the entire classifier. This level of importance is calculated as “mean decrease impurity", which is defined as the total decrease in node impurity (weighted by the probability of reaching that node, which is approximated by the proportion of samples reaching that node), averaged over all trees over the ensemble \citep{breiman2001}. In other words, how well each feature separates the sample into the expected classes (the decrease in class impurity). The values of the feature importances are then normalised, such that they all sum to one. 

\begin{figure}
\centering
\includegraphics[width=0.8\columnwidth, trim=1mm 1mm 1mm 0.5mm, clip]{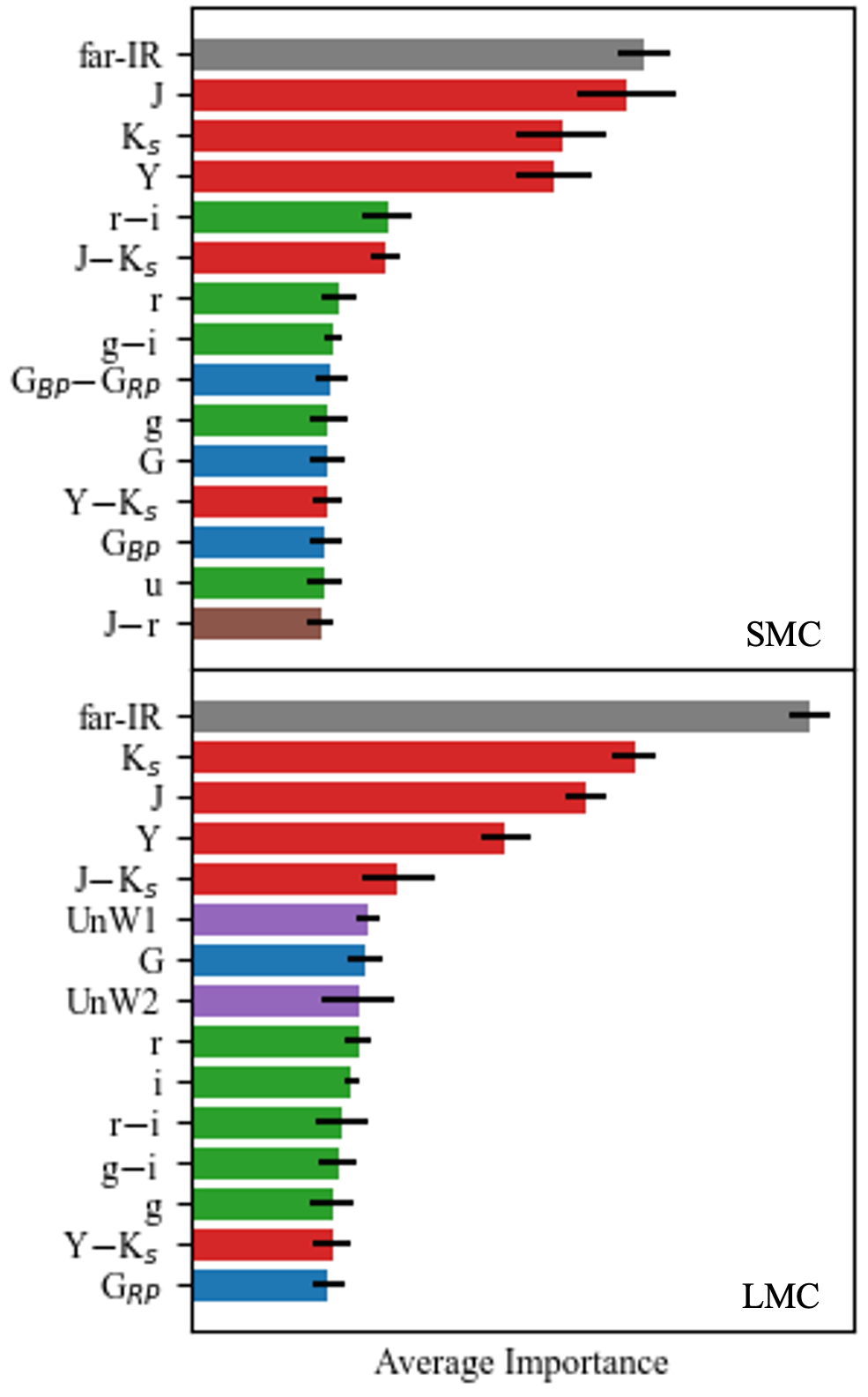}
\caption[Top 15 important features for the SMC and LMC classifier.]{Top 15 important features for the SMC (left) and LMC (right) classifier. Note that, while the top 15 features are in a different order, they are almost all the same features for both the SMC and LMC classifier. The bars are colour-coded to represent the survey origins, where green is SMASH, blue is \textit{Gaia}, red is VMC, purple is unWISE, grey is HERITAGE and brown represents a colour calculated from bands in different surveys.}
\label{featureimport}
\end{figure}

The classifiers are trained on the full datasets for SMC and LMC, from which the feature importances were calculated. This was then repeated ten times and the feature importances were then averaged for each feature. The top 15 ranked feature importances for both the SMC and LMC classifiers can be seen in Figure \ref{featureimport}, the full list of feature importances are available as a data product alongside the paper. For both the SMC and LMC classifiers the top 15, whilst in a different order, only have three features not in common. The feature importance plateaus with a slight decline after this towards the least important features. The full list of features and importances is available as online supplementary material. Note that feature importances are only calculated for all classes, and not individual classes.

VMC photometry and colours rank high amongst the feature importance, most likely due to all sources having at least one observation in the $YJK$\textsubscript{s} bands. However, it is unlikely that is the sole reason for their high importance, therefore the colours and photometry are providing good distinction between the sources as well, showing the near-IR is a powerful resource in separating different classes.

Far-IR background emission is either the top feature or close to it. As discussed in \ref{extinction} this is most likely due to far-IR being used as an analogue of extinction, which would affect the photometry in bluer bands for where far-IR flux density is higher. Also, far-IR is an indicator of how close to the centre of each of the Clouds a source is (higher far-IR nearer the centre), where stellar density is higher closer to the centre and thus a source is more likely to be a star rather than a background extragalactic object. Away from the Clouds, at the edges of the survey area, there is little to no far-IR emission from the Clouds and a source is more likely to be a background extragalactic object rather than a star. 

Because all of the sources used for training have classifications based on spectroscopy, this means that they tend to be bright enough to be observed with \textit{Gaia}, hence why the \textit{Gaia} colours and photometry rank quite highly. 
However, despite this high dependence on \textit{Gaia} photometry and colours, the proper motions in RA and DEC do not rank nearly as high, at ranks 85 (0.0027 $\pm$ 0.0002) and 67 (0.0036 $\pm$ 0.0004) for proper motion in RA for the SMC and LMC, respectively, and at ranks 48 (0.0046 $\pm$ 0.0004) and 110 (0.0017 $\pm$ 0.0001) for proper motion in DEC for the SMC and LMC, respectively. This is unexpected since proper motions would be the most obvious way of separating the high proper-motion stars and extragalactic sources from the Magellanic stellar sources. This could be due to the Clouds lying at the very limits of the usability of the \textit{Gaia} data \citep[e.g.][]{2018Vasiliev, 2021A&A...649A...7G} making proper motions have significant uncertainties at the distances of the Clouds, especially for the fainter sources ($G$ $>$ 18 mag). It should be noted that an increased uncertainty in a feature tends to lead to a decrease in class probability, especially for the more important features. 

The worst features are most likely due to an abundance of missing values for these features, brought upon by either a lack of coverage in certain areas and/or a lack of depth, such as for colours based on SMASH and AllWISE photometry (e.g. any SMASH photometry $-$ any AllWISE photometry), which are the least important features for both classifiers. Leaving these features in should not affect the accuracy of the classifiers as they have been deemed unimportant, and therefore unlikely to be relied upon to make a classification.


\section{Results}\label{Results}


The full VMC dataset for the SMC and LMC consists of 29,514,739 and 103,172,194 sources, respectively, where, of the sources not classed as Unknown, $\sim$ 9\% (SMC) and $\sim$ 6\% (LMC) are classed as extragalactic. Table \ref{PRFcounts} shows the distribution of classes for the entire SMC and LMC fields, as well as for sources with class probabilities (P$_{\rm class}$) $>$ 60\% and $>$ 80\%. This shows that the majority of sources (that are not classed as Unknown) in the SMC and LMC fields are classified as stars, as expected. OB and RGB stars have the highest number, which could be caused by other stellar sources that were not trained upon being classed as these classes, such as bluer stars for the OB class and redder stars for the RGB class. The majority of extragalactic sources are expected to be galaxies not hosting an AGN, however, galaxy counts tend to be lower than AGN. This could be explained by the host galaxies used to train the PRF being all low redshift sources, which could mean the higher redshift galaxies are being predicted to be other classes. It could also be because AGN can be detected out to higher redshifts than galaxies as the bright AGN continuum can be seen when the fainter continuum of a galaxy cannot, therefore we can find more AGN. High-z galaxies could potentially be classed as RGB stars if the galaxy is particularly red and dusty, but could also be classified as Unknown, which is where most of the fainter sources are expected to end up due to lack of faint sources in the training set. 

For the known (sources not classed as Unknown) sources with P$_{\rm class}$ $>$ 80\% in Table \ref{PRFcounts}, it can be seen that the SMC sources outnumber the LMC sources. This can be attributed to the RGB class, as most of the classes are larger in number in the LMC, except the RGB, which is less than half the SMC RGB number. This may be due to the training set, as the SMC RGB training set had fainter examples of RGB than the LMC training set. Therefore, these fainter RGBs are not being picked up by the LMC classifier, and the number of the RGBs is less for the LMC.

\begin{table}
\caption[Distribution of the classifications of sources in the SMC field.]{Distribution of the classifications of sources in the SMC and LMC fields for all sources, sources with P$_{\rm class}$ $>$ 60\% and $<$ 80\% and sources with P$_{\rm class}$ $>$ 80\%. }
    \centering
    \begin{tabular}{|l|r|r|r|}
        \hline\hline 
        Class & All \mbox{~~~~~}  &  60\% $<$ P$_{\rm class}$ $<$ 80\% & P$_{\rm class}$ $>$ 80\% 
  \\

       \hline
       SMC & 29,514,739 & 4,012,812 & 10,478,568\\
       SMC (Known) & 7,953,558 & 350,287 & 707,939\\
       \hline
         AGN & 680,721 & 112,070 & 7902 \\
         Galaxy & 67,522 & 17,158 & 3167 \\
         OB & 4,735,098 & 5006 & 8739\\
          RGB & 1,119,492 & 203,795 & 682,502\\
          PNe  & 1,112,906 & 442 & 48 \\
          Post-AGB/RGB   & 29,385 & 2907 & 89 \\
          AGB & 137,618 & 703 & 2382\\
         H\textsc{ii}/YSO & 38,365 & 324 & 89\\
         PM & 30,777 & 7676 & 2777\\
         RSG & 1674 & 206 & 244\\
         Unknown & 21,561,181 & 3,662,525 & 9,770,629 \\
         \hline
         LMC & 103,172,194 & 7,176,530 & 46,218,151 \\
         LMC (Known) & 30,889,945 & 580,880 & 397,899 \\
         \hline
         AGN & 1,593,270 & 403,382 & 42,605 \\
         Galaxy & 230,515 & 49,503 & 23,979 \\
         OB & 336,181 & 27,303 & 64,178 \\
         RGB & 628,388 & 61,794 & 237,841 \\
         PNe & 44,118 & 545 & 77 \\
         Post-AGB/RGB & 4411 & 20 & 33 \\
         AGB & 25,376,797 & 5218 & 15,892\\
         H\textsc{ii}/YSO & 214,791 & 19,601 & 3268 \\
         PM & 2,455,355 & 12,721 & 8898 \\
         RSG & 6119 & 793 & 1128 \\
         Unknown & 72,282,249 & 6,595,650 & 45,820,252 \\
         
       \hline
    \end{tabular}
    
    \label{PRFcounts}
\end{table}

A layout of the results table for the classification of all the sources can be seen in the Appendices in Table \ref{tab:results_table}. The catalogues of sources are separated into high-confidence sources (P$_{\rm class}$ $>$ 80\%), mid-confidence  sources (60\% $<$ P$_{\rm class}$ $<$ 80\%) and low-confidence sources (P$_{\rm class}$ $<$ 60\%). Some low/mid-confidence AGNs have the possibility of being moved up to the high-confidence catalogue if they are found to be associated with an X-ray and/or radio detection (See Sections \ref{Radio} \& \ref{Xray}), or if the combined AGN and galaxy probabilities put them into a higher threshold (See Section \ref{AGN-Gal}).

For the rest of this work, unless stated otherwise, we will be referring to the high-confidence sources when exploring their distributions and properties.


\subsection{PRF classification spatial distributions across the SMC and LMC fields}

\begin{figure*}
\centering
\begin{tabular}{c}
	\includegraphics[width=1\textwidth, trim=1mm 1mm 1mm 2mm, clip]{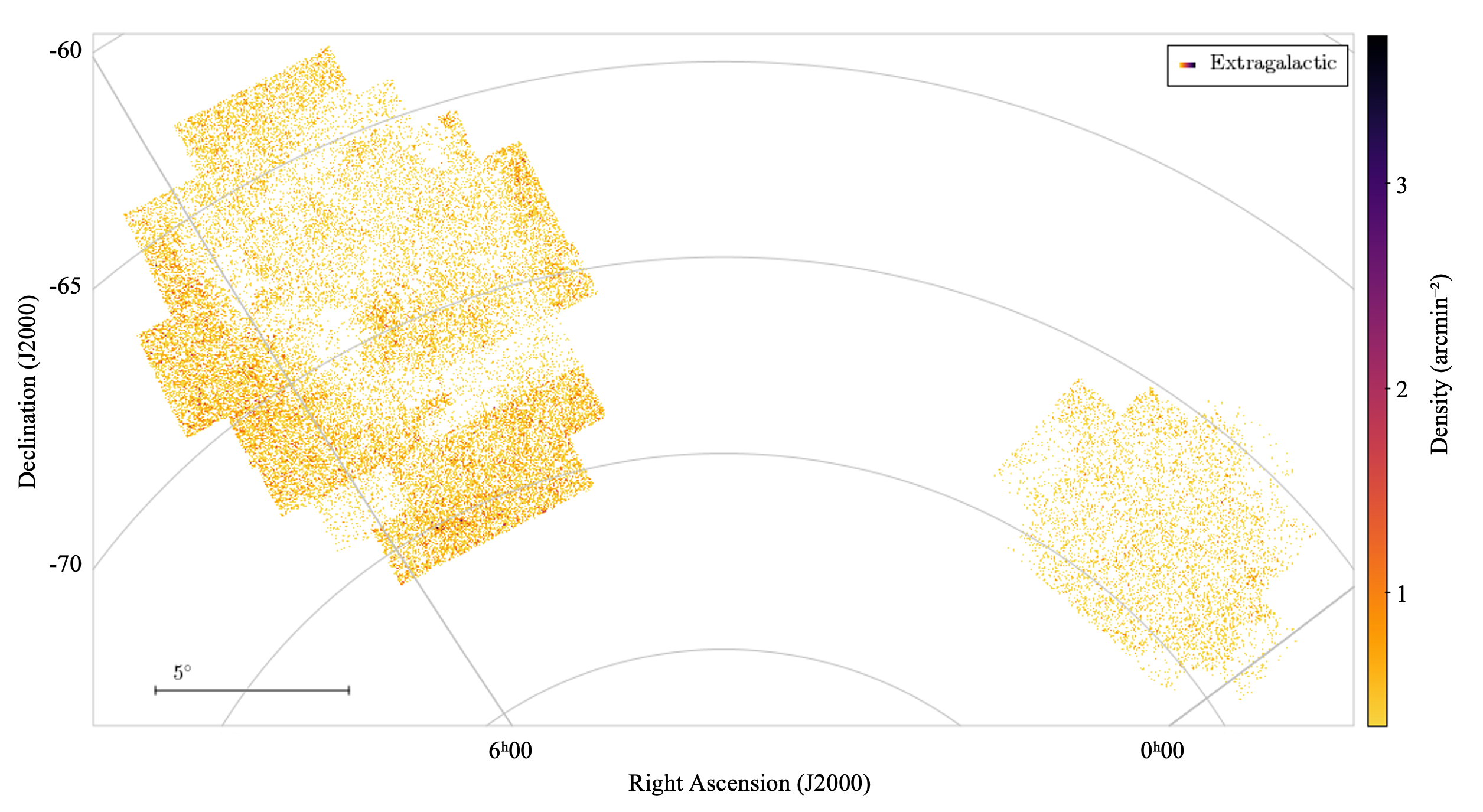}
 \end{tabular}
    \caption{The sky density of the combined AGN and galaxy sources with P$_{\rm class}$ $>$ 80\% for the LMC (left) and SMC (right). The density of sources identified as extragalactic is fairly uniform over the survey areas, with a slight increase away from the centres of the Magellanic Clouds (as expected due to extinction and source confusion) and with some visible structure due to the differing footprints of some of the datasets that enable robust identification of extragalactic sources.}
    \label{Dist_exgal80}
\end{figure*}

\begin{figure*}
\centering
\begin{tabular}{c}
	\includegraphics[width=1\textwidth, trim=1mm 1mm 1mm 2mm, clip]{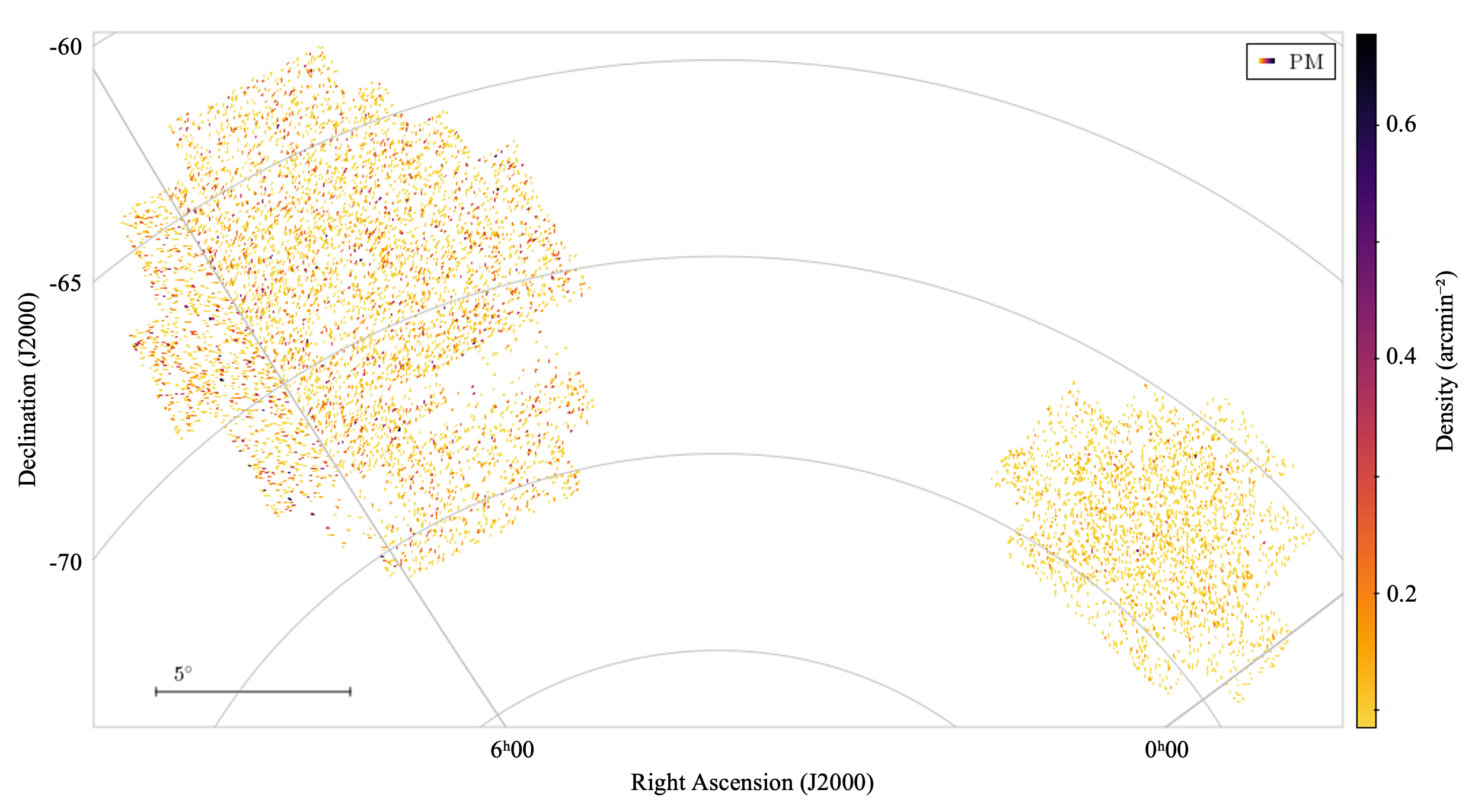}
 \end{tabular}
    \caption{The sky density of the foreground high proper-motion (PM) sources with P$_{\rm class}$ $>$ 80\% for the LMC (left) and SMC (right). The density of sources identified as foreground stars is fairly uniform over the survey areas, as expected.} 
    \label{Dist_PM80}
\end{figure*}

\begin{figure*}
\centering
\begin{tabular}{c}
	\includegraphics[width=1\textwidth, trim=1mm 1mm 1mm 2mm, clip]{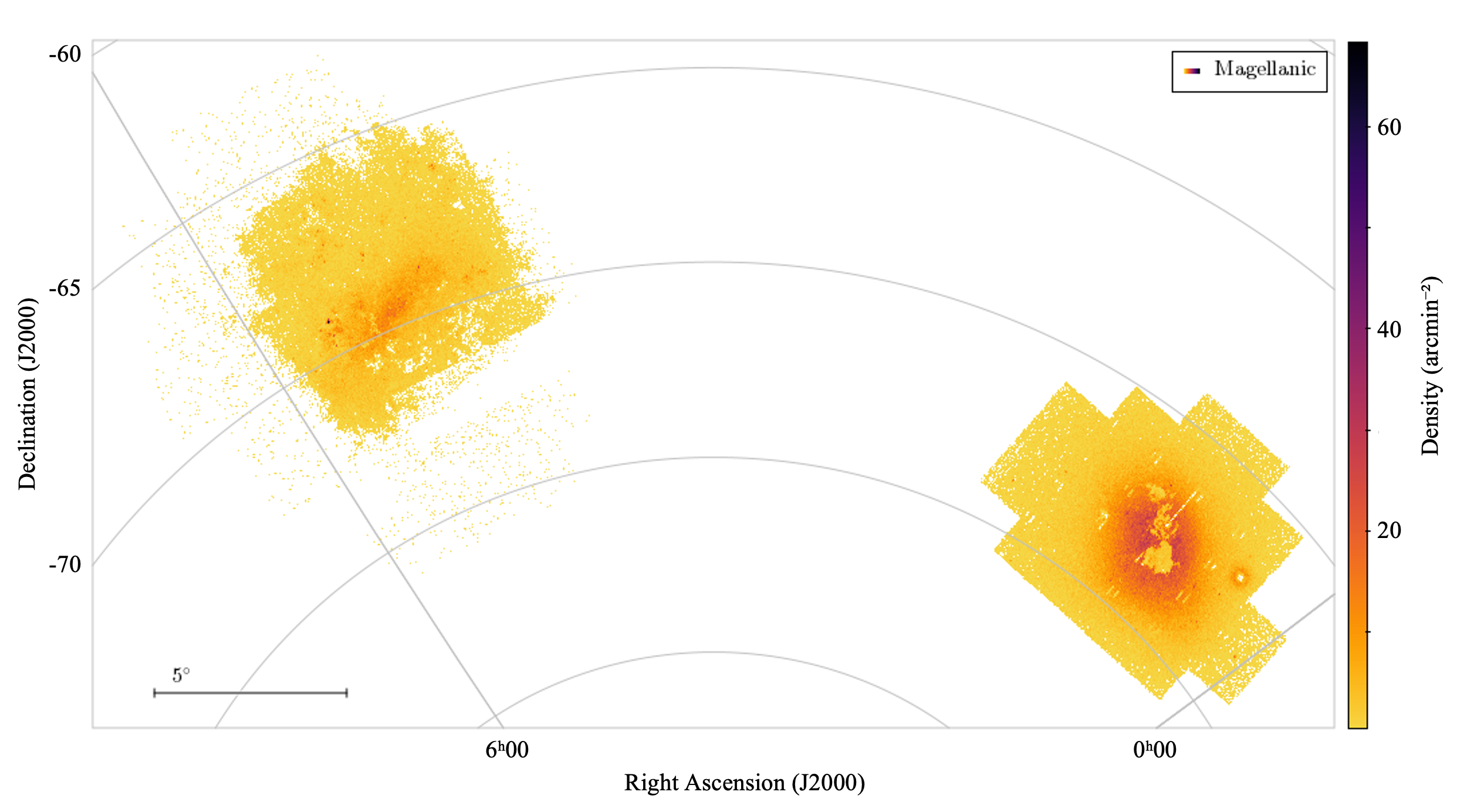}
 \end{tabular}
    \caption{The sky density of the combined stellar Magellanic sources with P$_{\rm class}$ $>$ 80\% for the LMC (left) and SMC (right). The distribution is dominated by intermediate-age/old RGB stars.}
    \label{Dist_mag80}
\end{figure*}

\begin{figure*}
\centering
\begin{tabular}{c}
	\includegraphics[width=1\textwidth, trim=1mm 1mm 1mm 2mm, clip]{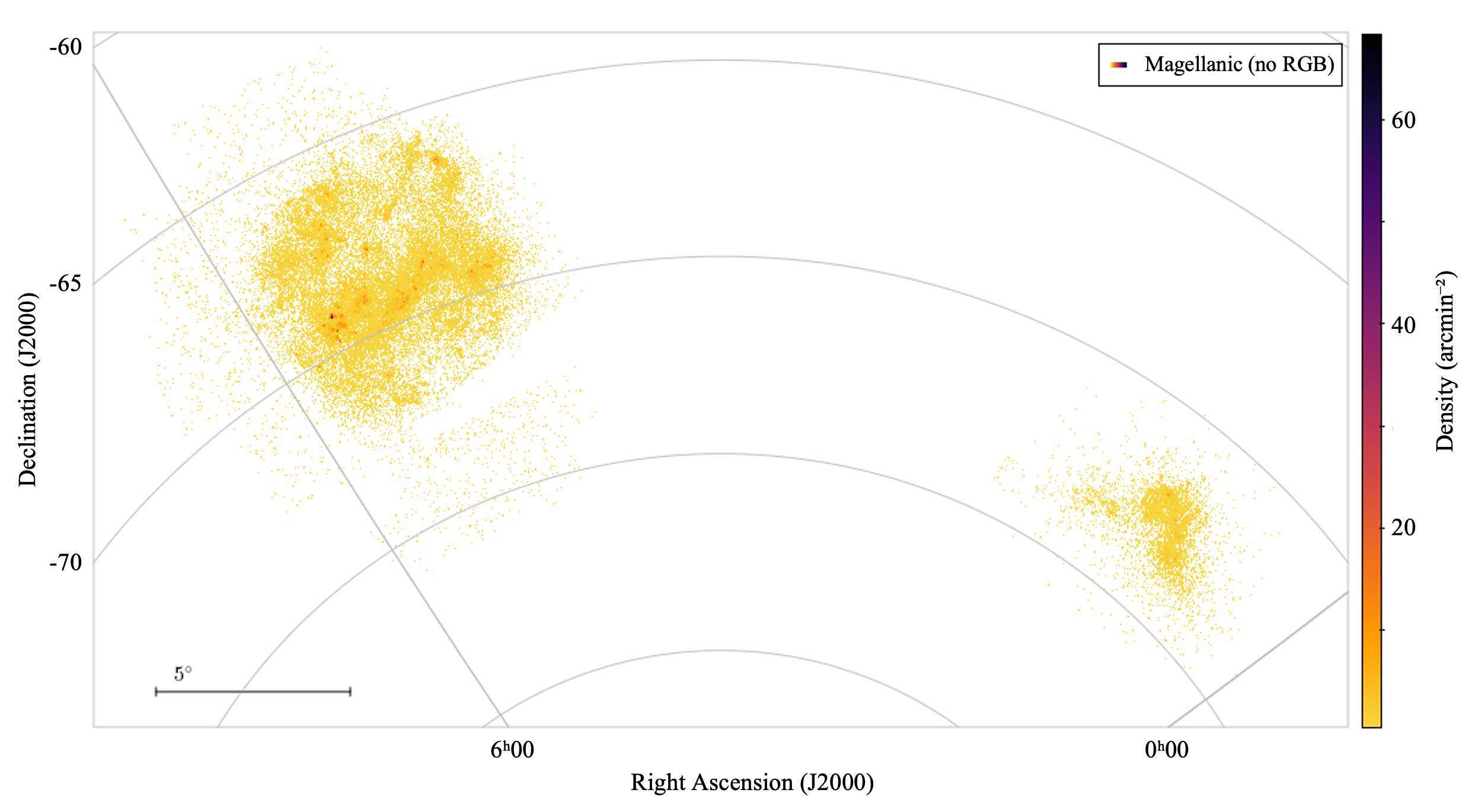}
 \end{tabular}
    \caption{The sky density of the combined stellar Magellanic sources with RGB sources removed and P$_{\rm class}$ $>$ 80\% for the LMC (left) and SMC (right).The sources identified as Magellanic concentrate in the centres of the Clouds with fewer sources on the outskirts of the survey area. The sources concentrate more in the bar structure of the Clouds and the distribution is dominated by young stars.}
    \label{Dist_magnorgb80}
\end{figure*}

The spatial distributions of each of the classes across the SMC and LMC fields can be seen in Figures \ref{Dist_exgal80} -- \ref{Dist_magnorgb80}. Note that the most confident class predictions tend to be in the areas where all the photometric surveys overlap (see Figure \ref{MCcoverage} for comparison), and therefore the PRF had access to the most complete dataset to classify with. 

The spatial distributions of the extragalactic sources are expected to be homogeneous when not looking through a nearby galaxy. In the presence of the SMC and LMC the spatial distribution is expected to be mostly homogeneous, but highest away from the centres of the SMC and LMC, and decrease as the stellar density increases towards the centres of the SMC and LMC, as stellar sources are more likely to be in the way of the background extragalactic sources and extinction becomes more prominent. The spatial distribution of the sources classed as AGN and galaxies, seen in Figure \ref{Dist_exgal80}, is as expected, the number of sources slightly decreases towards the centres of the Magellanic Clouds. The highest density areas are where there is overlap between SMASH and VMC datasets (see Fig. \ref{MCcoverage}), the combination of which would therefore allow for more confident classifications.

The spatial distribution of the foreground Milky Way stars can be seen in Figure \ref{Dist_PM80}. This is expected to be homogeneous across the sky and this is the spatial distribution that we see.

The spatial distribution of the combined Magellanic stellar sources can be seen in Figure \ref{Dist_mag80}. The spatial distribution for the LMC is as expected with the Magellanic sources concentrating in the centre. The spatial distribution of sources for the SMC is not as expected. Though the numbers do become fewer towards the edge of the survey region, the sources extend to the edges of the VMC survey area where extragalactic sources are expected to dominate. The majority of the sources causing this unexpected behaviour are classified as RGB stars.

After removing the dominating RGB class, we see the spatial distribution in Figure \ref{Dist_magnorgb80}, in which the sources concentrate in the centre of the Magellanic Clouds as expected. We can see the stellar structures of the Magellanic Clouds. The SMC is known to have a bar structure with an extension towards the East, which is what we are seeing here. We can also see the bar structure of the LMC \citep{2019Dalal} clearer after removing the RGBs. Since RGBs tend to be older, this could suggest that the bars are not an old structure.  



The spatial distribution maps of the individual classes (except PM stars) can be seen in Appendix Section \ref{App_dist}. 

The spatial distribution of galaxies is mostly homogeneous as expected, with a decrease to the highest stellar densities in the centres of the two Magellanic Clouds, which is a more obvious effect for the LMC catalogue. AGN on the other hand, are mostly homogeneous across the entire survey footprint, with no decrease towards the galactic centres. The lack of decrease towards the centres of the Clouds for the AGN spatial distribution could be due to brighter AGNs being easier to see through the Clouds than the fainter background galaxies.

The two most noticeable globular clusters, 47 Tucan\ae\ and NGC 362,  show noticeably fewer sources with probability $>$ 80\%. This is most likely because no sources from these clusters, which are closer to us than the SMC, were used in training. They can be traced mainly by the spatial distributions of RGB, Unknown and PM stars, as expected.

Overall, the different classes are distributed across the VMC fields of the LMC and SMC mostly as expected.

\subsection{AGN and Galaxy classifications}\label{AGN-Gal}

Some sources have a combined AGN and Galaxy probability that makes them a high confidence extragalactic source, if not necessarily a high confidence AGN or Galaxy. These sources tend to be obviously extended in VISTA images, proving their most likely extragalactic nature.

For this reason sources with a combination of AGN and Galaxy probabilities $\geq$ 80\% (56,134 and 140,212 for the SMC and LMC, respectively) are moved to the high confidence catalogue, and those that have $\geq$ 60\% and $<$ 80\% (126,514 and 302,530 for the SMC and LMC, respectively) are moved to the mid-confidence catalogue.

\section{Discussion}\label{Discussion}
\subsection{Classification of unusual dust-dominated AGN}\label{dustyAGN}
As part of the work done in \cite{2022MNRAS.515.6046P}, unusual AGN that appeared to showcase dust emission almost entirely from the AGN, that are often misidentified as stellar objects, were found using unsupervised machine learning and spectroscopically observed. This work was then continued into the LMC. Some of these sources were used in training. The PRF classifications that we obtain here for the full sample of candidate dust-dominated AGN from \cite{2022MNRAS.515.6046P} and the extended sample in the LMC field can be seen in Table \ref{SALTPRF}.

\begin{table}
\caption{Classifications of sources from \citet{2022MNRAS.515.6046P} and Pennock et al. (\textit{in prep.}) that were candidate AGN with little to no dust from the host galaxy and that also had a tendency to be misclassified as dusty stellar sources in the Magellanic Clouds. C star refers to carbon stars and Em* refers to emission-line stars. Note that some of these sources were used in the training set, indicated by the `T?' column, by either `Y' (yes) or `N' (no).}
    \centering
    \begin{tabular}{lllllcl}
    \hline\hline
    Name & RA & DEC & PRF & P$_{\rm class}$ & \llap{T}? & Spec.\\
    &  &  & Class &  &  & Class\\
    \hline
    SMCtSNE1 & \llap{0}0:36:16.99& \llap{$-$7}4:31:31.3& AGN & \llap{0}.99 & \llap{Y} & AGN\\
    SMCtSNE2 & \llap{0}1:13:37.08& \llap{$-$7}4:27:55.3& AGN & \llap{0}.97 & \llap{Y} & AGN\\
    SMCtSNE3 & \llap{0}0:31:56.89& \llap{$-$7}3:31:13.6& AGN & \llap{0}.96 & \llap{Y} & AGN\\
    SMCtSNE4 & \llap{0}0:26:02.54& \llap{$-$7}2:47:18.0& AGN & \llap{0}.98 & \llap{Y}& AGN\\
    SMCtSNE5 & \llap{0}0:48:25.70& \llap{$-$7}2:44:03.0& AGB & \llap{0}.99 & \llap{Y} & C star\\
    SMCtSNE6 & \llap{0}1:14:08.00& \llap{$-$7}2:32:43.3& AGN & \llap{0}.98 & \llap{Y} & AGN\\
    SMCtSNE7 & \llap{0}0:55:51.51& \llap{$-$7}3:31:10.0& AGN & \llap{0}.97 & \llap{Y} & AGN\\
    SMCtSNE8 & \llap{0}1:22:36.90& \llap{$-$7}3:10:16.7& AGN & \llap{0}.45 & \llap{N} & Em*\\
    SMCtSNE9 & \llap{0}1:21:08.40& \llap{$-$7}3:07:13.1& AGN & \llap{0}.99 & \llap{Y} & AGN\\
    SMCtSNE10 & \llap{0}1:15:34.09& \llap{$-$7}2:50:49.3& AGN & \llap{0}.94 & \llap{Y} & AGN\\
    SMCtSNE11 & \llap{0}0:39:10.78& \llap{$-$7}1:34:09.9& AGN & \llap{0}.98 & \llap{Y} & AGN\\
    SMCtSNE12 & \llap{0}0:51:16.95& \llap{$-$7}2:16:51.5& AGN & \llap{0}.98 & \llap{Y}& AGN\\
    SMCtSNE13 & \llap{0}0:57:32.80& \llap{$-$7}2:13:02.0& AGN & \llap{0}.99 & \llap{Y} & AGN\\
    SMCtSNE15 & \llap{0}0:34:05.30& \llap{$-$7}0:25:52.3& AGN & \llap{0}.82 & \llap{Y} & AGN\\
    SMCtSNE16 & \llap{0}0:49:52.50& \llap{$-$6}9:29:56.0& AGN & \llap{0}.34 & \llap{Y} & AGN\\
    LMCtSNE2 & \llap{0}6:15:04.01& \llap{$-$6}6:17:16.4& AGN & \llap{0}.69 & \llap{Y} & AGN \\
    LMCtSNE3 & \llap{0}5:33:57.69& \llap{$-$6}4:20:24.9& AGN & \llap{0}.87 & \llap{Y} & Galaxy\\
    LMCtSNE4 & \llap{0}5:01:10.84& \llap{$-$7}3:36:35.0& AGN & \llap{0}.92 & \llap{Y} & AGN \\
    LMCtSNE5 & \llap{0}5:41:12.99& \llap{$-$6}4:11:53.7& AGN & \llap{0}.69 & \llap{N} & ? \\
    LMCtSNE6 & \llap{0}5:45:05.73& \llap{$-$6}4:11:19.3& AGN & \llap{0}.60 & \llap{N} & AGN\\
    LMCtSNE7 & \llap{0}5:20:19.84& \llap{$-$7}3:55:37.3& AGN & \llap{0}.31 & \llap{N} & AGN \\
    LMCtSNE8 & \llap{0}5:32:10.38& \llap{$-$7}3:57:22.3& AGN & \llap{0}.36 & \llap{N} & AGN\\
    LMCtSNE9 & \llap{0}4:38:50.67& \llap{$-$7}2:17:12.6& AGN & \llap{0}.50 & \llap{N} & AGN\\
    LMCtSNE10 & \llap{0}5:14:17.90& \llap{$-$7}2:20:19.2& AGN & \llap{0}.56 & \llap{Y} & AGN\\
    LMCtSNE11 & \llap{0}4:51:38.41& \llap{$-$7}1:02:06.1& AGN & \llap{0}.95 & \llap{Y} & AGN\\
    LMCtSNE12 & \llap{0}5:40:55.08& \llap{$-$7}0:34:46.9& OB & \llap{0}.84 & \llap{N} & Em*\\
    LMCtSNE13 & \llap{0}5:22:52.28& \llap{$-$6}9:50:42.6& H\textsc{ii}/YS\rlap{O} & \llap{0}.21 & \llap{N} & Em*\\
    LMCtSNE14 & \llap{0}5:51:43.28& \llap{$-$6}8:45:43.0& Galaxy & \llap{0}.58 & \llap{Y} & AGN\\
    LMCtSNE15 & \llap{0}5:22:30.52& \llap{$-$6}7:54:43.6& OB & \llap{0}.71 & \llap{N} & Em*\\
    LMCtSNE16 & \llap{0}5:31:48.96& \llap{$-$6}7:21:33.8& H\textsc{ii}/YS\rlap{O} & \llap{0}.41 & \llap{N}& Star\\
    LMCtSNE17 & \llap{0}5:31:54.44& \llap{$-$6}8:26:40.4& H\textsc{ii}/YS\rlap{O} & \llap{0}.99 & \llap{Y} & Em*\\
    LMCtSNE18 & \llap{0}5:48:22.29& \llap{$-$6}7:58:53.3& AGB & \llap{0}.44 & \llap{N} & Em*\\
    LMCtSNE19 & \llap{0}5:04:47.16& \llap{$-$6}6:40:30.7& H\textsc{ii}/YS\rlap{O} & \llap{0}.69 & \llap{N} & Em*\\
    LMCtSNE20 & \llap{0}5:53:57.48& \llap{$-$6}6:50:01.6& AGN & \llap{0}.90 & \llap{N} & AGN\\
    LMCtSNE21 & \llap{0}6:10:52.23& \llap{$-$6}6:30:11.5& AGN & \llap{0}.62 & \llap{N} & AGN\\
    LMCtSNE22 & \llap{0}5:19:42.45& \llap{$-$6}5:02:16.8& AGN & \llap{0}.88 & \llap{Y} & AGN\\
    LMCtSNE23 & \llap{0}5:49:13.47& \llap{$-$6}4:29:29.2& AGN & \llap{0}.60 & \llap{N} & AGN\\
    LMCtSNE24 & \llap{0}5:43:34.33& \llap{$-$6}4:22:58.2& AGN & \llap{0}.83 & \llap{N} & AGN\\
    \hline
    \end{tabular}
\label{SALTPRF}
\end{table}

For the SMC sources, most are confident AGN, except the two known stars. SMCtSNE5 is confidently classed as an AGB star, which is not unexpected as it is a carbon star. SMCtSNE8, which is a long-period variable AGB star with a H$\alpha$ emission line from shock dissipation, is classed as a low confidence AGN ($\sim$ 54\% probability) with a $\sim$ 15\% probability of being a pAGB/RGB and $\sim$ 12\% probability of being an AGB, as the next highest possibilities. 

For the LMC sources, 15 sources are classed as AGN with varying confidences. LMCtSNE14 (AGN, $z$ $\sim$ 0.4) is classed as a galaxy with a $\sim$ 30\% chance of being an AGN, which, considering its visible host galaxy in VMC images, is not that surprising. We find that, although not predicted confidently as AGN, dusty AGB stars and YSOs are the main contaminants of AGN-dust dominated samples. 

This shows that unusual dust-dominated AGN that have often been mistaken for dusty Magellanic objects are being classified correctly by the PRF, most likely helped by the inclusion of similar sources in the training set. This does, however, show that emission line stars have a chance of being classed as AGN, though with possibly low probabilities.

\subsection{Unseen stellar classes}\label{Unseenclasses}
One way to ascertain the performance of the classifier in separating extragalactic from stellar sources is to test it on classes it has not seen before. The classifications from SAGE-spec \citep{Ruffle2015} were used as part of the training set for the classifier. Not all the classes in this dataset were added to the training set as they were deemed too few in number for training purposes and/or not a well defined class (e.g. emission-line stars). The SAGE-spec dataset for the SMC has 18 sources that were not used. The predicted probabilities and classes of these 18 sources were extracted from the full SMC dataset and can be seen in Table \ref{SAGEspecPRF}.

\begin{table}
\caption[Classifications of sources from \citet{Ruffle2015} that were not used in the training of the PRF classifier.]{Sources from \citet{Ruffle2015} that were not used in the training of the PRF classifier. The SAGE classes are Wolf-Rayet stars (WR), R Coron\ae\ Borealis variable stars (RCrB), Blue supergiants (BSG), S-type stars (S star), Symbiotic stars (Sym. star), and stars of indiscernible type (star).}
    \centering
    \begin{tabular}{|l|l|l|l|l|l|l|l|}
        \hline\hline 
        Source Nam\rlap{e} & SAGE  & RA & DEC & PRF  & P$_{\rm class}$   \\
         & Class & (J2000) & (J2000) & Class &  \\
       \hline
       SMC-WR9 & WR & 00:54:32.2 & \llap{$-$7}2:44:36 & OB  & 0.99\\
       SMC-WR12 & WR & 01:02:52.2 & \llap{$-$7}2:06:52 & OB  & 0.96  \\
       GSC\,09141$-$056\rlap{31} & WR & 00:43:42.2 & \llap{$-$7}3:28:54 & OB & 0.99 \\
       SMC-WR2 & WR & 00:48:31.0 & \llap{$-$7}3:15:45 & OB  & 0.99  \\
       SMC-WR3 & WR & 00:49:59.3 & \llap{$-$7}3:22:14 & OB  & 0.99  \\
       SMC-WR4 & WR & 00:50:43.4 & \llap{$-$7}3:27:05 & OB & 0.98 \\
       RMC\,31 & WR & 01:03:25.2 & \llap{$-$7}2:06:44 & OB & 0.80  \\
       SMC-WR11 & WR & 00:52:07.5 & \llap{$-$7}2:35:38 & OB & 0.99  \\
       MSX\,SMC\,014 & RCrB & 00:46:16.4 & \llap{$-$7}4:11:13 & AGB & 0.74 \\
       MSX\,SMC\,155 & RCrB & 00:57:18.2 & \llap{$-$7}2:42:35 & AGB & 0.56  \\
       AzV\,404 & Star & 01:06:29.4 & \llap{$-$7}2:22:09 & OB & 0.52  \\
       BFM\,1 & S star & 00:47:19.3 & \llap{$-$7}2:40:04 & AGB & 0.97 \\
       AzV\,456 & Star & 01:10:55.8 & \llap{$-$7}2:42:57 & OB & 0.84 \\
       AzV\,23 & Star & 00:47:38.9 & \llap{$-$7}3:22:54 & OB & 0.50 \\
       OGLE\,SMC-SC1\rlap{0} & RCrB & 01:04:53.0 & \llap{$-$7}2:04:04 & AGB & 0.43  \\
       \mbox{~~}107856  & & & & & \\
       MSX\,SMC\,185 & Sym. st\rlap{ar} & 00:54:20.0 & \llap{$-$7}2:29:09 & PNe & 0.53 \\
       HD\,5980 & WR & 00:59:26.7 & \llap{$-$7}2:09:54 & OB & 0.62 \\
       HD\,6884 & BSG & 01:07:18.1 & \llap{$-$7}2:28:04 & AGB & 0.41 \\
       \hline
    \end{tabular}
    
    \label{SAGEspecPRF}
\end{table}

\begin{table}
\centering
\caption[Classifications of sources from \citet{Jones2017} that were not used in the training of the PRF classifier.]{Sources from \citet{Jones2017} that were not used in the training of the PRF classifier. The SAGE classes are Wolf-Rayet stars (WR), R Coron\ae\ Borealis variable stars (RCrB), Blue supergiants (BSG), Yellow supergiants (YSG), S-type stars (S star), Symbiotic stars (Sym. star), RV Tauri stars (RVTau), supernova remnants (SNR), unknown (UNK), luminous blue variable stars (LBV) and stars of indiscernible type (star).}
\begin{tabular}{l l l l l l}   
\hline\hline
Source Name & SAGE & \llap{RA} & \llap{DE}C & PRF & P\rlap{$_{\rm class}$} \\
 & Class & \llap{(J2}000) & \llap{(J2}000) & Class &  \\
\hline
LHA 120-N 82 & WR & \llap{04}:53:30.30 & \llap{$-$6}9:17:49\rlap{.2} & H\textsc{ii}/Y\rlap{SO} & 0.39 \\
HD 268813 & STAR & \llap{04}:54:23.23 & \llap{$-$7}0:26:56\rlap{.8} & RSG & 0.39 \\
RP 1631 & RCrB & \llap{05}:00:35.35 & \llap{$-$7}0:52:00\rlap{.5} & AGB & 0.60 \\
HV 2281 & RVTau & \llap{05}:03:05.05 & \llap{$-$6}8:40:25\rlap{.0} & pA/R\rlap{GB} & 0.82 \\
LMC-BM 11-19 & STAR & \llap{05}:03:43.43 & \llap{$-$6}7:59:19\rlap{.0} & AGB & 0.86 \\
RP 1878 & UNK & \llap{05}:04:34.34 & \llap{$-$6}7:52:21\rlap{.4} & AGN & 0.39 \\
 & BSG & \llap{05}:06:39.39 & \llap{$-$6}8:22:09\rlap{.5} & OB & 0.99 \\
HV 915 & RVTau & \llap{05}:14:18.18 & \llap{$-$6}9:12:35\rlap{.3} & pA/R\rlap{GB} & 0.41 \\
 & STAR & \llap{05}:15:26.26 & \llap{$-$6}7:51:27\rlap{.0} & AGB & 0.99 \\
KDM 3196 & STAR & \llap{05}:18:08.08 & \llap{$-$7}1:51:53\rlap{.6} & AGB & 0.65 \\
HV 2444 & RVTau & \llap{05}:18:45.45 & \llap{$-$6}9:03:22\rlap{.0} & AGB & 0.58 \\
 & STAR & \llap{05}:19:45.45 & \llap{$-$6}9:30:00\rlap{.0} & AGB & 0.88 \\
HV 942 & RCrB & \llap{05}:21:48.48 & \llap{$-$7}0:09:57\rlap{.2} & AGB & 0.46 \\
HV 5829 & RVTau & \llap{05}:25:19.19 & \llap{$-$7}0:54:10\rlap{.1} & AGB & 0.56 \\
MACHO 82.8405.\rlap{15} & RVTau & \llap{05}:31:51.51 & \llap{$-$6}9:11:46\rlap{.3} & pA/R\rlap{GB} & 0.57 \\
 & STAR & \llap{05}:32:07.07 & \llap{$-$7}0:10:25\rlap{.0} & AGB & 0.94 \\
SHP LMC 256 & UNK & \llap{05}:34:44.44 & \llap{$-$6}7:37:50\rlap{.5} & AGN & 0.78 \\
KDM 5345 & UNK & \llap{05}:38:24.24 & \llap{$-$6}6:09:00\rlap{.4} & AGB & 0.99 \\
MACHO 81.9728.\rlap{14} & RVTau & \llap{05}:40:01.01 & \llap{$-$6}9:42:14\rlap{.8} & OB & 0.31 \\
 & UNK & \llap{05}:45:46.46 & \llap{$-$6}7:32:39\rlap{.1} & AGB & 0.40 \\
KDM 6247 & STAR & \llap{05}:47:57.57 & \llap{$-$6}8:14:57\rlap{.1} & AGB & 0.91 \\
HV 2862 & RVTau & \llap{05}:51:23.23 & \llap{$-$6}9:53:51\rlap{.4} & pA/R\rlap{GB} & 0.84 \\
PMP 133 & STAR & \llap{05}:52:53.53 & \llap{$-$6}9:30:35\rlap{.3} & PM & 0.96 \\
HD 270754 & STAR & \llap{04}:47:05.05 & \llap{$-$6}7:06:53\rlap{.3} & OB & 0.97 \\
HD 32402 & WR & \llap{04}:57:24.24 & \llap{$-$6}8:23:56\rlap{.8} & OB & 0.87 \\
HD 269187 & STAR & \llap{05}:14:04.04 & \llap{$-$6}7:15:50\rlap{.8} & PM & 0.78 \\
S Dor & LBV & \llap{05}:18:14.14 & \llap{$-$6}9:15:01\rlap{.4} & PM & 0.35 \\
OGLE LMC-RCB-\rlap{10} & RCrB & \llap{05}:20:48.48 & \llap{$-$7}0:12:13\rlap{.0} & AGB & 0.91 \\
HD 36402 & WR & \llap{05}:26:04.04 & \llap{$-$6}7:29:57\rlap{.1} & OB & 0.72 \\
W Men & RCrB & \llap{05}:26:25.25 & \llap{$-$7}1:11:11\rlap{.8} & AGB & 0.40 \\
HD 269662 & LBV & \llap{05}:30:52.52 & \llap{$-$6}9:02:58\rlap{.9} & PM & 0.53 \\
HV 12620 & STAR & \llap{05}:33:00.00 & \llap{$-$7}0:41:23\rlap{.6} & AGB & 0.82 \\
HV 2671 & RCrB & \llap{05}:33:49.49 & \llap{$-$7}0:13:23\rlap{.5} & H\textsc{ii}/Y\rlap{SO} & 0.40 \\
SN 1987A & SNR & \llap{05}:35:28.28 & \llap{$-$6}9:16:11\rlap{.3} & H\textsc{ii}/Y\rlap{SO} & 0.36 \\
W61 27-27 & STAR & \llap{05}:36:04.04 & \llap{$-$6}9:01:30\rlap{.4} & OB & 0.91 \\
 & WR & \llap{05}:36:44.44 & \llap{$-$6}9:29:46\rlap{.0} & OB & 0.98 \\
SNR B0540-69.3 & SNR & \llap{05}:40:11.11 & \llap{$-$6}9:19:54\rlap{.5} & H\textsc{ii}/Y\rlap{SO} & 0.76 \\
HD 269953 & YSG & \llap{05}:40:12.12 & \llap{$-$6}9:40:04\rlap{.8} & RSG & 0.39 \\
IRAS 05413-6934 & UNK & \llap{05}:40:54.54 & \llap{$-$6}9:33:18\rlap{.7} & H\textsc{ii}/Y\rlap{SO} & 0.99 \\
MSX LMC 1795 & RCrB & \llap{05}:42:22.22 & \llap{$-$6}9:02:59\rlap{.6} & AGB & 0.90 \\
LHA 120-S 61 & WR & \llap{05}:45:52.52 & \llap{$-$6}7:14:25\rlap{.8} & OB & 0.80 \\
HD 270422 & STAR & \llap{05}:56:48.48 & \llap{$-$6}6:39:05\rlap{.0} & RSG & 0.35 \\
HD 270467 & STAR & \llap{05}:58:12.12 & \llap{$-$6}6:20:23\rlap{.6} & PM & 0.81 \\
WOH G 642 & STAR & \llap{05}:59:21.21 & \llap{$-$6}6:31:56\rlap{.6} & PM & 0.99 \\
HD 41466 & STAR & \llap{06}:00:19.19 & \llap{$-$6}6:13:27\rlap{.5} & PM & 0.34 \\
HD 270485 & STAR & \llap{06}:00:53.53 & \llap{$-$6}6:55:48\rlap{.0} & PM & 0.99 \\
HD 271776 & STAR & \llap{06}:01:38.38 & \llap{$-$6}6:35:20\rlap{.0} & PM & 0.92 \\
\hline
\end{tabular}

\end{table}

Overall, for the SMC classifier, all of the sources were classed as stellar sources. 
All the sources are predicted to have a $<$ 10\% probability of being an AGN and $<$ 5\% probability of being a galaxy. This implies that similar stellar sources of these natures would most likely not be predicted to be extragalactic. RCrB are often associated with post-AGB stars so it is not surprising that two out of three were classed as AGB stars. WR can be similar to O type stars so would be expected to be predicted as an OB star, which all but one are. The WR star predicted as an RGB star could possibly be a case of a dusty WR star. 

For the LMC classifier, two sources were classified as UNK in the SAGE-spec catalogue, but were classed by the PRF as AGN. The rest of the sources were predicted to be one of the stellar classes. Just as for the SMC, the RCrB were mostly classed as AGB, 
WR were mostly predicted to be OB stars and the one BSG was predicted to be an OB star. The two SNRs were predicted to be H\textsc{ii}/YSOs, the two LBVs were predicted as PM stars, the seven RVTau were predicted as post-AGB/RGB and AGB stars and the YSG was predicted as an RSG. These are all unsurprising as these stellar objects share properties with the stellar classes they have been classed as.

These results are promising for AGN and galaxy classifications, as this shows that Magellanic classes that have not been trained upon are classified as one of the other Magellanic classes.

\subsection{Colour--magnitude selections of PRF classified sources}\label{CCD}

In this section we explore the class distributions across colour--colour and colour--magnitude diagrams in the optical, near-IR and mid-IR.  Here we focus on the sources with P$_{\rm class}$ $>$ 80\%.

\subsubsection{Optical}
\begin{figure*}
\centering
\begin{tabular}{c}
	\includegraphics[width=1\textwidth]{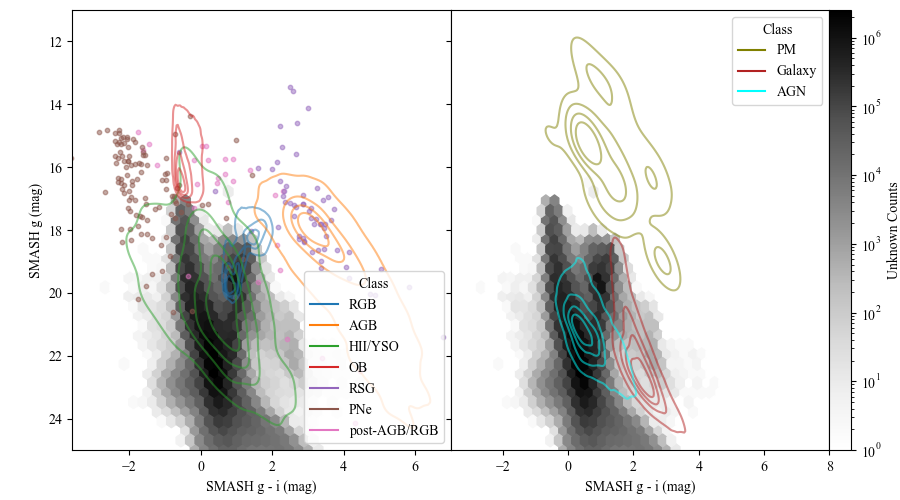}
 \end{tabular}
    \caption{SMASH colour--magnitude diagrams of the sources classified as Magellanic sources (left), foreground stars and extragalactic (right) in the SMC and LMC fields. The contours represent a probability distribution in intervals of 0.2. The sources identified as Unknown with P$_{\rm class}$ $>$ 80\% are represented as a 2D histogram in the background of both plots.}
    \label{SMASH_CM}
\end{figure*}

From the optical colour--magnitude diagram seen in Figure \ref{SMASH_CM}, we can see in the Unknown sources the structure of stellar sequences \citep{Nidever2017}. The main sequence stars are expected to start from the bottom of the diagram and fork to the left, which is what we see in the Unknowns. It is not surprising that main sequence stars would be picked up as Unknowns as they were not a class that was trained on. The right fork of the Unknowns is expected to be supergiants and RGB stars. The sources classed as RGB stars and RSG stars do follow this fork, so its likely that not all the RGB and RSG stars were classified, possibly due to mismatches between the different photometry and/or missing data. AGN can be mostly found in the middle of the expected stellar sequence, meaning they would be hard to classify in just optical alone. Galaxies tend to concentrate just to the right of the stellar sequences, and it is possible the Unknowns in this region are also galaxies that were not accounted for in the training set.

\subsubsection{Near-IR}
\begin{figure*}
\centering
\begin{tabular}{c}
	\includegraphics[width=1\textwidth]{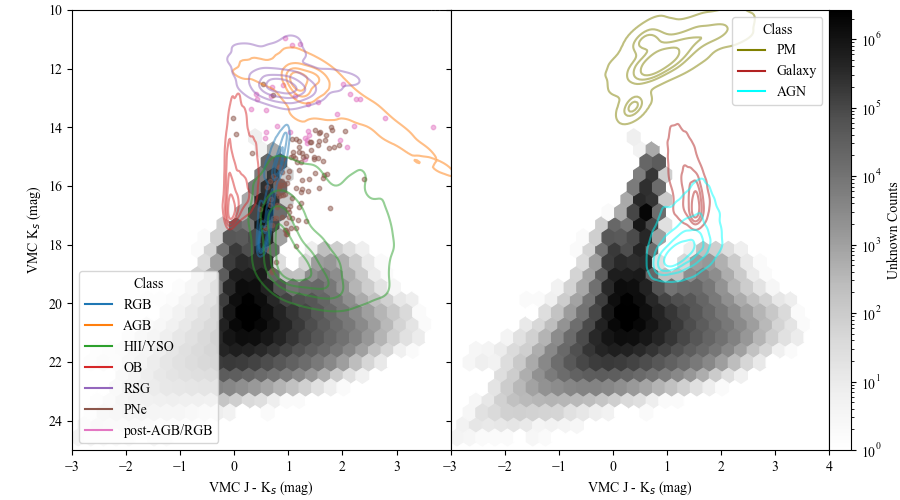}
 \end{tabular}
    \caption{VISTA colour--magnitude diagrams of the sources classified as Magellanic sources (left), foreground stars and extragalactic (right) in the SMC and LMC fields. The contours represent a probability distribution in intervals of 0.2. The sources identified as Unknown with P$_{\rm class}$ $>$ 80\% are represented as a 2D histogram in the background of both plots.}
    \label{VMC_CM}
\end{figure*}

The near-IR VISTA colour--magnitude diagram ($J$ $-$ $K$$_{\rm s}$ vs $K$$_{\rm s}$) can be seen in Figure \ref{VMC_CM}. Most sources that have $J$ $-$ $K_{\rm s}$ $>$ 1 mag and $K_{\rm s}$ $>$ 12 mag are expected to be background galaxies and quasars \citep[see region L in ][]{cioni2014, Cioni2016}, and from Figure \ref{VMC_CM} (right panel) we see that the AGN and galaxy populations follow this. \citet{cioni2014} also states that a minority of RGB stars could also be scattered to this region due to larger extinctions. Fainter Unknown sources in this region are most likely extragalactic sources at higher redshift than in the training set.

The stellar classes in Figure \ref{VMC_CM} mostly avoid the extragalactic classes. The ones that do not, YSOs, PNe, post-AGB/RGB and AGB stars, are most likely reddened due to dust. The RGB stars that are $K_{\rm s}$ $>$ 16 are in the region expected for RGB stars, and the RGB stars brighter than this are in the area for dusty AGB stars.

\subsubsection{Mid-IR}\label{midIRCC}
\begin{figure*}
\centering
\begin{tabular}{c}
	\includegraphics[width=1\textwidth]{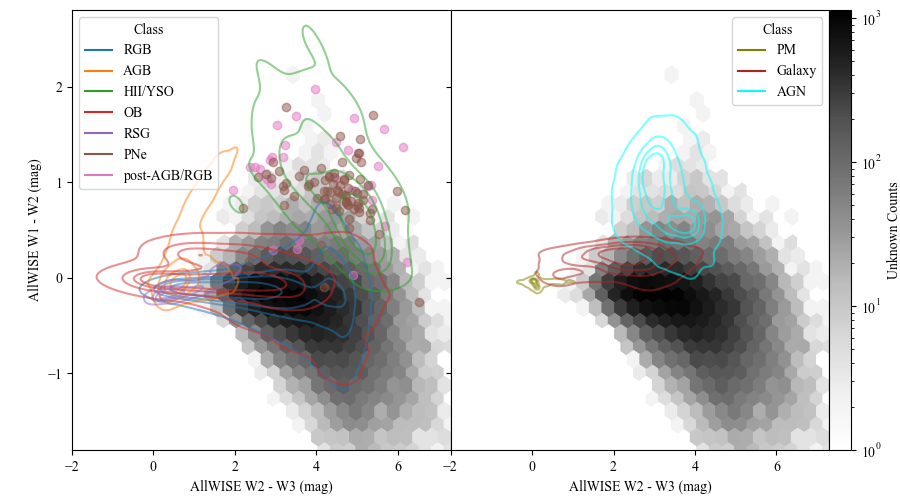}
 \end{tabular}
    \caption{AllWISE colour--colour diagrams of the sources classified as Magellanic sources (left), foreground stars and extragalactic (right) in the SMC and LMC fields. The contours represent a probability distribution in intervals of 0.2. The sources identified as Unknown with P$_{\rm class}$ $>$ 80\% are represented as a 2D histogram in the background of both plots.}
    \label{WISEPRF}
\end{figure*}

The distributions of extragalactic and stellar sources are plotted in AllWISE colour--colour diagrams \citep[for expected distributions see, e.g.][]{Stern2012,Assef2013,Nikutta2014}. The extragalactic sources can be seen in Figure \ref{WISEPRF} (right panel) where both AGN and galaxies occupy expected regions in colour--colour space. The galaxy class occupies the region expected for both star-forming and elliptical galaxies, whilst AGN occupy the region of QSOs and Seyferts. The foreground stars (PM) are also plotted here and are nicely centred on (0,0) in Vega colours, as expected. The “Unknown\textquotedblright\, sources are shown to overlap with several of the classes, but spread further to the bottom right than the other classes. Note that this is where the WISE sources with low signal to noise (S/N $<$ 3) tend to end up.

The distributions of the stellar Magellanic sources across the AllWISE colour--colour diagram can be seen in Figure \ref{WISEPRF} (left panel). The AGB sequence can be clearly seen. The populations of RGB, OB and RSG stars tend to concentrate below W1 -- W2 $\sim$ 0, unlike the extragalactic sources which tend to concentrate above W1 $-$ W2 $\sim$ 0. The PNe and YSO and post-AGB/RGB  are the classes that show the most cross-over with the extragalactic classes, which is not unexpected as they are known to be hard to differentiate from extragalactic sources in colour--colour diagrams. It is noted that sources below W1 $-$ W2 $\sim-1$ mag are mostly found within the higher density regions (centre of the SMC). 
This could be due to \textit{WISE} photometry being affected by blends, making the longer wavelength, poorer angular resolution data appear brighter.

\begin{figure}
\centering
	\includegraphics[width=1\columnwidth]{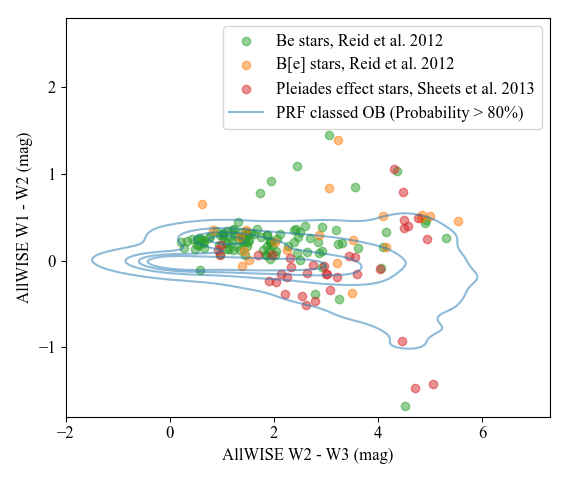}
    \caption{AllWISE colour--colour diagram of the sources classified as OB stars (blue contours). The contours represent a probability distribution in intervals of 0.2. Over-plotted are stars that exhibit the Pleiades effect (red circles), Be stars (green circles) and B[e] stars (orange circles).}
    \label{WISEOB}
\end{figure}

The OB stars appear to have two populations in the AllWISE colour--colour diagram. One population that concentrates below W1 $-$ W2 $\sim$ 0 as expected, and a smaller one that concentrates just above this, a redder population, which also happens to be the expected area for galaxies. Some possibilities are that it is either the star lighting up surrounding ISM \citep[“Pleiades effect", e.g.][]{2013Sheets,2013Adams}, a nascent star (B[e] star), or a mature, Be star with an excretion disc, in which case the red W1 $-$ W2 colour is caused by free-free emission from the circumstellar ionised gas (rather than dust). Spectroscopically observed stars undergoing the Pleiades effect \citep{2013Sheets}, as well as B[e] and Be stars \citep{2012Reid} have been plotted on the AllWISE colour--colour diagram in Figure \ref{WISEOB}. From this we can see that redder sources are most likely Be or B[e] stars, with near-IR excess most likely due to free-free emission.

Overall, the colour--colour and colour--magnitude diagrams for the different wavelength regimes show that the majority of sources are being separated where expected and that even in areas where multiple classes can be found the PRF is still capable of separating the sources. For comparisons between the distributions of the training set vs the classed sources see Appendix Section \ref{trainvsout}.

\subsection{Classifications of the radio population}\label{Radio}

Cross-matching with the radio ASKAP SMC \citep{Joseph2019} and LMC \citep{2021MNRAS.506.3540P} catalogues with a 2$^{\prime\prime}$ search radius, gives 1047/7736 and 8120/54612 sources for SMC and LMC, respectively, which have a P$_{\rm class}$ $>$80\%. The numbers of radio sources per class can be seen in Table \ref{tab:radioPRF}. A search radius of 2$^{\prime\prime}$ was used, as when cross-matching with a larger search radius of 5$^{\prime\prime}$ it was seen that the AGN class, the sources that are the most likely true counterparts, peaked at a separation radius of $\sim$1$^{\prime\prime}$. Doubling this to a search radius of 2$^{\prime\prime}$ was used to include the majority of AGN matches whilst reducing the number of mismatches.

\begin{table}
    \caption{The number of sources per class that were cross-matched with ASKAP within a 2$^{\prime\prime}$ search radius. }
    \centering
    \begin{tabular}{lrrr}
        \hline\hline
         Class &  All & 60\%$<$ P$_{\rm class}$ $<$80\% & P$_{\rm class}$ $>$80\%\\
         \hline
         SMC & 4469 & 965 & 1047\\
         \hline
         Unk& 2208  & 544 & 813\\
         AGN& 790  & 192 & 98\\
         Galaxy&  431 & 189 & 50\\
         RGB& 153  & 33 & 76\\
         OB& 540  & 1 & 4\\
         H\textsc{ii}/YSO & 36  & 4 & 1\\
         AGB& 18  & 0 & 3\\
         PNe& 285 & 1 & 2\\
         RSG& 0  & 0 & 0\\
         Post-AGB/RGB& 5 & 1 & 0\\
         PM& 3 & 0 & 0\\
         \hline
          LMC & 37375 & 6450 & 8120\\
         \hline
         Unk&  15752 & 2002 & 5660\\
         AGN& 10399  & 3012 & 1658\\
         Galaxy & 4224  & 1361 & 609\\
         RGB&  197 & 25 & 57\\
         OB&  1324 & 16 & 63\\
         H\textsc{ii}/YSO&  206 & 23 & 28\\
         AGB&  3880 & 1 & 1\\
         PNe&  292 & 7 & 38\\
         RSG& 1  & 1 & 0\\
         Post-AGB/RGB & 6 & 0 & 2\\
         PM& 1154 & 2 & 4\\
         \hline
    \end{tabular}
    \label{tab:radioPRF}
\end{table}

The majority (78\% for the SMC and 70\% for the LMC) of these sources are classed as Unknown. From the other classes, the class with the highest number is AGN, followed by galaxies, as expected as such sources are often radio bright. However, for both the SMC and LMC, RGBs number $>$50. RGBs are not expected to be radio sources, so this implies that there were some misclassification, or that these RGB are not the true counterparts to the radio sources. Due to the significant differences in resolution between both data sets, mismatching is not unexpected. The separation between the VMC coordinates and ASKAP coordinates for the RGB class is in general larger than AGN ($\sim$1.35$^{\prime\prime}$ for RGB compared to $\sim$1$^{\prime\prime}$ for AGN), which implies that these RGB sources could be merely mismatches.

Of the other stellar sources, H\textsc{ii}/YSO and PNe are expected to be associated with a radio detection, and foreground stars are close enough that a radio detection is possible. One of the brightest and well known radio sources in the LMC is supernova SN\,1987A, which matched with 2 sources in the PRF catalogue within 1 arcsec. Both of which had a classification of HII/YSO  with probability 31-35\%, with the next highest class as AGN with probability 22-27\%. So the classifier did not know what to class it as and did not put it in the Unknown class, proving that it is quite a unique source.

In relation to the full catalogue of sources with P$_{\rm class}$ $>$ 80\%, the fraction of AGN with a radio detection is $\sim$1.24\% and $\sim$3.89\% and for galaxies is $\sim$1.58\% and $\sim$2.54\%, for the SMC and LMC, respectively. The expected radio loud population is about 10\%, but ours is limited to the likeliest AGN, i.e. those that are well sampled in the training data, so its possible that the missing fraction is in the lower confident AGN population and/or the Unknown class.

\begin{figure*}
\centering
\begin{tabular}{c}
	\includegraphics[width=1\textwidth]{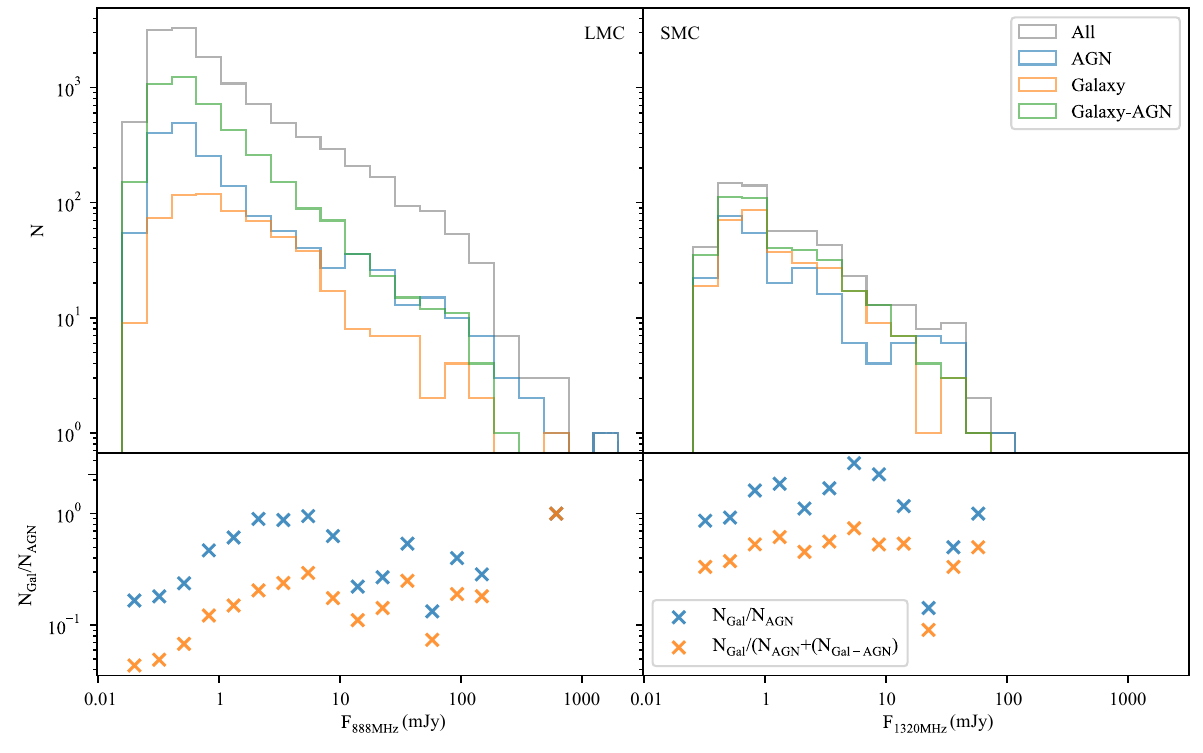}
 \end{tabular}
    \caption{Radio LMC ASKAP 888 MHz (left) and SMC ASKAP 1320 MHz (right) flux density distributions and the ratio of galaxy to AGN counts (bottom panels) of the predicted AGN and galaxy sources with P$_{\rm class}$ $>$ 80\%. A galaxy-AGN subset has also been included where the individual AGN and galaxy probabilities are $<$ 80\%, but the combined probabilities are $>$ 80\%.}
    \label{RadioPRF}
\end{figure*}

It has been seen that there is an upturn in the number of sources towards fainter flux densities, representing the beginning of the faint galaxy population, as well as the radio quiet AGN population \citep[e.g.][]{2021MNRAS.506.3540P}. Therefore, it is expected that the number of radio detected galaxies will have increased towards the fainter fluxes. Looking at the radio flux density distribution at 888 MHz (LMC) and 1320 MHz (SMC), and the ratio of galaxy to AGN counts in Figure \ref{RadioPRF}, it can be seen at the brightest fluxes, there are less galaxies compared to AGN, as expected, and towards lower flux densities the number of galaxies compared to AGN increases, also as expected. However, at about F$_{\rm888MHz}$, F$_{\rm1320MHz}$ $<$ 7 mJy, the number of galaxies compared to AGN starts deccreasing towards lower flux densities, which is unexpected, as at lower flux densities we would expect an increase in faint galaxies, but it could be that the faint AGN population is easier to classify than the faint galaxy population. The trend is less clear for the SMC where source statistics are poorer. 

It should be noted that the AGN and galaxies being compared here only represent the more confident classifications of the PRF, a fraction of the true number in this field that will have been detected with ASKAP. More AGN than galaxies are also found, most likely due to them being easier to detect at higher redshifts, which could account for the increase towards lower radio flux densities, as we are identifying the radio quiet AGN, but not the fainter star-forming galaxies. It should also be noted that a source classed as a galaxy with a radio association could indicate AGN activity that is not visible at other wavelengths.

The 1320 MHz band was used over the 960 MHz for the SMC, as the observation at 960 MHz did not use the full ASKAP array, but the 1320 MHz observation did, and would therefore have similar depths to the LMC 888 MHz observation.

\subsection{Classifications of the X-ray population}\label{Xray}

Cross-matching with the XMM-Newton catalogues for the SMC \citep{sturm2013} and all-sky \citep{2020XMM} catalogues at a cross-matching radius less than the positional error in the X-ray coordinates for each source yields a total of 627 and 4794 X-ray sources with reliable PRF classifications (P$_{\rm class}$ $>$80\%) for the SMC and LMC, respectively. The number of sources per class can be seen in Table \ref{tab:Xraycounts}. Most of the sources are Unknown. Of the other classifications, the highest count is AGN, as expected.

\begin{table}
    \caption{The number of sources per class that were cross-matched with XMM-Newton within the positional error of each source's coordinates.}
    \centering
    \begin{tabular}{lrrr}
        \hline\hline
         Class&   All & 60\%$<$ P$_{\rm class}$ $<$80\% & P$_{\rm class}$ $>$80\%\\
         \hline
         SMC & 1607 & 381 & 627\\
         \hline
         Unk&  823 & 235 & 411\\
         AGN&  364 & 109 & 118\\
         Galaxy&  25 & 6 & 4\\
         RGB&  103 & 23 & 39\\
         OB&  173 & 3 & 36\\
         H\textsc{ii}/YSO &  9 & 0 & 0\\
         AGB&  17 & 0 & 1\\
         PNe&  44 & 0 & 1\\
         RSG&  12 & 1 & 1\\
         Post-RGB/AGB& 4 & 2 & 0\\
         PM& 33 & 2 & 16\\
         \hline
         LMC & 10139 & 1499 & 4794\\
         \hline
         Unk&  7157 & 1066 & 4127\\
         AGN& 1277  & 324 & 436\\
         Galaxy& 84  & 17 & 7\\
         RGB&  192 & 30 & 105\\
         OB& 278  & 11 & 65\\
         H\textsc{ii}/YSO &  213 & 21 & 4\\
         AGB&  727 & 4 & 7\\
         PNe&  14 & 0 & 0\\
         RSG& 13  & 0 & 2\\
         Post-RGB/AGB & 0 & 0 & 0\\
         PM& 184 & 26 & 41\\
         \hline
    \end{tabular}
    \label{tab:Xraycounts}
\end{table}

From the cross-matching it can be found that for sources with P$_{\rm class}$ $>$ 80\%, $\sim$1.49\% and $\sim$1.02\% of all AGN are X-ray detected, for the SMC and LMC respectively. However, the XMM-Newton surveys of the SMC and LMC do not cover the full VMC survey areas, and concentrate mainly in the centres of the Clouds, so the true percentages are most likely much higher.

\begin{figure*}
\centering
\begin{tabular}{c}
	\includegraphics[width=1\textwidth]{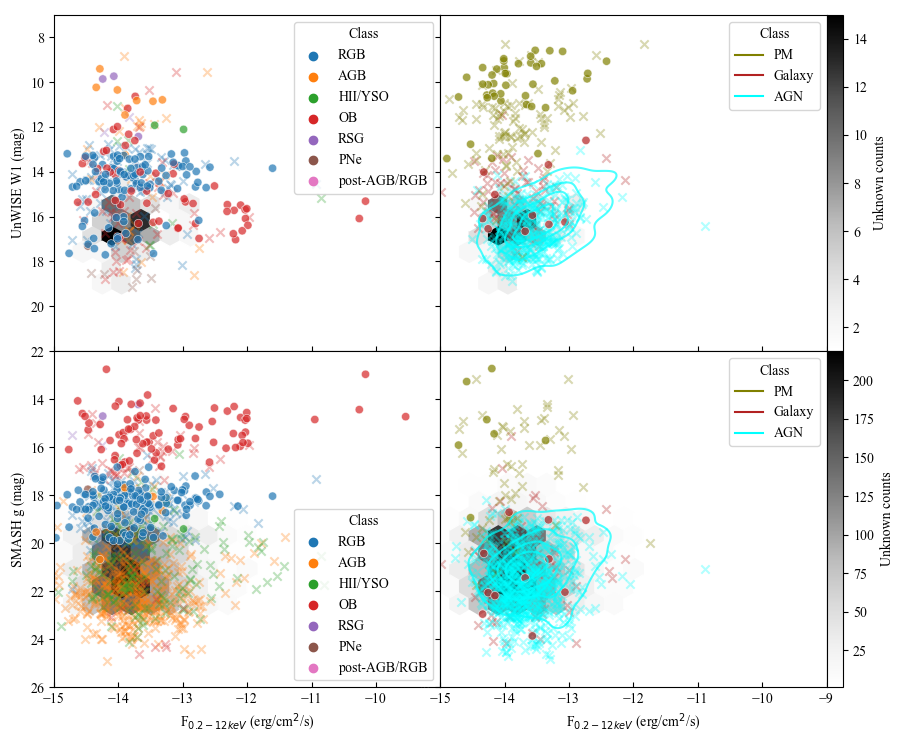}
 \end{tabular}
    \caption{X-ray flux vs unWISE W1 (top) and SMASH g (bottom) bands for the Magellanic (left) and extragalactic and foreground sources (right). The high confidence (P$_{\rm class}$ $>$ 80\%) Unknowns are plotted in the background of all plots. Contours and filled circles refer to sources with P$_{\rm class}$ $>$ 80\%, whilst crosses refer to sources with P$_{\rm class}$ $<$ 80\%. The contours represent a probability distribution in intervals of 0.2.}
    \label{XrayPRF}
\end{figure*}

In Figure \ref{XrayPRF}, we plot the unWISE W1 and SMASH g band magnitudes as a function of X-ray flux for Magellanic (left) and extragalactic and foreground stars (right). In these plots the extragalactic and stellar sources tend to separate. The extragalactic tend to concentrate at the fainter optical/IR magnitudes for a given X-ray flux \citep[e.g.][]{2001Hornschemeier, 2015Nandra, 2015Civano, 2018Salvato}.

Sources classified as Unknown which have X-ray detections are mainly concentrated on the extragalactic side of the plots. This side of the plots is mainly occupied by AGN, so a tentative AGN label can be assigned to these Unknowns. These sources tend to be the fainter sources, where there are fewer spectroscopic observations for the classifier to learn from. When plotted on an AllWISE colour--colour diagram, such as the one seen in Figure \ref{WISEPRF}, the Unknown sources (P$_{\rm class}$ $>$ 80\%) that are bright enough to be detected in the W3 band (12 sources) tend to concentrate where galaxies are expected, with a few falling into the region where the AGN area overlaps the galaxy area. It is possible that these are obscured AGN, which the training set for AGN would have been biased against. Ultraluminous infrared galaxies (ULIRGs) tend to be found in the lower right corner where AGN and galaxies overlap, which supports that these sources are more obscured.

The X-ray catalogue for the SMC from \cite{sturm2013} also provides classifications within their catalogue. PRF classed OB stars that have cross-matches within the SMC X-ray catalogue indicate that sources classed as OB by the PRF are a mix of foreground stars and high mass X-ray binaries (HMXB), which appear to separate in the unWISE W1 band, with foreground stars tending to be brighter, and HMXBs fainter and closer to extragalactic magnitudes. The OB stars that are classed as HMXB in \cite{sturm2013}, among which is SMC\,X-1, are grouped with the other stellar sources in optical, however. This could indicate that X-ray sources that are grouped with the extragalactic at IR magnitudes but grouped with the stellar sources at optical magnitudes could be given a tentative HMXB classification.

Furthermore, 53 out of 54 HMXBs found in the LMC \citep{2019Maitra, 2021aMaitra, 2021bMaitra, 2022Haberl, 2023Haberl} were also classified as OB by the PRF. These inlcude the well known LMC\,X-1 and LMC\,X-4 with P$_{OB}$ > 96\%. Of these sources, those with the lowest probabilities  (<70\%) for being an OB star were found to have a ~10-20\% probability of being a YSO or PNe (LMC\,X-3, 4XMM J052546.5-694451 and 4XMM J052417.1$-$692533). Of these three sources, LMC\,X-3 is atypical as it is a Roche Lobe filling Black hole binary in the LMC with a B3V/B5V companion \citep[e.g.][]{1983Cowley, 2001Soria}. 4XMM J052546.5-694451 and 4XMM J052417.1$-$692533 are questionable HMXB candidates due to lack of expected H$\alpha$ emission from a decretion disc \citep{2018Jaarsveld}, which makes their lower confidence classifications not unexpected. The source that could not be classified, RX J0512.6$-$6717, was a ROSAT selected candidate and had a large position error circle of 7$^{\prime\prime}$, within which the Gaia counterpart could not be identified. However, cross-matching with the PRF catalogue with a matching radius 7$^{\prime\prime}$ gave 11 sources, one of which (at 78.17215, $-$67.28993) was classed as an OB star at P$_{Class}$ $\sim$ 88\%. If confirmed to be the true counterpart, this shows the PRF has the potential to match X-ray detections to their optical/near-IR counterparts.

The true identity of the sources classed as RGB stars remains in contention. RGB stars should not have observable X-ray emission at Magellanic distances, so their inclusion in the X-ray detected sources is questionable. As with the radio detected sources, it is possible that if the RGB class is the correct classification, then they are not the true counterpart. In Figure \ref{XrayPRF} the RGB-classified sources are concentrated between the stellar and extragalactic, which also happens to be the location of some of the few galaxies with X-ray emission. It is possible that these RGB-classified sources are actually red dusty star-forming galaxies. However, when looking at the distribution at separation between VMC and XMM-Newton coordinates it can be seen that the AGN have a median of 0.59$^{\prime\prime}$ and 0.71$^{\prime\prime}$ for the SMC and LMC, respectively, whilst RGBs display a median of 0.97$^{\prime\prime}$ and 1.22$^{\prime\prime}$, which might indicate that these RGBs could be merely spurious alignments.

Overall, the X-ray detected sources are classified as expected, and that the sources classified as Unknown which have a corresponding X-ray detection can be a tentative extragalactic/AGN classification. We've also shown that sources classified by the PRF with an associated X-ray detection are HMXBs, and that the PRF classifier has great potential to find the optical/near-IR counterparts to X-ray detections.

\subsection{Quaia comparison}\label{Quaia}

The Quaia survey \citep{2023Quaia} is an all-sky spectroscopic quasar catalogue that used low-resolution BP/RP spectra from \textit{Gaia} to identify AGN candidates, and has had cuts applied based on \textit{Gaia} brightness ($G$ $<$ 20.5 mag), proper motions and unWISE colours to obtain a purer sample. Cross-matching Quaia with the VMC survey with a cross-matching radius of 1$^{\prime\prime}$ yields 4325 and 1906 sources in the LMC and SMC, respectively. This leaves 34 and 5 sources in Quaia that are inside the LMC and SMC VMC survey footprints, respectively, that were not matched with a source in the VMC. It is surprising that sources that are $G$ $<$ 20.5 mag are not picked up in the deeper VMC catalogue.

Of the sources in the VMC survey, 427 were in the training set. 422 of these were known AGN and two were known galaxies. However, there was one known pAGB/RGB from the LMC, SMP LMC 11, and one known AGB from the SMC, OGLE SMC-SC5 255936. These classes are known to be variable sources, which could have played a part in their misclassification in Quaia.

After removing the sources in the training sets there were 4072 and 1733 sources in the LMC and SMC left, respectively. The highest fraction of sources is, as expected, classed as AGN (3335/4072 and 1622/1694 for the LMC and SMC, respectively). Five sources from the LMC footprint are classed as galaxies, with the second likeliest classification of an AGN, so these are most likely galaxies with an AGN component.

There are four sources classed as AGB from the SMC and seven from the LMC. Seven of these have P$_{\rm class}$ $>$ 80\% of being an AGB. With the one known AGB being classed as an AGN by Quaia, it is possible these sources are actually AGB. The \textit{Gaia} proper motions would not be as capable of separating the stars in the Magellanic Clouds as in the Milky Way, unWISE colours (W1 $-$ W2, see Figure \ref{WISEPRF}) could have been in the AGN regime and the low-resolution spectra might not have provided good enough S/N to make a good classification. Furthermore, AGB are known to vary regularly and Simbad contains matches for 8/11 of the AGB, and they are all classed as variable stars. 

There are also 18 sources classed as H\textsc{ii}/YSOs in the Quaia sample. Only two have mid-confidence of being this class (P$_{\rm class}$ of $\sim$ 79\% and 75\%), the rest have P$_{\rm class}$ $<$ 60\%. The majority of these sources also have AGN as the next likeliest class. 

Five sources in the SMC were classed as RGB stars with P$_{\rm class}$ $<$ 80\%, with five sources with $<$ 60\%. One source in the LMC was classed as an OB star but this had a P$_{\rm class}$ $<$ 60\%, so is unlikely to be the true class.

Lastly, 99 and 709 sources from the SMC and LMC, respectively were classed as Unknown, suggesting that there is possibly a population of AGN that are being missed by the classifier. The sources the PRF classed as AGN had a magnitude range of 16.3 $<$ $G$ $<$ 20.5 mag, whilst the Unknown classed sources had a magnitude range of 18.8 $<$ $G$ $<$ 20.5 mag. This shows that the Unknown Gaia QSOs are fainter examples, for which there are less spectroscopically observed sources to train  the classifier upon.

Looking at the Quaia sources that had no match in the PRF catalogues, VISTA images revealed that at the majority of the Quaia coordinates there are multiple sources in close proximity at the expected coordinates. If there is a galaxy underneath an AGN a larger separation is possible, as centres for galaxies are more difficult to establish. The majority ($>$ 90\%) of Quaia sources that matched with the VMC had a match within $<$ 0.2$^{\prime\prime}$. We increased the cross-matching radius to 5$^{\prime\prime}$ and matched the Quaia sources with no match in the PRF catalogue again to find that all but one of the sources had matches between 1 -- 1.8$^{\prime\prime}$. Amongst the matches are 15 AGN, 4 galaxies, 14 Unknown and five low confidence stellar classes. The majority of these classes being extragalactic implies that these are the true counterparts, though may not be as reliable due to the distance between counterparts.

Overall, the majority ($\sim$86\%) of AGN found in Quaia are found by the classifier. This is not unexpected, as these are most likely the more obvious type I (broad-line) AGN, which make up the most of the AGN class in the training set. The type II (narrow-line) AGN, however, are harder to test against as spectroscopic samples tend to be biased towards type I, so the sources classed as Unknown are more likely the type II AGN.

\subsection{YSOs in the LMC}\label{YSOs}
YSOs are dusty sources that can exhibit emission lines, making them easily confused with AGN, therefore ascertaining that YSOs are not being misclassified as AGN or vice versa by the PRF is necessary. A recent study by \citet{2023Kokusho} has compiled all the YSOs in the area of the LMC, numbering 4097, the majority of which are candidates located using photometry and SED fitting. The number of H\textsc{ii}/YSOs the PRF detects is greater in the LMC than the SMC, and shows signs of structure, see Figure \ref{LMCYSO} (See Appendix Section \ref{App_dist}, Figure \ref{Dist_yso80} for the SMC). 

\begin{figure}
\centering
	\includegraphics[width=1\columnwidth, trim=5mm 1mm 4mm 2mm, clip]{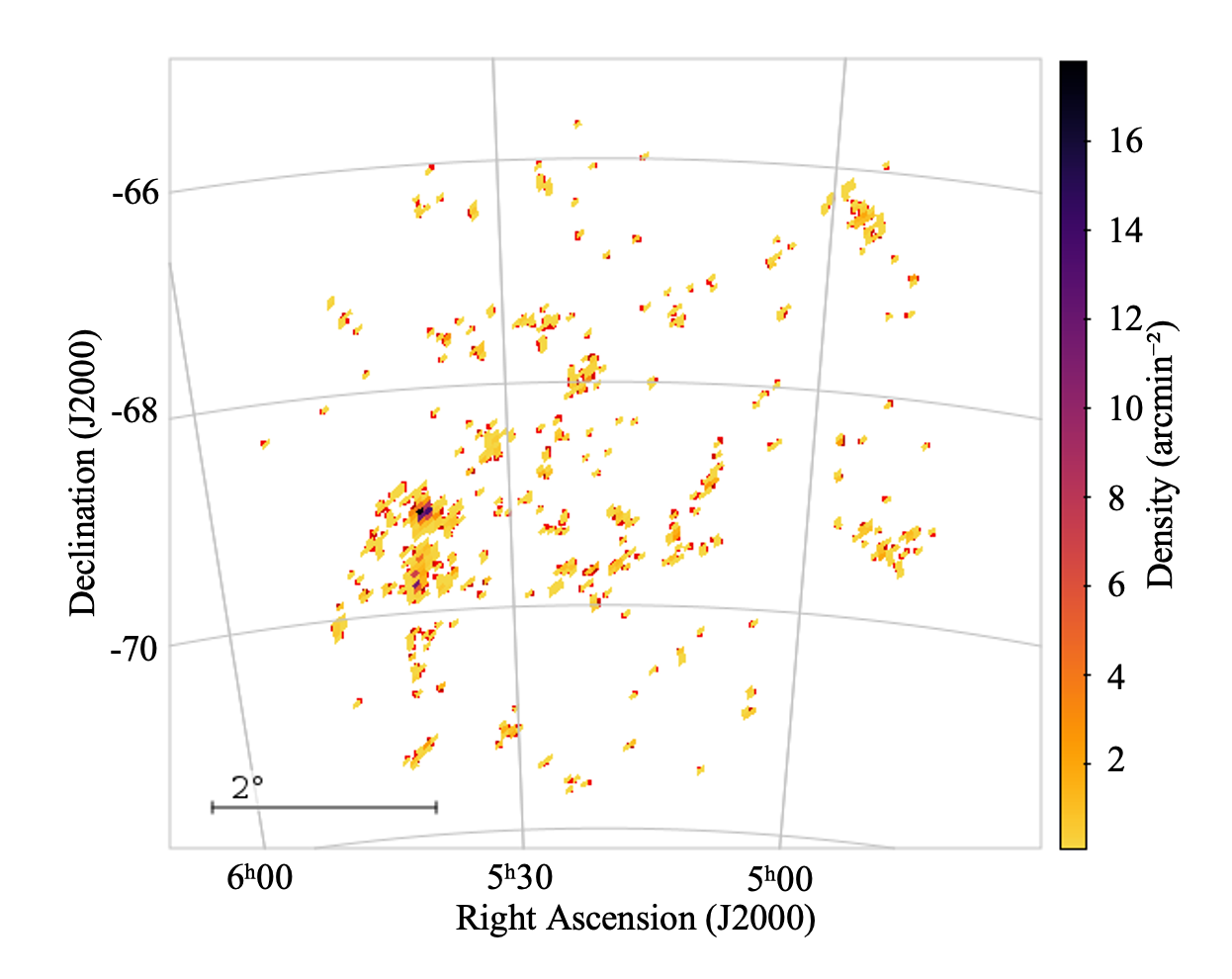}
    \caption{The sky density distribution of the H\textsc{ii}/YSOs with P$_{\rm class}$ $>$ 80\% in the LMC.}
    \label{LMCYSO}
\end{figure}

Figure \ref{LMCYSO} shows an expected distribution, with easily identifiable structures such as 30 Dor and the Southern molecular ridge below it. N11 is seen to the right and most of the Henize H\textsc{ii} regions scattered across the face of the LMC.

Cross-matching with the PRF results with a matching radius of 1$^{\prime\prime}$ yields 2715 matches, and restricting to those not in the training set, numbers 2274. Further restricting to only those with P$_{\rm class}$ $>$ 80\% gives 630 sources.

Of the 630 high-confidence sources, 226 are classified as H\textsc{ii}/YSOs and 117 are classed as Unknown, where the Unknown sources tend to be the fainter sources. 105 and 6 are classed as AGN and galaxies, respectively. The rest are classified as OB (58), AGB (49), PNe (8), RGB (5), post-AGB/RGB (1) and PM (1).

About a sixth of the YSOs have been classed as an extragalactic class. It should, however, be noted that of the sources from this study that are also in the PRF training set (441 sources), some are spectroscopically confirmed to be other classes. 297 of them are confirmed as H\textsc{ii}/YSOs. 47 and 11 are spectroscopically confirmed as AGN and  galaxies, respectively. The rest are classed as AGB (24), post-AGB/RGB (22), OB (20), PNe (18), RGB (1) and RSG (1). So it is not unexpected that not all the sources classed as YSOs in \citet{2023Kokusho} are classed as such here.

\subsection{The Unknown class}\label{Unknownclass}

The Unknown class represents all the sources that were not given a similar enough representative in the training set. The majority of these sources are the fainter sources that the photometry for is mostly missing and that we have little spectroscopy for due to the limits of ground based spectroscopy instruments. There are, however, still brighter sources amongst the Unknowns, that are made up of classes and/or sub-classes that were not accounted for in the training sets due to lack of available spectroscopy. 

The distribution of the Unknown class can be seen in the Appendix D in Figure C10. From this it can be seen that the Unknowns cover the entirety of the VMC LMC field, and the area covered by the SMASH survey for the VMC SMC field. For the SMC field this implies that the confident Unknown sources are those with coverage from most if not all the photometric surveys, indicating that these sources, specifically the brighter sources with fewer missing data, do not have a match in the training set, and that they are classes that are not trained on, rather than not having enough information. 

Selection criteria for selecting stars of different ages using colour--magnitude diagrams have been created for the SMC and LMC \citep{2014Cioni,2016Cioni,2019Dalal} for sources down to K$_{s}$ $>$ 19.8 mag and only those with a 70\% probability of being stars and with photometric uncertainties <0.1 mag. We apply the selection criteria from \cite{2019Dalal} to the confident Unknowns to provide a tentative classification and to discover the most underrepresented classes in the training set for the Magellanic sources.

After applying the selection criteria to the Unknowns,  91\% and 71\% of sources for the LMC and SMC, respectively, were too faint for the selection cut. The division of the sources that were bright enough can be seen in Table \ref{Unk_CMD}. See Appendix Section \ref{UNKselCMD} for a figure showing the selection criteria plotted on top of the Unknown distribution in near-IR CMD space.

\begin{table}
    \caption{The number of Unknowns with K$_{s}$ $<$ 19.8 mag (Vega)  and P$_{\rm class}$ $>$ 80\% that are categorised as different stellar populations based on the near-IR CMD selection from \citet{2019Dalal}, for only those sources with a 70\% probability of being stars and with photometric uncertainties <0.1 mag. Note that `All' represents all sources with K$_{s}$ $>$ 19.8 mag. The table of sources with the VMC colour--magnitude classification will be made public with the final VMC public release.}
    \centering
    \begin{tabular}{llrr}
        \hline\hline
         Region&   Dominant stellar & SMC & LMC\\
          & population & & \\
         \hline
         All & & 875,233 & 13,077,735  \\
         \hline
         A &  Main-sequence & 34,032 ($\mbox{~}$4\%) & 200,596 ($\mbox{~}$2\%)\\
         B & Main-sequence  & 136,738 (16\%) & 886,994 ($\mbox{~}$7\%)\\
         C & Main-sequence  & 150,780 (17\%) & 2,216,757 (17\%)\\
         D & Main-sequence  & 132,100 (15\%) & 3,063,970 (23\%)\\
          & and subgiants & & \\
         E & RGB  & 306,060 (35\%) & 4,459,540 (34\%)\\
         F &  Milky Way & 43,686 ($\mbox{~}$5\%) & 532,015 ($\mbox{~}$4\%)\\
         G & Supergiants and  & 1 ($<$1\%) & 0\\
          & giant stars & & \\
         H & Supergiants and  & 0 ($<$1\%) & 16 ($<$1\%)\\
          & giant stars & & \\
         I & Supergiants and  & 21 ($<$1\%) & 32,360 ($<$1\%) \\
          & giant stars & & \\
         J & Red clump stars & 420 ($<$1\%) & 1,325,249 (10\%)\\
         K & RGB & 7 ($<$1\%) & 296,100 ($\mbox{~}$2\%)\\
         L & Extragalactic & 71,378 ($\mbox{~}$8\%) & 64,138 ($<$1\%) \\
         \hline
    \end{tabular}
    \label{Unk_CMD}
\end{table}

Some of the stellar populations are separated into multiple regions to represent different average stellar ages (for full details see \citet{2019Dalal}). For example, the main-sequence stars in region A tend to be younger than the stars in region B.

From Table \ref{Unk_CMD} we can see that, as expected, the main sequence stars (populations A, B, C and D) make up the majority of the Unknowns, most likely due to their lack of corresponding class in the training sets, where only O and B main sequence stars are accounted for. RGBs (populations E and K) are the next largest population in the Unknowns, despite the majority of sources already being classified as RGB by the PRF, implying that we are not capturing the full scope of the RGB star class with our training sets for the PRF, despite it already being the largest class. 

Population L is where extragalactic sources can be found. From Table \ref{Unk_CMD} it can be seen that $\sim$ 135,000 sources across the Clouds can be given a tentative extragalactic classification. Additional information would be required to further separate the sources into AGN and galaxies. 

Another way of giving a tentative classification to sources is by using wavelengths that were not used in the PRF. A source being associated with an X-ray or radio detection tends to imply an extragalactic source rather than a stellar source, as seen in Sections \ref{Xray} and \ref{Radio}. This means we can give a tentative extragalactic classification, though with the caveat that cross-matching can lead to contamination from spurious alignments with Magellanic or foreground stars and that there are some stellar objects that do emit in the X-ray and radio, though they tend to be easy to pick out in the X-ray. 


To improve the success of this machine learning technique, more spectroscopy is needed to bolster the training set (for both stellar and extragalactic populations), especially for fainter sources and those detected in IR but not in the optical, among which would be higher redshift galaxies/AGN, as well as heavily dust reddened sources. Therefore, spectroscopy from telescopes such as the JWST \citep[][]{2006JWST}, 4MOST \citep{4MOST} and WEAVE \citep{WEAVE} would provide greater potential in identifying the fainter extragalactic population. The training set could potentially be augmented to reduce selection bias by using simulations/models, or by taking observational data of high redshift sources from a deep field survey, modelling the spectral energy distribution and then estimating what the surveys used in the area of the VMC would measure in each waveband. Another method would be to take the known sources and model how they would look at fainter magnitudes to regain some of the fainter sources from the Unknown class.

In terms of features, we have selected the best current survey data available with unique contributions. The VISTA data is deep and is complemented by the optical SMASH data, though the SMASH survey does not cover the entirety of the VMC fields (see Figure \ref{MCcoverage}). The future Legacy Survey of Space and Time \citep[LSST;][]{2019LSST} would be an improvement on the SMASH data as it would cover all of the VMC field and would reach comparable depths in its first data release. Furthermore, there is no current complementary mid-IR photometry that reaches the same depths. Mid-IR is a wavelength range that is particularly powerful in identifying AGN due to being sensitive to the emission from the dust surrounding the accretion disk. Therefore, photometry from an IR telescope which reaches greater depths than \textit{WISE} and \textit{Spitzer} would provide greater potential in identifying the extragalactic.

\section{Conclusions}\label{conclusions}
In summary, we trained a probabilistic random forest on the UV -- IR photometry of spectroscopically observed sources in the field of the VMC survey, augmented with AGN and galaxies from the SDSS observations of the GAMA09 field. This yielded overall accuracies of 0.79 $\pm$ 0.01 for the SMC classifier, and 0.87 $\pm$ 0.01 for the LMC classifier. For the extragalactic classifications the classifiers yielded accuracies of 0.93 $\pm$ 0.01 for both classifiers. When restricted to P$_{\rm class}$ $>$ 80\% the accuracy of the classifiers were 0.98 $\pm$ 0.01 and 0.90 $\pm$ 0.01 for the LMC and SMC, respectively. 

The classifiers were used on the entirety of the LMC and SMC PSF catalogues and the sources were separated into three catalogues with different ranges of probabilities of the classification being correct (low-confidence -- P$_{\rm class}$ $<$ 60\%, mid-confidence -- 60\% $<$ P$_{\rm class}$ $<$ 80\%, high-confidence -- P$_{\rm class}$ $>$ 80\%). At high-confidence we classify a total of 707,939 and 397,899 sources in the SMC and LMC, respectively, with a total of 50,507 AGN and 27,146 galaxies (>49,500 and >26,500 of which, respectively, are new candidates) across the two Clouds. Looking at the high-confidence classifications we find:
\begin{itemize}
    \item The spatial distributions of the different classes across the VMC fields of the SMC and LMC are as expected. The extragalactic and foreground sources being mostly homogeneous across the field and the Magellanic sources concentrating in the centres of the Clouds with fewer sources towards the edges of the fields.
    \item We tested the classifiers on stellar and extragalactic sources that are known to be confused with each other from \citet{2022MNRAS.515.6046P}. The results showed that all the AGN were classified correctly. 
    We showed that even sources that are often confused with another class are well classified by the classifiers. However, emission-line stars have the possibility of being classed as AGN, though not necessarily with high confidence.
    \item We tested the behaviour of the classifiers on classes which they have not trained upon \citep[65 sources from the SAGE-spec catalogues,][]{Ruffle2015, Jones2017}. 
    From this we found that for all stellar sources the classifiers classified them as another stellar class. For the two sources that were classified as an AGN their spectral classification was not known so they may have been AGN. This means that stellar classes that we have not trained upon are unlikely to be misclassified as extragalactic sources.
    \item Plotting the sources across optical, near-IR and mid-IR colour--colour and colour--magnitude diagrams showed that the classes separated as expected, and that where the classes do overlap the classifiers are still able to discern between the different classes. This shows that the large array of features from optical to far-IR is helping to separate sources that would have been otherwise hard to untangle in single colour--colour/magnitude diagrams.
    \item Investigating the sources that had a corresponding ASKAP 888/960/1320 MHz radio or XMM-Newton X-ray detection showed that, as expected, the majority of the radio/X-ray detected sources were (when restricting to sources not classed as Unknown) predominantly classed as extragalactic ($\sim$ 89\% and $\sim$ 64\%, respectively).
    \item The proportions of radio AGN and galaxies were found to vary with radio flux density. The brightest flux densities are dominated by AGN, then towards lower flux densities the fraction of galaxies increases as we start to pick up fainter emission from star-formation from galaxies, as expected. Unexpectedly, at about F$_{\rm888MHz}$, F$_{\rm1320MHz}$ $<$ 7 mJy, the number of galaxies compared to AGN starts decreasing again. We expect this could be accounted for by selection bias, where the faint AGN population is easier to classify than the faint galaxy population.
    \item Quaia survey \citep{2023Quaia} AGN candidates were predominantly classified as AGN ($\sim$ 85\%), as expected. Only $\sim$ 14\% of the Quaia AGN candidates were classed as Unknown. This implies that the sample of AGN in the Quaia catalogue are well represented by the AGN training sample for the PRF classifiers, which are mostly made up of the bright broad-line AGN. Those classed as Unknown are possibly the underrepresented narrow-line region AGN. 
    \item 
    VMC near-IR colour--colour magnitude diagrams of the brightest Unknown sources (K$_s$ < 19.8 mag) revealed that the main classes missing from the classifier are main-sequence stars and fainter examples (than are currently in the training set) of Milky Way stars (PM) and RGB stars.
    \item It is also possible to give a tentative extragalactic classification to Unknowns that have X-ray or radio counterparts. Plotting unWISE W1 and SMASH g bands against XMM-Newton flux showed that the majority of the Unknowns lie in the regions in these plots occupied by AGN and galaxies. However, the possibility of spurious alignments does lower the reliability of this.
\end{itemize}

The majority ($\sim$71\% for all sources, $\sim$98\% for sources with P$_{\rm class}$ $>$ 80\%) of sources are classed as Unknown. Whilst some of this is due to some missing classes such as main-sequence stars (other than O and B types) and fainter Milky Way stars, this is mostly due to these sources being fainter than the spectroscopically observed examples we provided the classifier to be trained upon. Therefore more spectroscopy of fainter sources is required. For the sources that are within the brightness range of the training set, but were still classed as Unknown, more classes are required, such as main sequence stars that are not of type O or B, as well as fainter examples of Milky Way stars. Deeper photometry from surveys such as optical LSST, as well as a complimentary mid-IR survey, would be preferable to reduce the amount of faint sources in the VMC catalogue with the majority of features missing. However, without the spectroscopy for sources at fainter magnitudes to train upon, the majority will remain Unknown.

\section*{Acknowledgements}
We thank the anonymous referee for their feedback, which helped improve the paper. C.M.P.\ and acknowledges funding from a STFC studentship and from a UKRI Future Leaders Fellowship (grant: MR/T020989/1). J.E.M.C.\  acknowledges funding from a STFC studentship. This research was supported in part by the Australian Research Council Centre of Excellence for All Sky Astrophysics in 3 Dimensions (ASTRO 3D), through project number CE170100013.

 This project has received funding from the European Research Council (ERC) under the European Union’s Horizon 2020 research and innovation programme (grant agreement no. 682115). We thank the Cambridge Astronomy Survey Unit (CASU) and the Wide Field Astronomy Unit (WFAU) in Edinburgh for providing the necessary data products under the support of the Science and Technology Facility Council (STFC) in the UK.

This paper uses observations made at the South African Astronomical Observatory (SAAO) 1.9m under programme Pennock-2019-05-74-inch-257, and with the Southern African Large Telescope (SALT) under programmes 2021-1-SCI-018 (PI: van Loon), 2021-1-SCI-029 (PI: van Loon), 2021-1-SCI-032 (PI: van Loon) and 2021-2-SCI-017 (PI: Anih). We would like to thank Francois van Wyk for his assistance in acquiring the observations at the 1.9m telescope during late-COVID lockdown.
 
This research made use of \textsc{Astropy},\footnote{\url{http://www.astropy.org}} a community-developed core Python package for Astronomy \citep{2013A&A...558A..33A, 2018AJ....156..123A}. We have made extensive use of the SIMBAD Database at CDS (Centre de Données astronomiques) Strasbourg, the NASA/IPAC Extragalactic Database (NED) which is operated by the Jet Propulsion Laboratory, CalTech, under contract with NASA, and of the VizieR catalog access tool, CDS, Strasbourg, France.

\section*{Data Availability}


The results table of source coordinates and corresponding classifications, as well as the VMC colour--magnitude classifications of the Unknowns, will be made available on the VSA\footnote{\url{http://vsa.roe.ac.uk}} (VISTA Science Archive) ESO archive\footnote{\url{http://archive.eso.org}} as part of VMC DR7, as well as on CDS\footnote{\url{https://cds.unistra.fr}} (Centre de Données astronomiques) when the paper is published. The training data for the SMC and LMC classifiers and the table of feature importances for the two classifiers created in this work are available as supplementary material to this article.



\bibliographystyle{mnras}
\bibliography{Ref} 

\begin{thebibliography}{}
\makeatletter
\relax
\def\mn@urlcharsother{\let\do\@makeother \do\$\do\&\do\#\do\^\do\_\do\%\do\~}
\def\mn@doi{\begingroup\mn@urlcharsother \@ifnextchar [ {\mn@doi@} {\mn@doi@[]}}
\def\mn@doi@[#1]#2{\def\@tempa{#1}\ifx\@tempa\@empty \href {http://dx.doi.org/#2} {doi:#2}\else \href {http://dx.doi.org/#2} {#1}\fi \endgroup}
\def\mn@eprint#1#2{\mn@eprint@#1:#2::\@nil}
\def\mn@eprint@arXiv#1{\href {http://arxiv.org/abs/#1} {{\tt arXiv:#1}}}
\def\mn@eprint@dblp#1{\href {http://dblp.uni-trier.de/rec/bibtex/#1.xml} {dblp:#1}}
\def\mn@eprint@#1:#2:#3:#4\@nil{\def\@tempa {#1}\def\@tempb {#2}\def\@tempc {#3}\ifx \@tempc \@empty \let \@tempc \@tempb \let \@tempb \@tempa \fi \ifx \@tempb \@empty \def\@tempb {arXiv}\fi \@ifundefined {mn@eprint@\@tempb}{\@tempb:\@tempc}{\expandafter \expandafter \csname mn@eprint@\@tempb\endcsname \expandafter{\@tempc}}}

\bibitem[\protect\citeauthoryear{{Adams} et~al.,}{{Adams} et~al.}{2013}]{2013Adams}
{Adams} J.~J.,  et~al., 2013, \mn@doi [\apj] {10.1088/0004-637X/771/2/112}, \href {https://ui.adsabs.harvard.edu/abs/2013ApJ...771..112A} {771, 112}

\bibitem[\protect\citeauthoryear{{Ahumada} et~al.,}{{Ahumada} et~al.}{2020}]{2020SDSS}
{Ahumada} R.,  et~al., 2020, \mn@doi [\apjs] {10.3847/1538-4365/ab929e}, \href {https://ui.adsabs.harvard.edu/abs/2020ApJS..249....3A} {249, 3}

\bibitem[\protect\citeauthoryear{{Assef} et~al.,}{{Assef} et~al.}{2013}]{Assef2013}
{Assef} R.~J.,  et~al., 2013, \mn@doi [\apj] {10.1088/0004-637X/772/1/26}, \href {https://ui.adsabs.harvard.edu/abs/2013ApJ...772...26A} {772, 26}

\bibitem[\protect\citeauthoryear{{Assef}, {Stern}, {Noirot}, {Jun}, {Cutri}  \& {Eisenhardt}}{{Assef} et~al.}{2018}]{2018Assef}
{Assef} R.~J.,  {Stern} D.,  {Noirot} G.,  {Jun} H.~D.,  {Cutri} R.~M.,   {Eisenhardt} P.~R.~M.,  2018, \mn@doi [\apjs] {10.3847/1538-4365/aaa00a}, \href {https://ui.adsabs.harvard.edu/abs/2018ApJS..234...23A} {234, 23}

\bibitem[\protect\citeauthoryear{{Astropy Collaboration} et~al.,}{{Astropy Collaboration} et~al.}{2013}]{2013A&A...558A..33A}
{Astropy Collaboration} et~al., 2013, \mn@doi [\aap] {10.1051/0004-6361/201322068}, \href {https://ui.adsabs.harvard.edu/abs/2013A&A...558A..33A} {558, A33}

\bibitem[\protect\citeauthoryear{{Astropy Collaboration} et~al.,}{{Astropy Collaboration} et~al.}{2018}]{2018AJ....156..123A}
{Astropy Collaboration} et~al., 2018, \mn@doi [\aj] {10.3847/1538-3881/aabc4f}, \href {https://ui.adsabs.harvard.edu/abs/2018AJ....156..123A} {156, 123}

\bibitem[\protect\citeauthoryear{{Bell} et~al.,}{{Bell} et~al.}{2019}]{Bell2019}
{Bell} C. P.~M.,  et~al., 2019, \mn@doi [\mnras] {10.1093/mnras/stz2325}, \href {https://ui.adsabs.harvard.edu/abs/2019MNRAS.489.3200B} {489, 3200}

\bibitem[\protect\citeauthoryear{{Bell} et~al.,}{{Bell} et~al.}{2020}]{Bell2020}
{Bell} C. P.~M.,  et~al., 2020, \mn@doi [\mnras] {10.1093/mnras/staa2786}, \href {https://ui.adsabs.harvard.edu/abs/2020MNRAS.499..993B} {499, 993}

\bibitem[\protect\citeauthoryear{{Bell} et~al.,}{{Bell} et~al.}{2022}]{Bell2022}
{Bell} C. P.~M.,  et~al., 2022, \mn@doi [\mnras] {10.1093/mnras/stac1545}, \href {https://ui.adsabs.harvard.edu/abs/2022MNRAS.516..824B} {516, 824}

\bibitem[\protect\citeauthoryear{{Blanton} et~al.,}{{Blanton} et~al.}{2017}]{2017SDSS}
{Blanton} M.~R.,  et~al., 2017, \mn@doi [\aj] {10.3847/1538-3881/aa7567}, \href {https://ui.adsabs.harvard.edu/abs/2017AJ....154...28B} {154, 28}

\bibitem[\protect\citeauthoryear{Breiman}{Breiman}{2001}]{breiman2001}
Breiman L.,  2001, \mn@doi [Machine Learning] {10.1023/A:1010933404324}, 45

\bibitem[\protect\citeauthoryear{{Buckley}, {Swart}  \& {Meiring}}{{Buckley} et~al.}{2006}]{2006SPIE.6267E..0ZB}
{Buckley} D. A.~H.,  {Swart} G.~P.,   {Meiring} J.~G.,  2006, in {Stepp} L.~M.,  ed.,  Society of Photo-Optical Instrumentation Engineers (SPIE) Conference Series Vol. 6267, Society of Photo-Optical Instrumentation Engineers (SPIE) Conference Series. p. 62670Z, \mn@doi{10.1117/12.673750}

\bibitem[\protect\citeauthoryear{{Burgh}, {Nordsieck}, {Kobulnicky}, {Williams}, {O'Donoghue}, {Smith}  \& {Percival}}{{Burgh} et~al.}{2003}]{2003SPIE.4841.1463B}
{Burgh} E.~B.,  {Nordsieck} K.~H.,  {Kobulnicky} H.~A.,  {Williams} T.~B.,  {O'Donoghue} D.,  {Smith} M.~P.,   {Percival} J.~W.,  2003, in {Iye} M.,  {Moorwood} A. F.~M.,  eds,  Society of Photo-Optical Instrumentation Engineers (SPIE) Conference Series Vol. 4841, Instrument Design and Performance for Optical/Infrared Ground-based Telescopes. pp 1463--1471, \mn@doi{10.1117/12.460312}

\bibitem[\protect\citeauthoryear{{Carrera}, {Conn}, {No{\"e}l}, {Read}  \& {L{\'o}pez S{\'a}nchez}}{{Carrera} et~al.}{2017}]{2017MAGIC}
{Carrera} R.,  {Conn} B.~C.,  {No{\"e}l} N. E.~D.,  {Read} J.~I.,   {L{\'o}pez S{\'a}nchez} {\'A}.~R.,  2017, \mn@doi [\mnras] {10.1093/mnras/stx1932}, \href {https://ui.adsabs.harvard.edu/abs/2017MNRAS.471.4571C} {471, 4571}

\bibitem[\protect\citeauthoryear{{Choudhury} et~al.,}{{Choudhury} et~al.}{2021}]{2021RGB}
{Choudhury} S.,  et~al., 2021, \mn@doi [\mnras] {10.1093/mnras/stab2446}, \href {https://ui.adsabs.harvard.edu/abs/2021MNRAS.507.4752C} {507, 4752}

\bibitem[\protect\citeauthoryear{{Cioni} et~al.,}{{Cioni} et~al.}{2011a}]{2011A&A...527A.116C}
{Cioni} M. R.~L.,  et~al., 2011a, \mn@doi [\aap] {10.1051/0004-6361/201016137}, \href {https://ui.adsabs.harvard.edu/abs/2011A&A...527A.116C} {527, A116}

\bibitem[\protect\citeauthoryear{{Cioni} et~al.,}{{Cioni} et~al.}{2011b}]{Cioni2011}
{Cioni} M. R.~L.,  et~al., 2011b, \mn@doi [\aap] {10.1051/0004-6361/201016137}, \href {https://ui.adsabs.harvard.edu/abs/2011A&A...527A.116C} {527, A116}

\bibitem[\protect\citeauthoryear{{Cioni} et~al.,}{{Cioni} et~al.}{2013}]{2013Cioni}
{Cioni} M. R.~L.,  et~al., 2013, \mn@doi [\aap] {10.1051/0004-6361/201219696}, \href {https://ui.adsabs.harvard.edu/abs/2013A&A...549A..29C} {549, A29}

\bibitem[\protect\citeauthoryear{Cioni, Girardi, Moretti, Piffl, Ripepi  \& Rubele}{Cioni et~al.}{2014a}]{cioni2014}
Cioni M.,  Girardi L.,  Moretti M.,  Piffl T.,  Ripepi V.,   Rubele S.,  2014a, \mn@doi [\aa] {10.1051/0004-6361/201322100}, 562

\bibitem[\protect\citeauthoryear{{Cioni} et~al.,}{{Cioni} et~al.}{2014b}]{2014Cioni}
{Cioni} M. R.~L.,  et~al., 2014b, \mn@doi [\aap] {10.1051/0004-6361/201322100}, \href {https://ui.adsabs.harvard.edu/abs/2014A&A...562A..32C} {562, A32}

\bibitem[\protect\citeauthoryear{{Cioni} et~al.,}{{Cioni} et~al.}{2016a}]{Cioni2016}
{Cioni} M.-R.~L.,  et~al., 2016a, \mn@doi [\aap] {10.1051/0004-6361/201527004}, \href {https://ui.adsabs.harvard.edu/abs/2016A&A...586A..77C} {586, A77}

\bibitem[\protect\citeauthoryear{{Cioni} et~al.,}{{Cioni} et~al.}{2016b}]{2016Cioni}
{Cioni} M.-R.~L.,  et~al., 2016b, \mn@doi [\aap] {10.1051/0004-6361/201527004}, \href {https://ui.adsabs.harvard.edu/abs/2016A&A...586A..77C} {586, A77}

\bibitem[\protect\citeauthoryear{{Civano} et~al.,}{{Civano} et~al.}{2015}]{2015Civano}
{Civano} F.,  et~al., 2015, \mn@doi [\apj] {10.1088/0004-637X/808/2/185}, \href {https://ui.adsabs.harvard.edu/abs/2015ApJ...808..185C} {808, 185}

\bibitem[\protect\citeauthoryear{{Cole}, {Tolstoy}, {Gallagher}  \& {Smecker-Hane}}{{Cole} et~al.}{2005}]{Cole2005}
{Cole} A.~A.,  {Tolstoy} E.,  {Gallagher} John~S. I.,   {Smecker-Hane} T.~A.,  2005, \mn@doi [\aj] {10.1086/428007}, \href {https://ui.adsabs.harvard.edu/abs/2005AJ....129.1465C} {129, 1465}

\bibitem[\protect\citeauthoryear{{Cowley}, {Crampton}, {Hutchings}, {Remillard}  \& {Penfold}}{{Cowley} et~al.}{1983}]{1983Cowley}
{Cowley} A.~P.,  {Crampton} D.,  {Hutchings} J.~B.,  {Remillard} R.,   {Penfold} J.~E.,  1983, \mn@doi [\apj] {10.1086/161267}, \href {https://ui.adsabs.harvard.edu/abs/1983ApJ...272..118C} {272, 118}

\bibitem[\protect\citeauthoryear{{Crause} et~al.,}{{Crause} et~al.}{2019}]{2019JATIS...5b4007C}
{Crause} L.~A.,  et~al., 2019, \mn@doi [Journal of Astronomical Telescopes, Instruments, and Systems] {10.1117/1.JATIS.5.2.024007}, \href {https://ui.adsabs.harvard.edu/abs/2019JATIS...5b4007C} {5, 024007}

\bibitem[\protect\citeauthoryear{{Crawford} et~al.,}{{Crawford} et~al.}{2010}]{SALTpipeline}
{Crawford} S.~M.,  et~al., 2010, in {Silva} D.~R.,  {Peck} A.~B.,   {Soifer} B.~T.,  eds,  Society of Photo-Optical Instrumentation Engineers (SPIE) Conference Series Vol. 7737, Observatory Operations: Strategies, Processes, and Systems III. p. 773725, \mn@doi{10.1117/12.857000}

\bibitem[\protect\citeauthoryear{{Cusano} et~al.,}{{Cusano} et~al.}{2021}]{2021RRL}
{Cusano} F.,  et~al., 2021, \mn@doi [\mnras] {10.1093/mnras/stab901}, \href {https://ui.adsabs.harvard.edu/abs/2021MNRAS.504....1C} {504, 1}

\bibitem[\protect\citeauthoryear{{Cutri} et~al.,}{{Cutri} et~al.}{2013}]{Cutri2013}
{Cutri} R.~M.,  et~al., 2013, {Explanatory Supplement to the AllWISE Data Release Products}, Explanatory Supplement to the AllWISE Data Release Products

\bibitem[\protect\citeauthoryear{{Dalton} et~al.,}{{Dalton} et~al.}{2006}]{2006vista}
{Dalton} G.~B.,  et~al., 2006, in {McLean} I.~S.,  {Iye} M.,  eds,  Society of Photo-Optical Instrumentation Engineers (SPIE) Conference Series Vol. 6269, Society of Photo-Optical Instrumentation Engineers (SPIE) Conference Series. p. 62690X, \mn@doi{10.1117/12.670018}

\bibitem[\protect\citeauthoryear{{Dalton} et~al.,}{{Dalton} et~al.}{2012}]{WEAVE}
{Dalton} G.,  et~al., 2012, in {McLean} I.~S.,  {Ramsay} S.~K.,   {Takami} H.,  eds,  Society of Photo-Optical Instrumentation Engineers (SPIE) Conference Series Vol. 8446, Ground-based and Airborne Instrumentation for Astronomy IV. p. 84460P, \mn@doi{10.1117/12.925950}

\bibitem[\protect\citeauthoryear{{De Bortoli}, {Parisi}, {Bassino}, {Geisler}, {Dias}, {Gimeno}, {Angelo}  \& {Mauro}}{{De Bortoli} et~al.}{2022}]{debortoli2022}
{De Bortoli} B.~J.,  {Parisi} M.~C.,  {Bassino} L.~P.,  {Geisler} D.,  {Dias} B.,  {Gimeno} G.,  {Angelo} M.~S.,   {Mauro} F.,  2022, \mn@doi [\aap] {10.1051/0004-6361/202243762}, \href {https://ui.adsabs.harvard.edu/abs/2022A&A...664A.168D} {664, A168}

\bibitem[\protect\citeauthoryear{{DeBoer} et~al.,}{{DeBoer} et~al.}{2009}]{DeBoer2009}
{DeBoer} D.~R.,  et~al., 2009, \mn@doi [IEEE Proceedings] {10.1109/JPROC.2009.2016516}, \href {https://ui.adsabs.harvard.edu/abs/2009IEEEP..97.1507D} {97, 1507}

\bibitem[\protect\citeauthoryear{{Dickey} et~al.,}{{Dickey} et~al.}{2013}]{Dickey2013}
{Dickey} J.~M.,  et~al., 2013, \mn@doi [\pasa] {10.1017/pasa.2012.003}, \href {https://ui.adsabs.harvard.edu/abs/2013PASA...30....3D} {30, e003}

\bibitem[\protect\citeauthoryear{{Dorigo Jones}, {Oey}, {Paggeot}, {Castro}  \& {Moe}}{{Dorigo Jones} et~al.}{2020}]{Jones2020}
{Dorigo Jones} J.,  {Oey} M.~S.,  {Paggeot} K.,  {Castro} N.,   {Moe} M.,  2020, \mn@doi [\apj] {10.3847/1538-4357/abbc6b}, \href {https://ui.adsabs.harvard.edu/abs/2020ApJ...903...43D} {903, 43}

\bibitem[\protect\citeauthoryear{{Driver} et~al.,}{{Driver} et~al.}{2011}]{2011GAMA}
{Driver} S.~P.,  et~al., 2011, \mn@doi [\mnras] {10.1111/j.1365-2966.2010.18188.x}, \href {https://ui.adsabs.harvard.edu/abs/2011MNRAS.413..971D} {413, 971}

\bibitem[\protect\citeauthoryear{{El Youssoufi} et~al.,}{{El Youssoufi} et~al.}{2019}]{2019Dalal}
{El Youssoufi} D.,  et~al., 2019, \mn@doi [\mnras] {10.1093/mnras/stz2400}, \href {https://ui.adsabs.harvard.edu/abs/2019MNRAS.490.1076E} {490, 1076}

\bibitem[\protect\citeauthoryear{{Emerson}, {McPherson}  \& {Sutherland}}{{Emerson} et~al.}{2006}]{VIRCAM}
{Emerson} J.,  {McPherson} A.,   {Sutherland} W.,  2006, The Messenger, \href {https://ui.adsabs.harvard.edu/abs/2006Msngr.126...41E} {126, 41}

\bibitem[\protect\citeauthoryear{{Esquej} et~al.,}{{Esquej} et~al.}{2013}]{Esquej2013}
{Esquej} P.,  et~al., 2013, \mn@doi [\aap] {10.1051/0004-6361/201218832}, \href {https://ui.adsabs.harvard.edu/abs/2013A&A...557A.123E} {557, A123}

\bibitem[\protect\citeauthoryear{{Evans} et~al.,}{{Evans} et~al.}{2015a}]{Evans2015b}
{Evans} C.~J.,  et~al., 2015a, \mn@doi [\aap] {10.1051/0004-6361/201424414}, \href {https://ui.adsabs.harvard.edu/abs/2015A&A...574A..13E} {574, A13}

\bibitem[\protect\citeauthoryear{{Evans}, {van Loon}, {Hainich}  \& {Bailey}}{{Evans} et~al.}{2015b}]{Evans2015a}
{Evans} C.~J.,  {van Loon} J.~T.,  {Hainich} R.,   {Bailey} M.,  2015b, \mn@doi [\aap] {10.1051/0004-6361/201525882}, \href {https://ui.adsabs.harvard.edu/abs/2015A&A...584A...5E} {584, A5}

\bibitem[\protect\citeauthoryear{{Finkbeiner} et~al.,}{{Finkbeiner} et~al.}{2004}]{2004AJ....128.2577F}
{Finkbeiner} D.~P.,  et~al., 2004, \mn@doi [\aj] {10.1086/425050}, \href {https://ui.adsabs.harvard.edu/abs/2004AJ....128.2577F} {128, 2577}

\bibitem[\protect\citeauthoryear{{Flesch}}{{Flesch}}{2019a}]{Flesch2019cat}
{Flesch} E.~W.,  2019a, VizieR Online Data Catalog, \href {https://ui.adsabs.harvard.edu/abs/2019yCat.7283....0F} {p. VII/283}

\bibitem[\protect\citeauthoryear{{Flesch}}{{Flesch}}{2019b}]{Flesch2019pap}
{Flesch} E.~W.,  2019b, arXiv e-prints, \href {https://ui.adsabs.harvard.edu/abs/2019arXiv191205614F} {p. arXiv:1912.05614}

\bibitem[\protect\citeauthoryear{{Gaia Collaboration} et~al.,}{{Gaia Collaboration} et~al.}{2016}]{Gaiamission}
{Gaia Collaboration} et~al., 2016, \mn@doi [\aap] {10.1051/0004-6361/201629272}, \href {https://ui.adsabs.harvard.edu/abs/2016A&A...595A...1G} {595, A1}

\bibitem[\protect\citeauthoryear{{Gaia Collaboration} et~al.,}{{Gaia Collaboration} et~al.}{2021a}]{Gaia2021}
{Gaia Collaboration} et~al., 2021a, \mn@doi [\aap] {10.1051/0004-6361/202039657}, \href {https://ui.adsabs.harvard.edu/abs/2021A&A...649A...1G} {649, A1}

\bibitem[\protect\citeauthoryear{{Gaia Collaboration} et~al.,}{{Gaia Collaboration} et~al.}{2021b}]{2021A&A...649A...7G}
{Gaia Collaboration} et~al., 2021b, \mn@doi [\aap] {10.1051/0004-6361/202039588}, \href {https://ui.adsabs.harvard.edu/abs/2021A&A...649A...7G} {649, A7}

\bibitem[\protect\citeauthoryear{{Gaia Collaboration} et~al.,}{{Gaia Collaboration} et~al.}{2022}]{GaiaDR3}
{Gaia Collaboration} et~al., 2022, arXiv e-prints, \href {https://ui.adsabs.harvard.edu/abs/2022arXiv220800211G} {p. arXiv:2208.00211}

\bibitem[\protect\citeauthoryear{{Gaia Collaboration} et~al.,}{{Gaia Collaboration} et~al.}{2023}]{GaiaEx}
{Gaia Collaboration} et~al., 2023, \mn@doi [\aap] {10.1051/0004-6361/202243232}, \href {https://ui.adsabs.harvard.edu/abs/2023A&A...674A..41G} {674, A41}

\bibitem[\protect\citeauthoryear{{Gardner} et~al.,}{{Gardner} et~al.}{2006}]{2006JWST}
{Gardner} J.~P.,  et~al., 2006, \mn@doi [\ssr] {10.1007/s11214-006-8315-7}, \href {https://ui.adsabs.harvard.edu/abs/2006SSRv..123..485G} {123, 485}

\bibitem[\protect\citeauthoryear{{Geha} et~al.,}{{Geha} et~al.}{2003}]{Geha2003}
{Geha} M.,  et~al., 2003, \mn@doi [\aj] {10.1086/344947}, \href {https://ui.adsabs.harvard.edu/abs/2003AJ....125....1G} {125, 1}

\bibitem[\protect\citeauthoryear{Gordon, Meixner, Meade, Whitney, Engelbracht  \& Bot}{Gordon et~al.}{2011}]{gordon2011}
Gordon K.~D.,  Meixner M.,  Meade M.,  Whitney B.~A.,  Engelbracht C.~W.,   Bot C.,  2011, \mn@doi [\aj] {10.1088/0004-6256/142/4/102}, 142

\bibitem[\protect\citeauthoryear{{Griffin} et~al.,}{{Griffin} et~al.}{2010}]{SPIRE}
{Griffin} M.~J.,  et~al., 2010, \mn@doi [\aap] {10.1051/0004-6361/201014519}, \href {https://ui.adsabs.harvard.edu/abs/2010A&A...518L...3G} {518, L3}

\bibitem[\protect\citeauthoryear{{Grin} et~al.,}{{Grin} et~al.}{2017}]{Grin2017}
{Grin} N.~J.,  et~al., 2017, \mn@doi [\aap] {10.1051/0004-6361/201629225}, \href {https://ui.adsabs.harvard.edu/abs/2017A&A...600A..82G} {600, A82}

\bibitem[\protect\citeauthoryear{{Groenewegen} \& {Blommaert}}{{Groenewegen} \& {Blommaert}}{1998}]{Groenewegen1998}
{Groenewegen} M.~A.~T.,  {Blommaert} J.~A.~D.~L.,  1998, \aap, \href {https://ui.adsabs.harvard.edu/abs/1998A&A...332...25G} {332, 25}

\bibitem[\protect\citeauthoryear{{Groenewegen} et~al.,}{{Groenewegen} et~al.}{2019}]{2019RGB}
{Groenewegen} M.~A.~T.,  et~al., 2019, \mn@doi [\aap] {10.1051/0004-6361/201833904}, \href {https://ui.adsabs.harvard.edu/abs/2019A&A...622A..63G} {622, A63}

\bibitem[\protect\citeauthoryear{{Groenewegen} et~al.,}{{Groenewegen} et~al.}{2020}]{2020AGB}
{Groenewegen} M.~A.~T.,  et~al., 2020, \mn@doi [\aap] {10.1051/0004-6361/201937271}, \href {https://ui.adsabs.harvard.edu/abs/2020A&A...636A..48G} {636, A48}

\bibitem[\protect\citeauthoryear{{Gullieuszik} et~al.,}{{Gullieuszik} et~al.}{2012}]{2012AGB}
{Gullieuszik} M.,  et~al., 2012, \mn@doi [\aap] {10.1051/0004-6361/201117493}, \href {https://ui.adsabs.harvard.edu/abs/2012A&A...537A.105G} {537, A105}

\bibitem[\protect\citeauthoryear{{Haberl}, {Maitra}, {Vasilopoulos}, {Maggi}, {Udalski}, {Monageng}  \& {Buckley}}{{Haberl} et~al.}{2022}]{2022Haberl}
{Haberl} F.,  {Maitra} C.,  {Vasilopoulos} G.,  {Maggi} P.,  {Udalski} A.,  {Monageng} I.~M.,   {Buckley} D.~A.~H.,  2022, \mn@doi [\aap] {10.1051/0004-6361/202243301}, \href {https://ui.adsabs.harvard.edu/abs/2022A&A...662A..22H} {662, A22}

\bibitem[\protect\citeauthoryear{{Haberl} et~al.,}{{Haberl} et~al.}{2023}]{2023Haberl}
{Haberl} F.,  et~al., 2023, \mn@doi [\aap] {10.1051/0004-6361/202245807}, \href {https://ui.adsabs.harvard.edu/abs/2023A&A...671A..90H} {671, A90}

\bibitem[\protect\citeauthoryear{{Hamuy}, {Suntzeff}, {Heathcote}, {Walker}, {Gigoux}  \& {Phillips}}{{Hamuy} et~al.}{1994}]{1994PASP..106..566H}
{Hamuy} M.,  {Suntzeff} N.~B.,  {Heathcote} S.~R.,  {Walker} A.~R.,  {Gigoux} P.,   {Phillips} M.~M.,  1994, \mn@doi [\pasp] {10.1086/133417}, \href {https://ui.adsabs.harvard.edu/abs/1994PASP..106..566H} {106, 566}

\bibitem[\protect\citeauthoryear{{Hickox} \& {Alexander}}{{Hickox} \& {Alexander}}{2018}]{2018Hickox}
{Hickox} R.~C.,  {Alexander} D.~M.,  2018, \mn@doi [\araa] {10.1146/annurev-astro-081817-051803}, \href {https://ui.adsabs.harvard.edu/abs/2018ARA&A..56..625H} {56, 625}

\bibitem[\protect\citeauthoryear{{Hony} et~al.,}{{Hony} et~al.}{2011}]{Hony2011}
{Hony} S.,  et~al., 2011, \mn@doi [\aap] {10.1051/0004-6361/201116845}, \href {https://ui.adsabs.harvard.edu/abs/2011A&A...531A.137H} {531, A137}

\bibitem[\protect\citeauthoryear{{Hopkins} et~al.,}{{Hopkins} et~al.}{2013}]{2013GAMAspec}
{Hopkins} A.~M.,  et~al., 2013, \mn@doi [\mnras] {10.1093/mnras/stt030}, \href {https://ui.adsabs.harvard.edu/abs/2013MNRAS.430.2047H} {430, 2047}

\bibitem[\protect\citeauthoryear{{Hornschemeier} et~al.,}{{Hornschemeier} et~al.}{2001}]{2001Hornschemeier}
{Hornschemeier} A.~E.,  et~al., 2001, \mn@doi [\apj] {10.1086/321420}, \href {https://ui.adsabs.harvard.edu/abs/2001ApJ...554..742H} {554, 742}

\bibitem[\protect\citeauthoryear{{Hotan} et~al.,}{{Hotan} et~al.}{2014}]{Hotan2014}
{Hotan} A.~W.,  et~al., 2014, \mn@doi [\pasa] {10.1017/pasa.2014.36}, \href {https://ui.adsabs.harvard.edu/abs/2014PASA...31...41H} {31, e041}

\bibitem[\protect\citeauthoryear{{Hotan} et~al.,}{{Hotan} et~al.}{2021}]{Hotan2021}
{Hotan} A.~W.,  et~al., 2021, \mn@doi [\pasa] {10.1017/pasa.2021.1}, \href {https://ui.adsabs.harvard.edu/abs/2021PASA...38....9H} {38, e009}

\bibitem[\protect\citeauthoryear{{Ivanov} et~al.,}{{Ivanov} et~al.}{2016}]{Ivanov2016}
{Ivanov} V.~D.,  et~al., 2016, \mn@doi [\aap] {10.1051/0004-6361/201527398}, \href {https://ui.adsabs.harvard.edu/abs/2016A&A...588A..93I} {588, A93}

\bibitem[\protect\citeauthoryear{{Ivezi{\'c}} et~al.,}{{Ivezi{\'c}} et~al.}{2019}]{2019LSST}
{Ivezi{\'c}} {\v{Z}}.,  et~al., 2019, \mn@doi [\apj] {10.3847/1538-4357/ab042c}, \href {https://ui.adsabs.harvard.edu/abs/2019ApJ...873..111I} {873, 111}

\bibitem[\protect\citeauthoryear{{Jarvis} et~al.,}{{Jarvis} et~al.}{2013}]{2013MNRAS.428.1281J}
{Jarvis} M.~J.,  et~al., 2013, \mn@doi [\mnras] {10.1093/mnras/sts118}, \href {https://ui.adsabs.harvard.edu/abs/2013MNRAS.428.1281J} {428, 1281}

\bibitem[\protect\citeauthoryear{{Johnston} et~al.,}{{Johnston} et~al.}{2008}]{Johnston2008}
{Johnston} S.,  et~al., 2008, \mn@doi [Experimental Astronomy] {10.1007/s10686-008-9124-7}, \href {https://ui.adsabs.harvard.edu/abs/2008ExA....22..151J} {22, 151}

\bibitem[\protect\citeauthoryear{{Jones}, {Saunders}, {Colless}, {Read}, {Parker}, {Watson}  \& {Campbell}}{{Jones} et~al.}{2005}]{6dfgs_v1}
{Jones} H.,  {Saunders} W.,  {Colless} M.,  {Read} M.,  {Parker} Q.,  {Watson} F.,   {Campbell} L.,  2005, in {Fairall} A.~P.,  {Woudt} P.~A.,  eds,  Astronomical Society of the Pacific Conference Series Vol. 329, Nearby Large-Scale Structures and the Zone of Avoidance. p.~11

\bibitem[\protect\citeauthoryear{{Jones} et~al.,}{{Jones} et~al.}{2009}]{Jones2009}
{Jones} D.~H.,  et~al., 2009, \mn@doi [\mnras] {10.1111/j.1365-2966.2009.15338.x}, \href {https://ui.adsabs.harvard.edu/abs/2009MNRAS.399..683J} {399, 683}

\bibitem[\protect\citeauthoryear{{Jones} et~al.,}{{Jones} et~al.}{2017}]{Jones2017}
{Jones} O.~C.,  et~al., 2017, \mn@doi [\mnras] {10.1093/mnras/stx1101}, \href {https://ui.adsabs.harvard.edu/abs/2017MNRAS.470.3250J} {470, 3250}

\bibitem[\protect\citeauthoryear{{Joseph} et~al.,}{{Joseph} et~al.}{2019}]{Joseph2019}
{Joseph} T.~D.,  et~al., 2019, \mn@doi [\mnras] {10.1093/mnras/stz2650}, \href {https://ui.adsabs.harvard.edu/abs/2019MNRAS.490.1202J} {490, 1202}

\bibitem[\protect\citeauthoryear{{Kamath}, {Wood}  \& {Van Winckel}}{{Kamath} et~al.}{2014}]{Kamath2014}
{Kamath} D.,  {Wood} P.~R.,   {Van Winckel} H.,  2014, \mn@doi [\mnras] {10.1093/mnras/stt2033}, \href {https://ui.adsabs.harvard.edu/abs/2014MNRAS.439.2211K} {439, 2211}

\bibitem[\protect\citeauthoryear{Khoshgoftaar, Golawala  \& Hulse}{Khoshgoftaar et~al.}{2007}]{2007Khosh}
Khoshgoftaar T.~M.,  Golawala M.,   Hulse J.~V.,  2007, in 19th IEEE International Conference on Tools with Artificial Intelligence(ICTAI 2007). pp 310--317, \mn@doi{10.1109/ICTAI.2007.46}

\bibitem[\protect\citeauthoryear{{Kinson}, {Oliveira}  \& {van Loon}}{{Kinson} et~al.}{2021}]{Kinson2021}
{Kinson} D.~A.,  {Oliveira} J.~M.,   {van Loon} J.~T.,  2021, \mn@doi [\mnras] {10.1093/mnras/stab2386}, \href {https://ui.adsabs.harvard.edu/abs/2021MNRAS.507.5106K} {507, 5106}

\bibitem[\protect\citeauthoryear{{Kinson}, {Oliveira}  \& {van Loon}}{{Kinson} et~al.}{2022}]{Kinson2022}
{Kinson} D.~A.,  {Oliveira} J.~M.,   {van Loon} J.~T.,  2022, \mn@doi [\mnras] {10.1093/mnras/stac2692}, \href {https://ui.adsabs.harvard.edu/abs/2022MNRAS.517..140K} {517, 140}

\bibitem[\protect\citeauthoryear{{Kobulnicky}, {Nordsieck}, {Burgh}, {Smith}, {Percival}, {Williams}  \& {O'Donoghue}}{{Kobulnicky} et~al.}{2003}]{2003SPIE.4841.1634K}
{Kobulnicky} H.~A.,  {Nordsieck} K.~H.,  {Burgh} E.~B.,  {Smith} M.~P.,  {Percival} J.~W.,  {Williams} T.~B.,   {O'Donoghue} D.,  2003, in {Iye} M.,  {Moorwood} A. F.~M.,  eds,  Society of Photo-Optical Instrumentation Engineers (SPIE) Conference Series Vol. 4841, Instrument Design and Performance for Optical/Infrared Ground-based Telescopes. pp 1634--1644, \mn@doi{10.1117/12.460315}

\bibitem[\protect\citeauthoryear{{Kokusho}, {Torii}, {Kaneda}, {Fukui}  \& {Tachihara}}{{Kokusho} et~al.}{2023}]{2023Kokusho}
{Kokusho} T.,  {Torii} H.,  {Kaneda} H.,  {Fukui} Y.,   {Tachihara} K.,  2023, \mn@doi [\apj] {10.3847/1538-4357/ace10e}, \href {https://ui.adsabs.harvard.edu/abs/2023ApJ...953..104K} {953, 104}

\bibitem[\protect\citeauthoryear{{Koz{\l}owski}, {Kochanek}  \& {Udalski}}{{Koz{\l}owski} et~al.}{2011}]{Kozlowski2011}
{Koz{\l}owski} S.,  {Kochanek} C.~S.,   {Udalski} A.,  2011, \mn@doi [\apjs] {10.1088/0067-0049/194/2/22}, 194

\bibitem[\protect\citeauthoryear{{Koz{\l}owski} et~al.,}{{Koz{\l}owski} et~al.}{2012}]{Kozlowski2012}
{Koz{\l}owski} S.,  et~al., 2012, \mn@doi [\apj] {10.1088/0004-637X/746/1/27}, \href {https://ui.adsabs.harvard.edu/abs/2012ApJ...746...27K} {746, 27}

\bibitem[\protect\citeauthoryear{{Koz{\l}owski} et~al.,}{{Koz{\l}owski} et~al.}{2013}]{Kozlowski2013}
{Koz{\l}owski} S.,  et~al., 2013, \mn@doi [\apj] {10.1088/0004-637X/775/2/92}, \href {https://ui.adsabs.harvard.edu/abs/2013ApJ...775...92K} {775, 92}

\bibitem[\protect\citeauthoryear{Kumar et~al.,}{Kumar et~al.}{2012}]{UVIT}
Kumar A.,  et~al., 2012, \mn@doi [Journal of Astronomical Telescopes, Instruments, and Systems] {10.1117/12.924507}, 8443

\bibitem[\protect\citeauthoryear{{Lacy} et~al.,}{{Lacy} et~al.}{2004}]{2004Lacy}
{Lacy} M.,  et~al., 2004, \mn@doi [\apjs] {10.1086/422816}, \href {https://ui.adsabs.harvard.edu/abs/2004ApJS..154..166L} {154, 166}

\bibitem[\protect\citeauthoryear{{Lamb}, {Oey}, {Segura-Cox}, {Graus}, {Kiminki}, {Golden-Marx}  \& {Parker}}{{Lamb} et~al.}{2016}]{Lamb2016}
{Lamb} J.~B.,  {Oey} M.~S.,  {Segura-Cox} D.~M.,  {Graus} A.~S.,  {Kiminki} D.~C.,  {Golden-Marx} J.~B.,   {Parker} J.~W.,  2016, \mn@doi [\apj] {10.3847/0004-637X/817/2/113}, \href {https://ui.adsabs.harvard.edu/abs/2016ApJ...817..113L} {817, 113}

\bibitem[\protect\citeauthoryear{{Maitra} et~al.,}{{Maitra} et~al.}{2019}]{2019Maitra}
{Maitra} C.,  et~al., 2019, \mn@doi [\mnras] {10.1093/mnras/stz2831}, \href {https://ui.adsabs.harvard.edu/abs/2019MNRAS.490.5494M} {490, 5494}

\bibitem[\protect\citeauthoryear{{Maitra}, {Haberl}, {Maggi}, {Kavanagh}, {Vasilopoulos}, {Sasaki}, {Filipovi{\'c}}  \& {Udalski}}{{Maitra} et~al.}{2021a}]{2021aMaitra}
{Maitra} C.,  {Haberl} F.,  {Maggi} P.,  {Kavanagh} P.~J.,  {Vasilopoulos} G.,  {Sasaki} M.,  {Filipovi{\'c}} M.~D.,   {Udalski} A.,  2021a, \mn@doi [\mnras] {10.1093/mnras/stab716}, \href {https://ui.adsabs.harvard.edu/abs/2021MNRAS.504..326M} {504, 326}

\bibitem[\protect\citeauthoryear{{Maitra}, {Haberl}, {Vasilopoulos}, {Ducci}, {Dennerl}  \& {Carpano}}{{Maitra} et~al.}{2021b}]{2021bMaitra}
{Maitra} C.,  {Haberl} F.,  {Vasilopoulos} G.,  {Ducci} L.,  {Dennerl} K.,   {Carpano} S.,  2021b, \mn@doi [\aap] {10.1051/0004-6361/202039468}, \href {https://ui.adsabs.harvard.edu/abs/2021A&A...647A...8M} {647, A8}

\bibitem[\protect\citeauthoryear{{Mauch}, {Murphy}, {Buttery}, {Curran}, {Hunstead}, {Piestrzynski}, {Robertson}  \& {Sadler}}{{Mauch} et~al.}{2003}]{Mauch2003}
{Mauch} T.,  {Murphy} T.,  {Buttery} H.~J.,  {Curran} J.,  {Hunstead} R.~W.,  {Piestrzynski} B.,  {Robertson} J.~G.,   {Sadler} E.~M.,  2003, \mn@doi [\mnras] {10.1046/j.1365-8711.2003.06605.x}, \href {https://ui.adsabs.harvard.edu/abs/2003MNRAS.342.1117M} {342, 1117}

\bibitem[\protect\citeauthoryear{{McConnell} et~al.,}{{McConnell} et~al.}{2016}]{McConnell2016}
{McConnell} D.,  et~al., 2016, \mn@doi [\pasa] {10.1017/pasa.2016.37}, \href {https://ui.adsabs.harvard.edu/abs/2016PASA...33...42M} {33, e042}

\bibitem[\protect\citeauthoryear{{McConnell} et~al.,}{{McConnell} et~al.}{2020}]{McConnell2020}
{McConnell} D.,  et~al., 2020, \mn@doi [\pasa] {10.1017/pasa.2020.41}, \href {https://ui.adsabs.harvard.edu/abs/2020PASA...37...48M} {37, e048}

\bibitem[\protect\citeauthoryear{{Meixner} et~al.,}{{Meixner} et~al.}{2006}]{Meixner2006}
{Meixner} M.,  et~al., 2006, \mn@doi [\aj] {10.1086/508185}, \href {https://ui.adsabs.harvard.edu/abs/2006AJ....132.2268M} {132, 2268}

\bibitem[\protect\citeauthoryear{{Meixner} et~al.,}{{Meixner} et~al.}{2010}]{Meixner2010}
{Meixner} M.,  et~al., 2010, \mn@doi [\aap] {10.1051/0004-6361/201014662}, \href {https://ui.adsabs.harvard.edu/abs/2010A&A...518L..71M} {518, L71}

\bibitem[\protect\citeauthoryear{More \& Rana}{More \& Rana}{2017}]{2017MoreRana}
More A.~S.,  Rana D.~P.,  2017, in 2017 1st International Conference on Intelligent Systems and Information Management (ICISIM). pp 72--78, \mn@doi{10.1109/ICISIM.2017.8122151}

\bibitem[\protect\citeauthoryear{Murphy et~al.,}{Murphy et~al.}{2010}]{Murphy2010}
Murphy T.,  et~al., 2010, \mn@doi [\mnras] {10.1111/j.1365-2966.2009.15961.x}, 402, 2403

\bibitem[\protect\citeauthoryear{{Nandra} et~al.,}{{Nandra} et~al.}{2015}]{2015Nandra}
{Nandra} K.,  et~al., 2015, \mn@doi [\apjs] {10.1088/0067-0049/220/1/10}, \href {https://ui.adsabs.harvard.edu/abs/2015ApJS..220...10N} {220, 10}

\bibitem[\protect\citeauthoryear{{Netzer}}{{Netzer}}{2015}]{2015Netzer}
{Netzer} H.,  2015, \mn@doi [\araa] {10.1146/annurev-astro-082214-122302}, \href {https://ui.adsabs.harvard.edu/abs/2015ARA&A..53..365N} {53, 365}

\bibitem[\protect\citeauthoryear{{Neugent}, {Levesque}, {Massey}, {Morrell}  \& {Drout}}{{Neugent} et~al.}{2020}]{Neugent2020}
{Neugent} K.~F.,  {Levesque} E.~M.,  {Massey} P.,  {Morrell} N.~I.,   {Drout} M.~R.,  2020, \mn@doi [\apj] {10.3847/1538-4357/ababaa}, \href {https://ui.adsabs.harvard.edu/abs/2020ApJ...900..118N} {900, 118}

\bibitem[\protect\citeauthoryear{{Nidever} et~al.,}{{Nidever} et~al.}{2017}]{Nidever2017}
{Nidever} D.~L.,  et~al., 2017, \mn@doi [\aj] {10.3847/1538-3881/aa8d1c}, \href {https://ui.adsabs.harvard.edu/abs/2017AJ....154..199N} {154, 199}

\bibitem[\protect\citeauthoryear{Nikutta, Hunt-Walker, Nenkova, Ivezi\'{c}  \& Elitzur}{Nikutta et~al.}{2014}]{Nikutta2014}
Nikutta R.,  Hunt-Walker N.,  Nenkova M.,  Ivezi\'{c} v.,   Elitzur M.,  2014, \mn@doi [\mnras] {10.1093/mnras/stu1087}, 442, 3361

\bibitem[\protect\citeauthoryear{{No{\"e}l}, {Conn}, {Carrera}, {Read}, {Rix}  \& {Dolphin}}{{No{\"e}l} et~al.}{2013}]{2013MAGIC}
{No{\"e}l} N.~E.~D.,  {Conn} B.~C.,  {Carrera} R.,  {Read} J.~I.,  {Rix} H.~W.,   {Dolphin} A.,  2013, \mn@doi [\apj] {10.1088/0004-637X/768/2/109}, \href {https://ui.adsabs.harvard.edu/abs/2013ApJ...768..109N} {768, 109}

\bibitem[\protect\citeauthoryear{{No{\"e}l}, {Conn}, {Read}, {Carrera}, {Dolphin}  \& {Rix}}{{No{\"e}l} et~al.}{2015}]{2015MAGIC}
{No{\"e}l} N.~E.~D.,  {Conn} B.~C.,  {Read} J.~I.,  {Carrera} R.,  {Dolphin} A.,   {Rix} H.~W.,  2015, \mn@doi [\mnras] {10.1093/mnras/stv1614}, \href {https://ui.adsabs.harvard.edu/abs/2015MNRAS.452.4222N} {452, 4222}

\bibitem[\protect\citeauthoryear{Norris, Hopkins, Afonso, Brown  \& Condon}{Norris et~al.}{2011}]{norris2011}
Norris R.~P.,  Hopkins A.~M.,  Afonso J.,  Brown S.,   Condon J.~J.,  2011, \mn@doi [\pasa] {10.1071/AS11021}, 28

\bibitem[\protect\citeauthoryear{{Oliveira} et~al.,}{{Oliveira} et~al.}{2011}]{Oliveira2011}
{Oliveira} J.~M.,  et~al., 2011, \mn@doi [\mnras] {10.1111/j.1745-3933.2010.00990.x}, \href {https://ui.adsabs.harvard.edu/abs/2011MNRAS.411L..36O} {411, L36}

\bibitem[\protect\citeauthoryear{{Oliveira} et~al.,}{{Oliveira} et~al.}{2013}]{Oliveira2013}
{Oliveira} J.~M.,  et~al., 2013, \mn@doi [\mnras] {10.1093/mnras/sts250}, \href {https://ui.adsabs.harvard.edu/abs/2013MNRAS.428.3001O} {428, 3001}

\bibitem[\protect\citeauthoryear{Oliveira et~al.,}{Oliveira et~al.}{2019}]{Oliveira2019}
Oliveira J.~M.,  et~al., 2019, \mn@doi [\mnras] {10.1093/mnras/stz2810}, 490, 3909

\bibitem[\protect\citeauthoryear{{Padovani} et~al.,}{{Padovani} et~al.}{2017}]{padovani2017}
{Padovani} P.,  et~al., 2017, \mn@doi [\aapr] {10.1007/s00159-017-0102-9}, \href {https://ui.adsabs.harvard.edu/abs/2017A&ARv..25....2P} {25, 2}

\bibitem[\protect\citeauthoryear{{Parisi}, {Grocholski}, {Geisler}, {Sarajedini}  \& {Clari{\'a}}}{{Parisi} et~al.}{2009}]{Parisi2009}
{Parisi} M.~C.,  {Grocholski} A.~J.,  {Geisler} D.,  {Sarajedini} A.,   {Clari{\'a}} J.~J.,  2009, \mn@doi [\aj] {10.1088/0004-6256/138/2/517}, \href {https://ui.adsabs.harvard.edu/abs/2009AJ....138..517P} {138, 517}

\bibitem[\protect\citeauthoryear{{Parisi}, {Geisler}, {Grocholski}, {Clari{\'a}}  \& {Sarajedini}}{{Parisi} et~al.}{2010}]{Parisi2010}
{Parisi} M.~C.,  {Geisler} D.,  {Grocholski} A.~J.,  {Clari{\'a}} J.~J.,   {Sarajedini} A.,  2010, \mn@doi [\aj] {10.1088/0004-6256/139/3/1168}, \href {https://ui.adsabs.harvard.edu/abs/2010AJ....139.1168P} {139, 1168}

\bibitem[\protect\citeauthoryear{{Parisi}, {Gramajo}, {Geisler}, {Dias}, {Clari{\'a}}, {Da Costa}  \& {Grebel}}{{Parisi} et~al.}{2022}]{Parisi2022}
{Parisi} M.~C.,  {Gramajo} L.~V.,  {Geisler} D.,  {Dias} B.,  {Clari{\'a}} J.~J.,  {Da Costa} G.,   {Grebel} E.~K.,  2022, \mn@doi [\aap] {10.1051/0004-6361/202142597}, \href {https://ui.adsabs.harvard.edu/abs/2022A&A...662A..75P} {662, A75}

\bibitem[\protect\citeauthoryear{Pedregosa et~al.,}{Pedregosa et~al.}{2011}]{scikit-learn}
Pedregosa F.,  et~al., 2011, Journal of Machine Learning Research, 12, 2825

\bibitem[\protect\citeauthoryear{{Pennock} et~al.,}{{Pennock} et~al.}{2021}]{2021MNRAS.506.3540P}
{Pennock} C.~M.,  et~al., 2021, \mn@doi [\mnras] {10.1093/mnras/stab1858}, \href {https://ui.adsabs.harvard.edu/abs/2021MNRAS.506.3540P} {506, 3540}

\bibitem[\protect\citeauthoryear{{Pennock} et~al.,}{{Pennock} et~al.}{2022}]{2022MNRAS.515.6046P}
{Pennock} C.~M.,  et~al., 2022, \mn@doi [\mnras] {10.1093/mnras/stac2096}, \href {https://ui.adsabs.harvard.edu/abs/2022MNRAS.515.6046P} {515, 6046}

\bibitem[\protect\citeauthoryear{{Pilbratt} et~al.,}{{Pilbratt} et~al.}{2010}]{Herschel}
{Pilbratt} G.~L.,  et~al., 2010, \mn@doi [\aap] {10.1051/0004-6361/201014759}, \href {https://ui.adsabs.harvard.edu/abs/2010A&A...518L...1P} {518, L1}

\bibitem[\protect\citeauthoryear{{Poglitsch} et~al.,}{{Poglitsch} et~al.}{2010}]{PACS}
{Poglitsch} A.,  et~al., 2010, \mn@doi [\aap] {10.1051/0004-6361/201014535}, \href {https://ui.adsabs.harvard.edu/abs/2010A&A...518L...2P} {518, L2}

\bibitem[\protect\citeauthoryear{{Reid} \& {Parker}}{{Reid} \& {Parker}}{2012}]{2012Reid}
{Reid} W.~A.,  {Parker} Q.~A.,  2012, \mn@doi [\mnras] {10.1111/j.1365-2966.2012.21471.x}, \href {https://ui.adsabs.harvard.edu/abs/2012MNRAS.425..355R} {425, 355}

\bibitem[\protect\citeauthoryear{Reis, Baron  \& Shahaf}{Reis et~al.}{2018}]{reisprob}
Reis I.,  Baron D.,   Shahaf S.,  2018, \mn@doi [\aj] {10.3847/1538-3881/aaf101}, 157

\bibitem[\protect\citeauthoryear{{Rezaeikh}, {Javadi}, {Khosroshahi}  \& {van Loon}}{{Rezaeikh} et~al.}{2014}]{2014Rezaeikh}
{Rezaeikh} S.,  {Javadi} A.,  {Khosroshahi} H.,   {van Loon} J.~T.,  2014, \mn@doi [\mnras] {10.1093/mnras/stu1807}, \href {https://ui.adsabs.harvard.edu/abs/2014MNRAS.445.2214R} {445, 2214}

\bibitem[\protect\citeauthoryear{{Ripepi} et~al.,}{{Ripepi} et~al.}{2015}]{2015Cepheid}
{Ripepi} V.,  et~al., 2015, \mn@doi [\mnras] {10.1093/mnras/stu2260}, \href {https://ui.adsabs.harvard.edu/abs/2015MNRAS.446.3034R} {446, 3034}

\bibitem[\protect\citeauthoryear{{Roman-Duval} et~al.,}{{Roman-Duval} et~al.}{2019}]{RomDuv2019}
{Roman-Duval} J.,  et~al., 2019, \mn@doi [\apj] {10.3847/1538-4357/aaf8bb}, \href {https://ui.adsabs.harvard.edu/abs/2019ApJ...871..151R} {871, 151}

\bibitem[\protect\citeauthoryear{{Rubele} et~al.,}{{Rubele} et~al.}{2012}]{Rubele2012}
{Rubele} S.,  et~al., 2012, \mn@doi [\aap] {10.1051/0004-6361/201117863}, \href {https://ui.adsabs.harvard.edu/abs/2012A&A...537A.106R} {537, A106}

\bibitem[\protect\citeauthoryear{{Rubele} et~al.,}{{Rubele} et~al.}{2015}]{2015Rubele}
{Rubele} S.,  et~al., 2015, \mn@doi [\mnras] {10.1093/mnras/stv141}, \href {https://ui.adsabs.harvard.edu/abs/2015MNRAS.449..639R} {449, 639}

\bibitem[\protect\citeauthoryear{{Rubele} et~al.,}{{Rubele} et~al.}{2018}]{Rubele2018}
{Rubele} S.,  et~al., 2018, \mn@doi [\mnras] {10.1093/mnras/sty1279}, \href {https://ui.adsabs.harvard.edu/abs/2018MNRAS.478.5017R} {478, 5017}

\bibitem[\protect\citeauthoryear{{Ruffle} et~al.,}{{Ruffle} et~al.}{2015}]{Ruffle2015}
{Ruffle} P. M.~E.,  et~al., 2015, \mn@doi [\mnras] {10.1093/mnras/stv1106}, \href {https://ui.adsabs.harvard.edu/abs/2015MNRAS.451.3504R} {451, 3504}

\bibitem[\protect\citeauthoryear{{Salvato} et~al.,}{{Salvato} et~al.}{2018}]{2018Salvato}
{Salvato} M.,  et~al., 2018, \mn@doi [\mnras] {10.1093/mnras/stx2651}, \href {https://ui.adsabs.harvard.edu/abs/2018MNRAS.473.4937S} {473, 4937}

\bibitem[\protect\citeauthoryear{{Schlafly} \& {Finkbeiner}}{{Schlafly} \& {Finkbeiner}}{2011}]{2011Schlafly}
{Schlafly} E.~F.,  {Finkbeiner} D.~P.,  2011, \mn@doi [\apj] {10.1088/0004-637X/737/2/103}, \href {https://ui.adsabs.harvard.edu/abs/2011ApJ...737..103S} {737, 103}

\bibitem[\protect\citeauthoryear{{Schlafly}, {Meisner}  \& {Green}}{{Schlafly} et~al.}{2019}]{UnWISE}
{Schlafly} E.~F.,  {Meisner} A.~M.,   {Green} G.~M.,  2019, \mn@doi [\apjs] {10.3847/1538-4365/aafbea}, \href {https://ui.adsabs.harvard.edu/abs/2019ApJS..240...30S} {240, 30}

\bibitem[\protect\citeauthoryear{{Schlegel}, {Finkbeiner}  \& {Davis}}{{Schlegel} et~al.}{1998}]{1998Schlegel}
{Schlegel} D.~J.,  {Finkbeiner} D.~P.,   {Davis} M.,  1998, \mn@doi [\apj] {10.1086/305772}, \href {https://ui.adsabs.harvard.edu/abs/1998ApJ...500..525S} {500, 525}

\bibitem[\protect\citeauthoryear{{Schumacher}, {Mondaca}, {Warner}, {Martinez}, {Estay}  \& {Abbott}}{{Schumacher} et~al.}{2010}]{DECAM}
{Schumacher} G.,  {Mondaca} E.,  {Warner} M.,  {Martinez} M.,  {Estay} O.,   {Abbott} T. M.~C.,  2010, in {Radziwill} N.~M.,  {Bridger} A.,  eds,  Society of Photo-Optical Instrumentation Engineers (SPIE) Conference Series Vol. 7740, Software and Cyberinfrastructure for Astronomy. p. 77402H, \mn@doi{10.1117/12.857986}

\bibitem[\protect\citeauthoryear{{Seale}, {Looney}, {Chu}, {Gruendl}, {Brandl}, {Chen}, {Brandner}  \& {Blake}}{{Seale} et~al.}{2009}]{Seale2009}
{Seale} J.~P.,  {Looney} L.~W.,  {Chu} Y.-H.,  {Gruendl} R.~A.,  {Brandl} B.,  {Chen} C. H.~R.,  {Brandner} W.,   {Blake} G.~A.,  2009, \mn@doi [\apj] {10.1088/0004-637X/699/1/150}, \href {https://ui.adsabs.harvard.edu/abs/2009ApJ...699..150S} {699, 150}

\bibitem[\protect\citeauthoryear{{Secrest}, {Dudik}, {Dorland}, {Zacharias}, {Makarov}, {Fey}, {Frouard}  \& {Finch}}{{Secrest} et~al.}{2015}]{2015Secrest}
{Secrest} N.~J.,  {Dudik} R.~P.,  {Dorland} B.~N.,  {Zacharias} N.,  {Makarov} V.,  {Fey} A.,  {Frouard} J.,   {Finch} C.,  2015, \mn@doi [\apjs] {10.1088/0067-0049/221/1/12}, \href {https://ui.adsabs.harvard.edu/abs/2015ApJS..221...12S} {221, 12}

\bibitem[\protect\citeauthoryear{{Shaw}, {Stanghellini}, {Mutchler}, {Balick}  \& {Blades}}{{Shaw} et~al.}{2001}]{Shaw2001}
{Shaw} R.~A.,  {Stanghellini} L.,  {Mutchler} M.,  {Balick} B.,   {Blades} J.~C.,  2001, \mn@doi [\apj] {10.1086/319013}, \href {https://ui.adsabs.harvard.edu/abs/2001ApJ...548..727S} {548, 727}

\bibitem[\protect\citeauthoryear{{Sheets}, {Bolatto}, {van Loon}, {Sandstrom}, {Simon}, {Oliveira}  \& {Barb{\'a}}}{{Sheets} et~al.}{2013}]{2013Sheets}
{Sheets} H.~A.,  {Bolatto} A.~D.,  {van Loon} J.~T.,  {Sandstrom} K.,  {Simon} J.~D.,  {Oliveira} J.~M.,   {Barb{\'a}} R.~H.,  2013, \mn@doi [\apj] {10.1088/0004-637X/771/2/111}, \href {https://ui.adsabs.harvard.edu/abs/2013ApJ...771..111S} {771, 111}

\bibitem[\protect\citeauthoryear{{Skrutskie} et~al.,}{{Skrutskie} et~al.}{2006}]{2MASS}
{Skrutskie} M.~F.,  et~al., 2006, \mn@doi [\aj] {10.1086/498708}, \href {https://ui.adsabs.harvard.edu/abs/2006AJ....131.1163S} {131, 1163}

\bibitem[\protect\citeauthoryear{{Soria}, {Wu}, {Page}  \& {Sakelliou}}{{Soria} et~al.}{2001}]{2001Soria}
{Soria} R.,  {Wu} K.,  {Page} M.~J.,   {Sakelliou} I.,  2001, \mn@doi [\aap] {10.1051/0004-6361:20000065}, \href {https://ui.adsabs.harvard.edu/abs/2001A&A...365L.273S} {365, L273}

\bibitem[\protect\citeauthoryear{Stern, Eisenhardt, Gorjian  \& Kochanek}{Stern et~al.}{2005}]{stern2005}
Stern D.,  Eisenhardt P.,  Gorjian V.,   Kochanek C.,  2005, \mn@doi [\apj] {10.1086/432523}, 631

\bibitem[\protect\citeauthoryear{{Stern} et~al.,}{{Stern} et~al.}{2012}]{Stern2012}
{Stern} D.,  et~al., 2012, \mn@doi [\apj] {10.1088/0004-637X/753/1/30}, \href {https://ui.adsabs.harvard.edu/abs/2012ApJ...753...30S} {753, 30}

\bibitem[\protect\citeauthoryear{{Storey-Fisher}, {Hogg}, {Rix}, {Eilers}, {Fabbian}, {Blanton}  \& {Alonso}}{{Storey-Fisher} et~al.}{2023}]{2023Quaia}
{Storey-Fisher} K.,  {Hogg} D.~W.,  {Rix} H.-W.,  {Eilers} A.-C.,  {Fabbian} G.,  {Blanton} M.,   {Alonso} D.,  2023, \mn@doi [arXiv e-prints] {10.48550/arXiv.2306.17749}, \href {https://ui.adsabs.harvard.edu/abs/2023arXiv230617749S} {p. arXiv:2306.17749}

\bibitem[\protect\citeauthoryear{Sturm, Haberl, Pietsch, Ballet, Hatzidimitriou  \& Buckley}{Sturm et~al.}{2013a}]{sturm2013}
Sturm R.,  Haberl F.,  Pietsch W.,  Ballet J.,  Hatzidimitriou D.,   Buckley D. A.~H.,  2013a, \mn@doi [\aa] {10.3847/1538-4365/aa7053}, 558

\bibitem[\protect\citeauthoryear{{Sturm} et~al.,}{{Sturm} et~al.}{2013b}]{2013XMM}
{Sturm} R.,  et~al., 2013b, \mn@doi [\aap] {10.1051/0004-6361/201219935}, \href {https://ui.adsabs.harvard.edu/abs/2013A&A...558A...3S} {558, A3}

\bibitem[\protect\citeauthoryear{{Taylor}}{{Taylor}}{2005}]{Taylor2005}
{Taylor} M.~B.,  2005, in {Shopbell} P.,  {Britton} M.,   {Ebert} R.,  eds,  Astronomical Society of the Pacific Conference Series Vol. 347, Astronomical Data Analysis Software and Systems XIV. p.~29

\bibitem[\protect\citeauthoryear{{Thilker}, {Bianchi}  \& {Simons}}{{Thilker} et~al.}{2014}]{2014AAS...22335511T}
{Thilker} D.~A.,  {Bianchi} L.,   {Simons} R.,  2014, in American Astronomical Society Meeting Abstracts \#223. p. 355.11

\bibitem[\protect\citeauthoryear{{Tody}}{{Tody}}{1986}]{1986SPIE..627..733T}
{Tody} D.,  1986, in {Crawford} D.~L.,  ed.,  Society of Photo-Optical Instrumentation Engineers (SPIE) Conference Series Vol. 627, Instrumentation in astronomy VI. p.~733, \mn@doi{10.1117/12.968154}

\bibitem[\protect\citeauthoryear{{Tody}}{{Tody}}{1993}]{1993ASPC...52..173T}
{Tody} D.,  1993, in {Hanisch} R.~J.,  {Brissenden} R.~J.~V.,   {Barnes} J.,  eds,  Astronomical Society of the Pacific Conference Series Vol. 52, Astronomical Data Analysis Software and Systems II. p.~173

\bibitem[\protect\citeauthoryear{{Vasiliev}}{{Vasiliev}}{2018}]{2018Vasiliev}
{Vasiliev} E.,  2018, \mn@doi [\mnras] {10.1093/mnrasl/sly168}, \href {https://ui.adsabs.harvard.edu/abs/2018MNRAS.481L.100V} {481, L100}

\bibitem[\protect\citeauthoryear{{Walborn} et~al.,}{{Walborn} et~al.}{2014}]{Walborn2014}
{Walborn} N.~R.,  et~al., 2014, \mn@doi [\aap] {10.1051/0004-6361/201323082}, \href {https://ui.adsabs.harvard.edu/abs/2014A&A...564A..40W} {564, A40}

\bibitem[\protect\citeauthoryear{{Webb} et~al.,}{{Webb} et~al.}{2020a}]{XMMS}
{Webb} N.~A.,  et~al., 2020a, \mn@doi [\aap] {10.1051/0004-6361/201937353}, \href {https://ui.adsabs.harvard.edu/abs/2020A&A...641A.136W} {641, A136}

\bibitem[\protect\citeauthoryear{{Webb} et~al.,}{{Webb} et~al.}{2020b}]{2020XMM}
{Webb} N.~A.,  et~al., 2020b, \mn@doi [\aap] {10.1051/0004-6361/201937353}, \href {https://ui.adsabs.harvard.edu/abs/2020A&A...641A.136W} {641, A136}

\bibitem[\protect\citeauthoryear{{Wenger} et~al.,}{{Wenger} et~al.}{2000}]{Simbad}
{Wenger} M.,  et~al., 2000, \mn@doi [\aaps] {10.1051/aas:2000332}, \href {https://ui.adsabs.harvard.edu/abs/2000A&AS..143....9W} {143, 9}

\bibitem[\protect\citeauthoryear{{Whiting}}{{Whiting}}{2020}]{Whiting2020}
{Whiting} M.~T.,  2020, in {Ballester} P.,  {Ibsen} J.,  {Solar} M.,   {Shortridge} K.,  eds,  Astronomical Society of the Pacific Conference Series Vol. 522, Astronomical Data Analysis Software and Systems XXVII. p.~469

\bibitem[\protect\citeauthoryear{{Wright} et~al.,}{{Wright} et~al.}{2010}]{WISE}
{Wright} E.~L.,  et~al., 2010, \mn@doi [\aj] {10.1088/0004-6256/140/6/1868}, \href {https://ui.adsabs.harvard.edu/abs/2010AJ....140.1868W} {140, 1868}

\bibitem[\protect\citeauthoryear{{Zivkov} et~al.,}{{Zivkov} et~al.}{2018}]{2018YS}
{Zivkov} V.,  et~al., 2018, \mn@doi [\aap] {10.1051/0004-6361/201833951}, \href {https://ui.adsabs.harvard.edu/abs/2018A&A...620A.143Z} {620, A143}

\bibitem[\protect\citeauthoryear{{Zivkov} et~al.,}{{Zivkov} et~al.}{2020}]{2020YS}
{Zivkov} V.,  et~al., 2020, \mn@doi [\mnras] {10.1093/mnras/staa626}, \href {https://ui.adsabs.harvard.edu/abs/2020MNRAS.494..458Z} {494, 458}

\bibitem[\protect\citeauthoryear{{de Jong} et~al.,}{{de Jong} et~al.}{2012}]{4MOST}
{de Jong} R.~S.,  et~al., 2012, in {McLean} I.~S.,  {Ramsay} S.~K.,   {Takami} H.,  eds,  Society of Photo-Optical Instrumentation Engineers (SPIE) Conference Series Vol. 8446, Ground-based and Airborne Instrumentation for Astronomy IV. p. 84460T (\mn@eprint {arXiv} {1206.6885}), \mn@doi{10.1117/12.926239}

\bibitem[\protect\citeauthoryear{{de Jong} et~al.,}{{de Jong} et~al.}{2017}]{2017VIKING}
{de Jong} J. T.~A.,  et~al., 2017, \mn@doi [\aap] {10.1051/0004-6361/201730747}, \href {https://ui.adsabs.harvard.edu/abs/2017A&A...604A.134D} {604, A134}

\bibitem[\protect\citeauthoryear{{van Gelder} et~al.,}{{van Gelder} et~al.}{2020}]{vanGelder2020}
{van Gelder} M.~L.,  et~al., 2020, \mn@doi [\aap] {10.1051/0004-6361/201936361}, \href {https://ui.adsabs.harvard.edu/abs/2020A&A...636A..54V} {636, A54}

\bibitem[\protect\citeauthoryear{{van Jaarsveld}, {Buckley}, {McBride}, {Haberl}, {Vasilopoulos}, {Maitra}, {Udalski}  \& {Miszalski}}{{van Jaarsveld} et~al.}{2018}]{2018Jaarsveld}
{van Jaarsveld} N.,  {Buckley} D.~A.~H.,  {McBride} V.~A.,  {Haberl} F.,  {Vasilopoulos} G.,  {Maitra} C.,  {Udalski} A.,   {Miszalski} B.,  2018, \mn@doi [\mnras] {10.1093/mnras/stx3270}, \href {https://ui.adsabs.harvard.edu/abs/2018MNRAS.475.3253V} {475, 3253}

\bibitem[\protect\citeauthoryear{van Loon \& Sansom}{van Loon \& Sansom}{2015}]{vanLoon2015}
van Loon J.~T.,  Sansom A.,  2015, \mn@doi [\mnras] {10.1093/mnras/stv1787}, 453

\bibitem[\protect\citeauthoryear{{van Loon} et~al.,}{{van Loon} et~al.}{1998}]{vanLoon1998}
{van Loon} J.~T.,  et~al., 1998, \aap, \href {https://ui.adsabs.harvard.edu/abs/1998A&A...329..169V} {329, 169}

\bibitem[\protect\citeauthoryear{{van Loon}, {Zijlstra}  \& {Groenewegen}}{{van Loon} et~al.}{1999a}]{vanLoon1999a}
{van Loon} J.~T.,  {Zijlstra} A.~A.,   {Groenewegen} M.~A.~T.,  1999a, \aap, \href {https://ui.adsabs.harvard.edu/abs/1999A&A...346..805V} {346, 805}

\bibitem[\protect\citeauthoryear{{van Loon}, {Groenewegen}, {de Koter}, {Trams}, {Waters}, {Zijlstra}, {Whitelock}  \& {Loup}}{{van Loon} et~al.}{1999b}]{vanLoon1999b}
{van Loon} J.~T.,  {Groenewegen} M.~A.~T.,  {de Koter} A.,  {Trams} N.~R.,  {Waters} L.~B.~F.~M.,  {Zijlstra} A.~A.,  {Whitelock} P.~A.,   {Loup} C.,  1999b, \aap, \href {https://ui.adsabs.harvard.edu/abs/1999A&A...351..559V} {351, 559}

\bibitem[\protect\citeauthoryear{{van Loon}, {Cioni}, {Zijlstra}  \& {Loup}}{{van Loon} et~al.}{2005}]{vanLoon2005}
{van Loon} J.~T.,  {Cioni} M. R.~L.,  {Zijlstra} A.~A.,   {Loup} C.,  2005, \mn@doi [\aap] {10.1051/0004-6361:20042555}, \href {https://ui.adsabs.harvard.edu/abs/2005A&A...438..273V} {438, 273}

\bibitem[\protect\citeauthoryear{{van Loon}, {Marshall}, {Cohen}, {Matsuura}, {Wood}, {Yamamura}  \& {Zijlstra}}{{van Loon} et~al.}{2006}]{vanLoon2006}
{van Loon} J.~T.,  {Marshall} J.~R.,  {Cohen} M.,  {Matsuura} M.,  {Wood} P.~R.,  {Yamamura} I.,   {Zijlstra} A.~A.,  2006, \mn@doi [\aap] {10.1051/0004-6361:20054222}, \href {https://ui.adsabs.harvard.edu/abs/2006A&A...447..971V} {447, 971}

\bibitem[\protect\citeauthoryear{{van Loon}, {Cohen}, {Oliveira}, {Matsuura}, {McDonald}, {Sloan}, {Wood}  \& {Zijlstra}}{{van Loon} et~al.}{2008}]{vanLoon2008}
{van Loon} J.~T.,  {Cohen} M.,  {Oliveira} J.~M.,  {Matsuura} M.,  {McDonald} I.,  {Sloan} G.~C.,  {Wood} P.~R.,   {Zijlstra} A.~A.,  2008, \mn@doi [\aap] {10.1051/0004-6361:200810036}, \href {https://ui.adsabs.harvard.edu/abs/2008A&A...487.1055V} {487, 1055}

\makeatother
\end{thebibliography}



\appendix
\section{Sources spectroscopically observed}\label{specsec}
We list the sources we spectroscopically observed with SAAO 1.9m telescope in 2019 and the sources observed on my behalf in 2021 in Table \ref{SAAOspec}.

We list the sources that were spectroscopically observed with SALT in Table \ref{SALTobs} and \ref{SALTobs2}, where we list the proposals under which they were observed and the classifications that have been made and redshifts that have been measured.

\begin{table}
\caption{Sources spectroscopically observed using the SAAO 1.9m telescope during observing runs in 2019 and 2021.}
    \centering
    \begin{tabular}{|l|l|l|r|}
        \hline\hline
        Target Name & RA & DEC & Date Observed \\ 
        \hline
        C031 & 01:34:22.9 & $-$73:18:11 & 23-24/10/2019 \\ 
        C034 & 00:25:00.0 & $-$72:33:01 & 25-26/10/2019 \\ 
        C014 & 01:38:23.4 & $-$72:36:53 & 25-26/10/2019 \\ 
        C132 & 01:34:20.5 & $-$73:18:07 & 25-26/10/2019 \\ 
        C010 & 00:20:23.9 & $-$73:20:21 & 25-26/10/2019 \\ 
        C023 & 04:17:53.6 & $-$73:15:56 & 25-26/10/2019 \\ 
        C139 & 20:35:00.3 & $-$62:34:54 & 26-27/10/2019 \\ 
        C136 & 00:08:22.9 & $-$73:52:52 & 26-27/10/2019 \\ 
        MQ120 & 00:08:37.88 & $-$72:33:46.2 & 26-27/10/2019 \\
        C141 & 01:46:29.7 & $-$72:48:45 & 26-27/10/2019 \\ 
        B009 & 00:58:21.53 & $-$72:25:02.5 & 26-27/10/2019 \\ 
        A129 & 00:24:46.29 & $-$74:50:13.5 & 26-27/10/2019 \\ 
        MQ118 & 05:32:06.08 & $-$66:30:23 & 26-27/10/2019 \\ 
        C092 & 05:25:25.3 & $-$67:29:24 & 26-27/10/2019 \\ 
        C124 & 06:10:06 & $-$66:11:33 & 26-27/10/2019 \\ 
        C122 & 23:06:53.8 & $-$34:39:09 & 29-30/10/2019 \\ 
        A056 & 00:33:24.08 & $-$74:13:57.9 & 29-30/10/2019 \\ 
        C032 & 00:26:12.6 & $-$72:37:05 & 29-30/10/2019 \\ 
        C045 & 01:03:10.2 & $-$71:51:54 & 29-30/10/2019 \\ 
        A244 & 01:14:07.99 & $-$72:32:43.3 & 29-30/10/2019 \\ 
        C012 & 01:04:50 & $-$70:21:06 & 29-30/10/2019 \\ 
        C028 & 04:23:57.3 & $-$72:47:02 & 29-30/10/2019 \\ 
        C101 & 05:27:56 & $-$67:25:35 & 29-30/10/2019 \\ 
        C137 & 20:39:42.1 & $-$59:57:32 & 30-31/10/2019 \\ 
        A288 & 00:44:54.39 & $-$74:17:50 & 30-31/10/2019 \\ 
        A060 & 00:38:57.67 & $-$72:48:57.9 & 30-31/10/2019 \\ 
        A023 & 00:50:57.44 & $-$73:12:49.0 & 30-31/10/2019 \\ 
        A053 & 01:36:04.46 & $-$72:13:15.4 & 30-31/10/2019 \\ 
        C138 & 20:39:51.1 & $-$59:57:48 & 31/10-1/11/2019 \\ 
        A148 & 01:15:04.9 & $-$73:28:16.4 & 31/10-1/11/2019 \\
        A075 & 01:13:37.08 & $-$74:27:55.3 & 31/10-1/11/2019\\ 
        C158 & 03:40:08.52 & $-$12:49:05.9 & 31/10-1/11/2019 \\ 
        MQ108 & 05:12:55.5 & $-$72:53:10.6 & 31/10-1/11/2019 \\ 
        C127 & 06:10:51.4 & $-$65:24:08 & 31/10-1/11/2019 \\ 
        C128 & 06:09:49.9 & $-$65:28:03 & 31/10-1/11/2019 \\ 
        C042 & 00:36:01.6 & $-$72:21:12 & 1-2/11/2019 \\ 
        A133 (red) & 00:51:39.93 & $-$72:38:17.6 & 1-2/11/2019 \\ 
        A213 & 01:37:07.45 & $-$74:23:39.2 & 1-2/11/2019 \\
        MQ115 & 04:26:57.52 & $-$68:46:15.6 & 1-2/11/2019 \\ 
        C026 & 04:17:45.8 & $-$73:15:08 & 1-2/11/2019 \\ 
        C030 & 06:07:55 & $-$65:52:31 & 1-2/11/2019 \\ 
        C126 & 06:08:57.8 & $-$65:46:01 & 1-2/11/2019 \\ 
        C140 & 20:36:03.1 & $-$58:10:13 & 2-3/11/2019 \\ 
        A142 & 00:52:45.52 & $-$72:44:01.1 & 2-3/11/2019 \\ 
        A152 & 00:39:10.78 & $-$71:34:09.9 & 2-3/11/2019 \\ 
        A314 & 00:31:25.48 & $-$71:50:03 & 2-3/11/2019 \\ 
        C157 & 01:53:51.51 & $-$50:31:37.7 & 2-3/11/2019 \\ 
        C059 & 05:50:38.5 & $-$69:55:41 & 2-3/11/2019 \\ 
        C063 & 06:03:27 & $-$72:48:36 & 2-3/11/2019 \\ 
        C133 & 00:06:29.2 & $-$74:01:33 & 3-4/11/2019 \\ 
        A103 & 00:39:53.96 & $-$72:24:08.3 & 3-4/11/2019 \\ 
        A111 & 01:17:32.64 & $-$74:39:34.9 & 3-4/11/2019 \\ 
        A124 & 01:02:42.18 & $-$73:24:41.6 & 3-4/11/2019 \\ 
        D006 & 04:24:15.7 & $-$70:34:15.4 & 3-4/11/2019 \\ 
        D001 & 05:31:20.6 & $-$71:31:22.7 & 3-4/11/2019 \\ 
        C130 & 06:08:49.5 & $-$65:44:41 & 3-4/11/2019 \\ 
        C134 & 00:06:45.9 & $-$72:08:01 & 4-5/11/2019 \\
        C147 & 00:51:16.8 & $-$73:40:02 & 4-5/11/2019 \\ 
        C009 & 00:20:43.4 & $-$73:21:26 & 4-5/11/2019 \\ 
        C025 & 04:17:59.3 & $-$73:16:58 & 4-5/11/2019 \\ 
        C125 & 06:08:42.4 & $-$65:47:14 & 4-5/11/2019 \\ 
        C029 & 06:08:52 & $-$65:43:50 & 4-5/11/2019 \\
        \hline
     \end{tabular}
     \label{SAAOspec}
\end{table}

\begin{table}
\contcaption{}
    \centering
    \begin{tabular}{|l|l|l|r|}
        \hline\hline
        Target Name & RA & DEC & Date Observed \\ 
        \hline      
         
        C011 & 00:19:56.9 & $-$73:22:17 & 5-6/11/2019 \\ 
        A036 & 00:31:56.89 & $-$73:31:13.6 & 5-6/11/2019 \\ 
        A045 & 00:59:34.14 & $-$72:09:43.2 & 5-6/11/2019 \\ 
        MQ031 & 01:27:17.35 & $-$71:04:10.2 & 5-6/11/2019 \\ 
        C058 & 05:12:05.4 & $-$70:32:04 & 5-6/11/2019 \\ 
        C062a & 05:44:24 & $-$72:51:07 & 5-6/11/2019 \\ 
        C129 & 06:11:17 & $-$66:09:22 & 5-6/11/2019 \\ 
        C135 & 00:06:48.3 & $-$72:22:52 & 13-14/11/2019 \\ 
        C148 & 00:55:53.5 & $-$72:42:01 & 13-15/11/2019 \\ 
        A039 & 00:43:12.46 & $-$73:46:46.8 & 14-15/11/2019 \\ 
        C150 & 00:57:35.4 & $-$73:46:30 & 14-15/11/2019 \\ 
        A043 & 00:47:08.38 & $-$74:30:09.5 & 14-15/11/2019 \\ 
        C152 & 00:59:09.7 & $-$74:02:37 & 14-15/11/2019 \\ 
        C124 & 06:10:06 & $-$66:11:33 & 14-15/11/2019, \\
         & & & 23-24/11/2019 \\ 
        C174 & 05:08:11.4 & $-$68:06:11 & 14-15/11/2019 \\ 
        A144 & 00:41:25.33 & $-$70:57:43.8 & 15-16/11/2019 \\
        C047 & 01:13:18.3 & $-$72:25:19 & 15-16/11/2019 \\ 
        A038 & 00:58:38.17 & $-$73:34:49.6 & 15-16/11/2019 \\ 
        C155 & 01:13:40.6 & $-$71:32:19 & 15-16/11/2019 \\ 
        C171 & 04:17:28.3 & $-$68:37:30 & 15-16/11/2019 \\ 
        C184 & 05:34:44.18 & $-$67:37:50.1 & 15-16/11/2019 \\ 
        C162 & 05:19:42.4 & $-$65:02:16 & 15-16/11/2019 \\ 
        C177 & 06:23:36 & $-$64:34:41 & 15-16/11/2019 \\ 
        A138 & 00:26:02.54 & $-$72:47:18 & 16-17/11/2019 \\ 
        A040 & 01:17:34.7 & $-$72:50:42.5 & 17-18/11/2019 \\ 
        A030 & 01:06:00.03 & $-$73:31:25.9 & 17-18/11/2019 \\ 
        A111 & 01:17:32.64 & $-$74:39:34.9 & 17-18/11/2019 \\ 
        C182 & 04:55:01.5 & $-$64:49:57 & 17-18/11/2019 \\ 
        C202 & 05:15:09.8 & $-$63:46:07 & 17-18/11/2019 \\ 
        C209 & 05:25:04.4 & $-$64:40:50 & 17-18/11/2019 \\ 
        A045 & 00:59:34.14 & $-$72:09:43.2 & 18-19/11/2019 \\ 
        A021 & 00:35:18.24 & $-$73:18:26 & 18-19/11/2019 \\ 
        C201 & 04:52:49.8 & $-$64:46:23 & 18-19/11/2019 \\ 
        C186 & 05:05:24.35 & $-$67:34:35.4 & 18-19/11/2019 \\ 
        C168 & 06:08:21.8 & $-$65:11:44 & 18-19/11/2019 \\ 
        C208 & 05:25:02.9 & $-$64:40:15 & 18-19/11/2019 \\ 
        C123 & 23:52:40.1 & $-$55:35:23 & 19-21/11/2019 \\ 
        A201 & 00:57:32.75 & $-$72:13:02.3 & 19-20/11/2019 \\ 
        MQ077 & 01:26:10.53 & $-$71:14:10.1 & 19-20/11/2019 \\ 
        MQ123 & 00:39:49.40 & $-$66:07:26.3 & 19-20/11/2019 \\ 
        C187 & 04:51:38.41 & $-$71:02:06 & 19-20/11/2019 \\ 
        MQ119 & 05:20:34.68 & $-$68:35:18.2 & 19-20/11/2019 \\ 
        C216 & 06:17:24 & $-$70:40:49 & 19-20/11/2019 \\ 
        C208 & 05:25:02.9 & $-$64:40:15 & 19-20/11/2019 \\ 
        C019 & 00:52:47.9 & $-$71:15:32 & 20-21/11/2019 \\ 
        C185 & 04:38:31.19 & $-$68:12:00.4 & 20-21/11/2019 \\ 
        C169 & 06:11:46.5 & $-$65:59:02 & 20-21/11/2019 \\ 
        C196 & 04:35:22.5 & $-$64:26:29 & 20-21/11/2019 \\ 
        MQ112 & 00:53:27.4 & $-$73:45:48.9 & 21-22/11/2019 \\ 
        C036 & 00:24:07.8 & $-$72:41:13 & 21-22/11/2019 \\ 
        C035 & 00:26:17.2 & $-$72:33:02 & 21-22/11/2019 \\ 
        C178 & 04:33:04.1 & $-$67:52:56 & 21-22/11/2019 \\ 
        C113 & 05:27:56.9 & $-$67:24:09 & 21-22/11/2019 \\ 
        C167 & 05:14:17.9 & $-$72:20:19 & 21-22/11/2019,\\
         & &  & 23-24/11/2019 \\ 
        C061 & 06:02:12.8 & $-$72:27:36 & 21-22/11/2019 \\ 
        A201 & 00:57:32.75 & $-$72:13:02.3 & 22-23/11/2019 \\
        A286 & 00:21:08.27 & $-$72:17:42 & 22-23/11/2019 \\ 
        A138 & 00:26:02.54 & $-$72:47:18 & 22-23/11/2019 \\ 
        C200 & 04:51:36 & $-$63:54:56 & 22-23/11/2019 \\ 
        C176 & 05:58:43.2 & $-$65:15:53 & 22-23/11/2019 \\ 
        C199 & 04:48:14.7 & $-$64:10:45 & 22-23/11/2019 \\
        \hline    
        \end{tabular}
\end{table}

\begin{table}
\contcaption{}
    \centering
    \begin{tabular}{|l|l|l|r|}
        \hline\hline
        Target Name & RA & DEC & Date Observed \\ 
        \hline 
        A062 & 00:53:56.22 & $-$70:38:04.4 & 23-24/11/2019 \\ 
        MQ044 & 00:32:25.73 & $-$73:59:08.3 & 23-24/11/2019 \\ 
        MQ121 & 01:02:48.28 & $-$72:06:16 & 23-24/11/2019 \\ 
        C023 & 04:17:53.6 & $-$73:15:56 & 23-24/11/2019 \\ 
        C166 & 05:30:41.1 & $-$66:05:35 & 23-24/11/2019 \\ 
        C159 & 05:06:47.96 & $-$19:36:50.7 & 23-24/11/2019 \\ 
        A310 & 00:36:16.99 & $-$74:31:31.3 & 24-25/11/2019 \\ 
        A089 & 00:41:05.74 & $-$70:14:34.8 & 24-25/11/2019 \\ 
        A300 & 01:02:37.76 & $-$75:05:30.7 & 24-25/11/2019 \\ 
        C172 & 04:14:04.9 & $-$69:34:11 & 24-25/11/2019 \\ 
        C210 & 05:26:13.5 & $-$64:31:41 & 24-25/11/2019 \\ 
        C170 & 04:17:31.5 & $-$68:37:40 & 24-25/11/2019 \\ 
        C212 & 05:29:07.3 & $-$63:58:38 & 24-25/11/2019 \\ 
        L01 & 04:37:48.48 & $-$73:15:13.0 & 13-14/01/2021 \\ 
        P11 & 04:55:06.6 & $-$69:17:02.5 & 13-14/01/2021 \\ 
        P23 & 05:41:13.13 & $-$64:11:53.6 & 13-14/01/2021 \\ 
        P47 & 05:58:46.46 & $-$74:59:05.2 & 13-14/01/2021 \\ 
        P12 & 06:15:04.4 & $-$66:17:16.3 & 13-14/01/2021 \\ 
        P11 South & 04:55:06.6 & $-$69:17:02.5 & 14-15/01/2021 \\ 
        L05 & 05:58:49.49 & $-$67:08:00.0 & 14-15/01/2021 \\ 
        C008 (mouse) & 06:08:55.55 & $-$65:52:55.0 & 14-15/01/2021 \\ 
        P51 & 04:21:16.16 & $-$67:10:25.2 & 15-16/01/2021 \\ 
        S28 & 04:35:02.2 & $-$65:19:48.0 & 15-16/01/2021 \\ 
        P16 & 05:33:53.53 & $-$66:43:24.5 & 15-16/01/2021 \\ 
        C184 & 05:34:44.44 & $-$67:37:50.1 & 15-16/01/2021 \\ 
        P14 & 05:53:42.42 & $-$66:52:43.1 & 15-16/01/2021 \\ 
        P21 & 04:36:58.58 & $-$66:22:50.6 & 16-17/01/2021 \\ 
        L03 & 04:37:40.40 & $-$73:15:06.0 & 16-17/01/2021 \\ 
        L02 & 04:37:55.55 & $-$73:14:32.0 & 16-17/01/2021 \\ 
        L04 & 06:03:07.7 & $-$73:25:29.0 & 16-17/01/2021 \\ 
        C131 & 06:09:07.7 & $-$65:44:54.0 & 16-17/01/2021 \\ 
        S62 & 06:10:41.41 & $-$65:06:59.0 & 16-17/01/2021 \\ 
        S15 & 06:13:55.55 & $-$66:03:33.0 & 16-17/01/2021 \\ 
        P26 & 04:29:38.38 & $-$65:26:36.5 & 17-18/01/2021 \\ 
        P05 & 04:38:51.51 & $-$72:17:12.5 & 17-18/01/2021 \\
        C192 & 04:53:05.5 & $-$69:41:06.4 & 17-18/01/2021 \\ 
        C218 & 05:20:57.57 & $-$70:24:53.9 & 17-18/01/2021 \\ 
        C219 & 05:34:49.49 & $-$66:08:57.2 & 17-18/01/2021 \\ 
        S20 & 05:50:17.17 & $-$66:02:18.0 & 17-18/01/2021 \\ 
        S11 & 05:53:51.51 & $-$67:33:40.0 & 17-18/01/2021 \\ 
        P39 & 04:50:09.9 & $-$68:25:33.2 & 18-19/01/2021 \\ 
        P53 & 04:50:53.53 & $-$62:38:54.0 & 18-19/01/2021 \\ 
        P31 & 05:00:08.8 & $-$73:37:49.5 & 18-19/01/2021 \\ 
        P38 & 05:09:53.53 & $-$69:14:36.4 & 18-19/01/2021 \\ 
        P02 & 06:15:00.0 & $-$72:56:42.7 & 18-19/01/2021 \\ 
        P01 & 06:21:06.6 & $-$74:09:26.1 & 18-19/01/2021 \\ 
        P34 & 06:23:16.16 & $-$69:27:36.5 & 18-19/01/2021 \\ 
        P37 & 05:05:59.59 & $-$69:39:53.5 & 19-20/01/2021 \\ 
        P27 & 05:08:44.44 & $-$64:28:31.6 & 19-20/01/2021 \\ 
        P24 & 05:33:58.58 & $-$64:20:24.8 & 19-20/01/2021 \\ 
        P45 & 05:43:34.34 & $-$64:22:58.3 & 19-20/01/2021 \\ 
        P10 & 05:44:05.5 & $-$68:27:21.4 & 19-20/01/2021 \\ 
        P20 & 05:49:13.13 & $-$64:29:29.1 & 19-20/01/2021 \\ 
        P36 & 05:51:23.23 & $-$70:03:13.8 & 19-20/01/2021 \\
        \hline
    \end{tabular}
\end{table}

\begin{table}
\caption{SALT observations made over the course of this work.}
\begin{tabular}{|l|c|r|}
\hline\hline
Name & Proposal & Date observed \\
\hline
LMCtSNE2 & 2022-1-SCI-022 & 07,17/10/2022\\
LMCtSNE1 & 2022-1-SCI-022 & 05/10/2022\\
LMCtSNE3 & 2022-1-SCI-022 & 04/10/2022\\
LMCtSNE4 & 2022-1-SCI-022 & 27/09/2022\\
Source 9 & 2022-1-SCI-022 & 09/09/2022\\
Source 14 & 2022-1-SCI-022 & 07/09/2022\\
Source 6 & 2022-1-SCI-023 & 20/07/2022\\
Source 8 & 2022-1-SCI-023 & 20/07/2022\\
SAGE0534AGN & 2021-2-SCI-017 & 17/03/2022\\
SAGE0534AGN & 2021-2-SCI-018 & 20/11/2021\\
OGLE source & 2021-2-SCI-017 & 15/11/2021\\
C217 & 2021-2-SCI-017 & 12/11/2021\\
C175+FRII & 2021-1-SCI-029 & 23,25/09/2021\\
C175+PAGB & 2021-1-SCI-029 & 23,24/09/2021\\
TSNE16 & 2021-1-SCI-032 & 01/09/2021\\
TSNE12 & 2021-1-SCI-032 & 01/09/2021\\
TSNE11 & 2021-1-SCI-032 & 01/09/2021\\
TSNE10 & 2021-1-SCI-032 & 01/09/2021\\
TSNE13 & 2021-1-SCI-029 & 14,23/07/2021\\
Source 5 & 2021-1-SCI-029 & 17/07/2021\\
TSNE15 & 2021-1-SCI-032 & 13/06/2021\\
ER1 & 2020-2-SCI-025 & 25/11/2020\\
ER6 & 2020-2-SCI-025 & 09/11/2020\\
SA12 & 2020-1-SCI-028 & 29/06/2020\\
Helicopter & 2019-2-SCI-045 & 12/01/2020\\
FuzzyOrange & 2019-2-SCI-041 & 05,07/12/2019\\
Mouse & 2019-2-SCI-045 & 04/12/2019\\
Blazar & 2019-2-SCI-041 & 03/11/2019\\
FuzzyOrange & 2019-2-SCI-041 & 02/11/2019\\
\hline
\end{tabular}
\label{SALTobs}
\end{table}

\begin{table}
\caption{Coordinates of the SALT observations made over the course of this work.}
\begin{tabular}{|l|l|l|l|}
\hline\hline
Name & RA (J2000) & Dec. (J2000) & Notes\\
\hline
LMCtSNE2 & 06:15:04.01 & $-$66:17:16.4 & AGN, $z\sim0.18$\\
LMCtSNE1 & 05:58:48.50 & $-$67:08:00.1 & AGN, $z\sim1.26$\\
LMCtSNE3 & 05:33:57.69 & $-$64:20:24.9 & AGN, $z\sim0.063$\\
LMCtSNE4 & 05:01:10.84 & $-$73:36:35.0 & AGN, $z\sim1.37$\\
Source 9 & 01:21:08.43 & $-$73:07:13.1 & AGN, $z\sim0.99$\\
Source 14 & 01:36:04.20 & $-$72:13:15.4 & AGN, $z\sim0.41$\\
Source 6 & 01:14:08.00 & $-$72:32:43.4 & AGN, $z\sim1.06$\\
Source 8 & 01:22:36.94 & $-$73:10:16.7 & Emission-line star\\
SAGE0534AGN & 05:34:44.17 & $-$67:37:50.1 & AGN, $z\sim1.01$\\
OGLE source & 05:49:30.13 & $-$70:35:43.6 & Variable M-type star \\
& & & (oxygen-rich AGB)\\
C217 & 06:08:43.60 & $-$70:50:07.7 & AGN, $z\sim0.60$\\
C175+FRII & 05:19:00.90 & $-$68:01:58.0 & \\
C175+PAGB &  05:19:00.90 & $-$68:01:58.0 & \\
TSNE16 & 00:49:52.50 & $-$69:29:56.0 & \\
TSNE12 & 01:13:37.10 & $-$74:27:55.0 & \\
TSNE11 & 01:35:05.10 & $-$75:07:03.0 & \\
TSNE10 & 01:27:17.30 & $-$71:04:10.0 & \\
TSNE13 & 00:57:32.8 & $-$72:13:02.0 & \\
Source 5 & 00:48:25.70 & $-$72:44:03.0 & Carbon star\\
TSNE15 & 01:19:13.4 & $-$71:08:51.0 & \\
ER1 & 00:36:59.04 & $-$71:38:10.9 & AGN, $z\sim0.46$\\
ER6 & 01:36:34.37 & $-$71:38:38.7 & AGN, $z\sim0.60$\\
SA12 & 00:36:59.25 & $-$71:38:13.6 & AGN, $z\sim0.46$\\
Helicopter & 06:02:54.1 & $-$71:03:10.0 & Galaxy, $z\sim0.08$\\
FuzzyOrange & 01:21:28.50 & $-$73:20:04.0 & \\
Mouse & 06:08:54.7 & $-$65:52:55.0 & Galaxy, $z\sim0.037$\\
Blazar & 00:57:16.0 & $-$70:40:46.0 & Not a blazar. \\
& & & AGN, $z\sim0.59$\\
\hline
\end{tabular}
\label{SALTobs2}
\end{table}


\section{Training set balancing methods}\label{sampling}

This section shows how different methods to balance the classes in the training sets for the LMC and SMC classifiers effect the precision, recall and F1 score (the harmonic mean of precision and recall, where F1 = $2\frac{\rm precision \times recall}{\rm precision + recall}$) of the trained classifiers that are trained on 75\% of the data and tested on the remaining 25\%.

The precision, recall and F1 score of using no balancing method ('None'), downsampling and upsampling the training sets are shown in Table \ref{sample}. This shows that upsampling allows the classifier to perform better for classes where the training set has few in number (e.g. AGB, H\textsc{ii}/YSO).

Table \ref{sample80} shows the same again but only for sources in the test set classed with P$_{Class} \geq$ 80\%. Also added are the fraction of the overall test set per class that have been classified with P$_{Class} \geq$ 80\%. This shows that upsampling increases the number of sources that are classified with higher confidence, especially for those classes with fewer numbers in the training set.

Overall, upsampling is an improvement over using an unbalanced data set as it increases the number of confident classifications, whilst also increasing the overall performance of the classifier for the classes in the training set that are not as well represented.

\begin{table*}
    \centering
    \caption{Precision, recall and F1 values for the SMC and LMC classifiers trained on 75\% of the data and tested on the remaining 25\%. All values have uncertainties of $\leq$ 0.01. In bold are the values of precision, recall and F1 for the classifier trained on upsampled training data that are equal to or better than downsampling and the 'None' configuration (no upsampling or downsampling).}
    \begin{tabular}{l|lll|lll|lll}
        \hline\hline
        Class & \multicolumn{3}{c}{None} & \multicolumn{3}{c}{Downsampling} & \multicolumn{3}{c}{Upsampling}\\
         & Precision & Recall & F1 & Precision & Recall & F1 & Precision & Recall & F1\\
        \hline
         SMC & 0.94 & 0.94  & 0.94 & 0.89 & 0.74 & 0.78 & 0.93 & 0.83 & 0.86 \\
        \hline
        AGN & 0.92 &  0.82 & 0.87 & 0.78 & 0.83 & 0.80 & 0.85 & \textbf{0.91} & \textbf{0.88} \\
        Galaxy & 0.95 & 0.97 & 0.96 & 0.93 & 0.93 & 0.93 & \textbf{0.97} & 0.95  & \textbf{0.96}\\
        OB & 0.94 & 0.94 & 0.94 & 0.11 & 0.93 & 0.19 & 0.19 & \textbf{0.95} & 0.31\\
        RGB & 0.90 & 0.77 & 0.83 & 0.46 & 0.84 & 0.58 & 0.74 & \textbf{0.87} & 0.80\\
        PNe & 1.00 & 0.68 & 0.81 & 0.18 & 0.73 & 0.27 & 0.12 & \textbf{0.73} & 0.19\\
        Post-AGB/RGB & 0.67 & 0.19 & 0.29 & 0.14 & 0.59 & 0.23 & 0.65 & 0.34 & \textbf{0.45}\\
        AGB & 0.93 & 0.80 & 0.86 & 0.65 & 0.79 & 0.71 & 0.62 & \textbf{0.84} & 0.71\\
        H\textsc{ii}/YSO & 0.74 & 0.54 & 0.62 & 0.24 & 0.86 & 0.37 & 0.67 & 0.73 & \textbf{0.70}\\
        PM & 0.94 & 0.86 & 0.90 & 0.96 & 0.84 & 0.90 & 0.94 & \textbf{0.92} & \textbf{0.93}\\
        RSG & 0.92 & 0.34 & 0.49 & 0.51 & 0.81 & 0.62 & 0.82 & 0.60 & \textbf{0.69}\\
        \hline
        LMC & 0.95 & 0.95 & 0.95 & 0.90 & 0.79 & 0.82 & 0.95 & 0.91 & 0.93 \\
        \hline
        AGN & 0.93 & 0.83 & 0.88 & 0.71 & 0.84 & 0.87 & 0.83 & \textbf{0.90} & 0.87 \\
        Galaxy & 0.95 & 0.97 & 0.96 & 0.95 & 0.92 & 0.96 & \textbf{0.97} & 0.95 & \textbf{0.96}\\
        OB & 0.95 & 0.98 & 0.96 & 0.82 & 0.96 & 0.97 & \textbf{0.96} & \textbf{0.98} & \textbf{0.97}\\
        RGB & 0.92 & 0.98 & 0.95 & 0.12 & 0.99 & 0.69 & 0.54 & \textbf{0.99} & 0.69\\
        PNe & 0.96 & 0.62 & 0.73 & 0.09 & 0.74 & 0.09 & 0.05 & 0.44 & 0.09\\
        Post-AGB/RGB & 0.50 & 0.29 & 0.35 & 0.15 & 0.77 & 0.34 & 0.31 & 0.44 & 0.34\\
        AGB & 0.91 & 0.85 & 0.88 & 0.80 & 0.81 & 0.80 & \textbf{0.92} & \textbf{0.92} & \textbf{0.92}\\
        H\textsc{ii}/YSO & 0.93 & 0.71 & 0.80 & 0.76 & 0.87 & 0.90 & 0.88 & \textbf{0.92} & \textbf{0.90}\\
        PM & 0.96 & 0.81 & 0.88 & 0.95 & 0.82 & 0.96 & \textbf{0.96} & \textbf{0.96} & \textbf{0.96}\\
        RSG & 0.92 & 0.34 & 0.49 & 0.51 & 0.81 & 0.62 & 0.82 & 0.60 & \textbf{0.69}\\
        \hline
    \end{tabular}
    \label{sample}
\end{table*}

\begin{table*}
    \centering
    \caption{Precision, recall and F1 values for the SMC and LMC classifiers trained on 75\% of the data and tested on the remaining 25\%, for those with P$_{Class} \geq$ 80\%. 'Test Fraction' indicates the fraction of the overall test sample for each class that have been classified with P$_{Class} \geq$ 80\% and bold values indicate where the upsampled training set are equal to or better than the other training sets. All values of precision, recall and F1 have uncertainties of $\leq$ 0.01. In bold are the values of precision, recall and F1 for the classifier trained on upsampled training data that are equal or better than downsampling and the 'None' configuration (no upsampling or downsampling).}
    \begin{tabular}{l|llll|llll|llll}
        \hline\hline
        Class & \multicolumn{4}{c}{None} & \multicolumn{4}{c}{Downsampling} & \multicolumn{4}{c}{Upsampling}\\
         & Precision & Recall & F1 & Test & Precision & Recall & F1 & Test & Precision & Recall & F1 & Test\\
         &  &  & & Fraction & & & & Fraction & & & & Fraction\\
        \hline
         SMC & 0.98 & 0.98 & 0.98 & & 0.99 & 0.99 & 0.99 & & 0.98 & 0.98 & 0.98 & --\\
        \hline
        AGN & 0.98 & 0.90 & 0.94 & 0.70$\pm$0.01  & 0.99  & 0.92 & 0.95 & 0.07$\pm$0.01 & 0.97 & \textbf{0.95} & \textbf{0.96} & \textbf{0.79$\pm$0.01} \\
        Galaxy & 0.98 & 0.99 & 0.99 & 0.94$\pm$0.02 & 0.99 & 0.99 & 0.99 & 0.27$\pm$0.07 & 0.98 & \textbf{0.99} & \textbf{0.99} & 0.86$\pm$0.01\\
        OB & 0.98 & 0.99 & 0.98 & 0.91$\pm$0.02 & 0.99 & 1.00 & 0.99 & 0.78$\pm$0.03 & 0.98 & 0.99 & \textbf{0.99} & 0.89$\pm$0.02\\
        RGB & 0.98 & 0.85 & 0.91 & 0.58$\pm$0.05 & 1.00 & 0.98 & 0.99 & 0.28$\pm$0.01 & 0.87 & 0.94 & 0.90 & \textbf{0.76$\pm$0.08}\\
        PNe & 1.00 & 0.87 & 0.93 & 0.47$\pm$0.05 & 1.00 & 1.00 & 1.00 & 0.47$\pm$0.05 & \textbf{1.00} & 0.94 & 0.97 & \textbf{0.63$\pm$0.09}\\
        Post-AGB/RGB & 0.00 & 0.00 & 0.00 & 0.14$\pm$0.06 & 0.44 & 0.67 & 0.50 & 0.05$\pm$0.03 & \textbf{1.00} & 0.42 & \textbf{0.57} & \textbf{0.21$\pm$0.06}\\
        AGB & 1.00 & 0.92 & 0.96 & 0.66$\pm$0.03 & 1.00 & 1.00 & 1.00 & 0.45$\pm$0.02 & \textbf{1.00} & \textbf{1.00} & \textbf{1.00} & 0.63$\pm$0.06\\
        H\textsc{ii}/YSO & 0.00 & 0.00 & 0.00 & 0.08$\pm$0.03 & 0.33 & 0.33 & 0.33 & 0.02$\pm$0.02 & \textbf{1.00} & \textbf{0.61} & \textbf{0.76} & \textbf{0.25$\pm$0.05}\\
        PM & 0.99 & 0.99 & 0.99 & 0.56$\pm$0.07 & 1.00 & 1.00 & 1.00 & 0.33$\pm$0.11 & 0.99 & \textbf{1.00} & 0.99 & \textbf{0.74$\pm$0.05} \\
        RSG & 1.00 &  1.00 & 1.00 & 0.04$\pm$0.04 & 0.33 & 0.33 & 0.33 & 0.30$\pm$0.05 & \textbf{1.00} & \textbf{1.00} & \textbf{1.00} & 0.26$\pm$0.05\\
        \hline
        LMC & 0.99 & 0.99 & 0.99 & & 0.99 & 0.99 & 0.99 & & 0.99 & 0.99 & 0.99 & --\\
        \hline
        AGN & 0.98 & 0.90 & 0.94 & 0.68$\pm$0.01 & 1.00 & 0.96 & 0.98 & 0.16$\pm$0.06 & 0.97 & 0.95 & 0.96 & \textbf{0.78$\pm$0.03} \\
        Galaxy & 0.98 & 1.00 & 0.99 & 0.95$\pm$0.02 & 0.99 & 1.00 & 0.99 & 0.30$\pm$0.10 & \textbf{0.99} & 0.99 & \textbf{0.99} & 0.87$\pm$0.02\\
        OB & 0.99 & 1.00 & 1.00 & 0.93$\pm$0.05 & 1.00 & 1.00 & 1.00 & 0.60$\pm$0.01 & 0.99 & 1.00 & 0.99 & \textbf{0.93$\pm$0.04}\\
        RGB & 0.98 & 1.00 & 0.99 & 0.89$\pm$0.02 & 0.94 & 1.00 & 0.96 & 0.86$\pm$0.03 & 0.95 & 0.99 & 0.97 & \textbf{1.00$\pm$0.02}\\
        PNe & 0.98 & 0.83 & 0.88 & 0.07$\pm$0.06 & 1.00 & 1.00 & 1.00 & 0.26$\pm$0.15 & 0.94 & 0.88 & 0.91 & \textbf{0.45$\pm$0.23} \\
        Post-AGB/RGB & 0.00 & 0.00 & 0.00 & 0.00$\pm$0.00 & 0.44 & 0.67 & 0.50 & 0.00$\pm$0.00 & \textbf{0.47} & 0.50 & 0.41 & \textbf{0.27$\pm$0.09} \\
        AGB & 0.97 & 0.99 & 0.98 & 0.47$\pm$0.07 & 1.00 & 0.98 & 0.99 & 0.27$\pm$0.13 & 0.98 & 0.98 & 0.98 & \textbf{0.68$\pm$0.10}\\
        H\textsc{ii}/YSO & 1.00 & 0.94 & 0.97 & 0.18$\pm$0.01 & 1.00 & 0.96 & 0.98 & 0.05$\pm$0.01 & 0.98 & \textbf{0.99} & \textbf{0.99} & \textbf{0.64$\pm$0.02}\\
        PM & 1.00 & 1.00 & 1.00 & 0.43$\pm$0.02 & 1.00 & 1.00 & 1.00 & 0.37$\pm$0.01 & \textbf{1.00} & \textbf{1.00} & \textbf{1.00} & \textbf{0.71$\pm$0.10}\\
        RSG & 0.67 & 0.67 & 0.67 & 0.53$\pm$0.02 & 0.98 & 1.00 & 0.99 & 0.50$\pm$0.01 & 0.47 & 0.50 & 0.41 & \textbf{0.71$\pm$0.01} \\
        \hline
    \end{tabular}
    \label{sample80}
\end{table*}

\section{Results table}\label{RT}
In Table \ref{tab:results_table} the format of the PRF class predictions table can be seen.

\begin{table}
\caption{Table format of the table of the PRF class predictions. All probabilities are within the range 0--1. The table is available at VSA\textsuperscript{a} (VISTA Science Archive) and CDS\textsuperscript{b} (Centre de Données astronomiques).}

\begin{tabular}{ll}
\hline\hline
          Column & Column \\
          name & description \\
          \hline
          VISTA ID & \llap{U}nique identifier in the VSA.\\
          VMC RA & \llap{R}ight Ascension in degrees from the VMC survey.\\
          VMC DEC & \llap{D}eclination in degrees from the VMC survey.\\
          AGN & \llap{P}robability of the source being an AGN.\\
          AGN err & \llap{E}rror on probability of the source being an AGN\\
          Galaxy & \llap{P}robability of the source being a Galaxy.\\
          Galaxy err & \llap{E}rror on probability of the source being a Galaxy.\\
          OB & \llap{P}robability of the source being an OB.\\
          OB err & \llap{E}rror on probability of the source being an OB.\\
          RGB & \llap{P}robability of the source being an RGB.\\
          RGB err & \llap{E}rror on probability of the source being a RGB.\\
          AGB & \llap{P}robability of the source being an AGB.\\
          AGB err & \llap{E}rror on probability of the source being an AGB.\\
          H\textsc{ii}/YSO & \llap{P}robability of the source being an H\textsc{ii}/YSO.\\
          H\textsc{ii}/YSO err & \llap{E}rror on probability of the source being a H\textsc{ii}/YSO err.\\
          PNe & \llap{P}robability of the source being an PNe.\\
          PNe err & \llap{E}rror on probability of the source being a PNe.\\
          pAGB/pRGB & \llap{P}robability of the source being a pAGB/RGB.\\
          pAGB/pRGB err & \llap{E}rror on probability of the source being a pAGB/pRGB.\\
          RSG & \llap{P}robability of the source being an RSG.\\
          RSG err & \llap{E}rror on probability of the source being a RSG.\\
          PM & \llap{P}robability of the source being a PM.\\
          PM err & \llap{E}rror on probability of the source being a PM.\\
          Unk & \llap{P}robability of the source being Unknown.\\
          Unk err & \llap{E}rror on probability of the source being Unknown.\\
          PRF Class & \llap{P}redicted Class. Class with highest probability.\\
          PRF P$_{\rm class}$ & \llap{P}robability of the source being the predicted Class.\\
          PRF P$_{\rm extragalactic}$ & \llap{C}ombined AGN and Galaxy probabilities.\\
          Cat flag & \llap{F}lag indicating which catalogue the source belongs to. \\
          & `H' = High-confidence, `M' = Mid-confidence and\\
        
          & `L' = Low-confidence.\\
          XR flag & \llap{S}tates whether there is an X-ray detected source within \\
          & the positional error for each X-ray detection (X) or a \\
           & radio detected source within 2$^{\prime\prime}$ (R), or both (B). \\
\hline
\end{tabular}
\label{tab:results_table}
\small\textsuperscript{a}: \url{http://vsa.roe.ac.uk}\\
\small\textsuperscript{b}: \url{https://cds.unistra.fr}
\end{table}

\newpage

\section{Spatial distribution of individual classes}\label{App_dist}

This section describes the spatial distribution of the individual classes across the sky for sources with probability $>$ 80\%.

\begin{figure*}
\centering
\begin{tabular}{c}
	\includegraphics[width=1\textwidth, trim=1mm 1mm 1mm 2mm, clip]{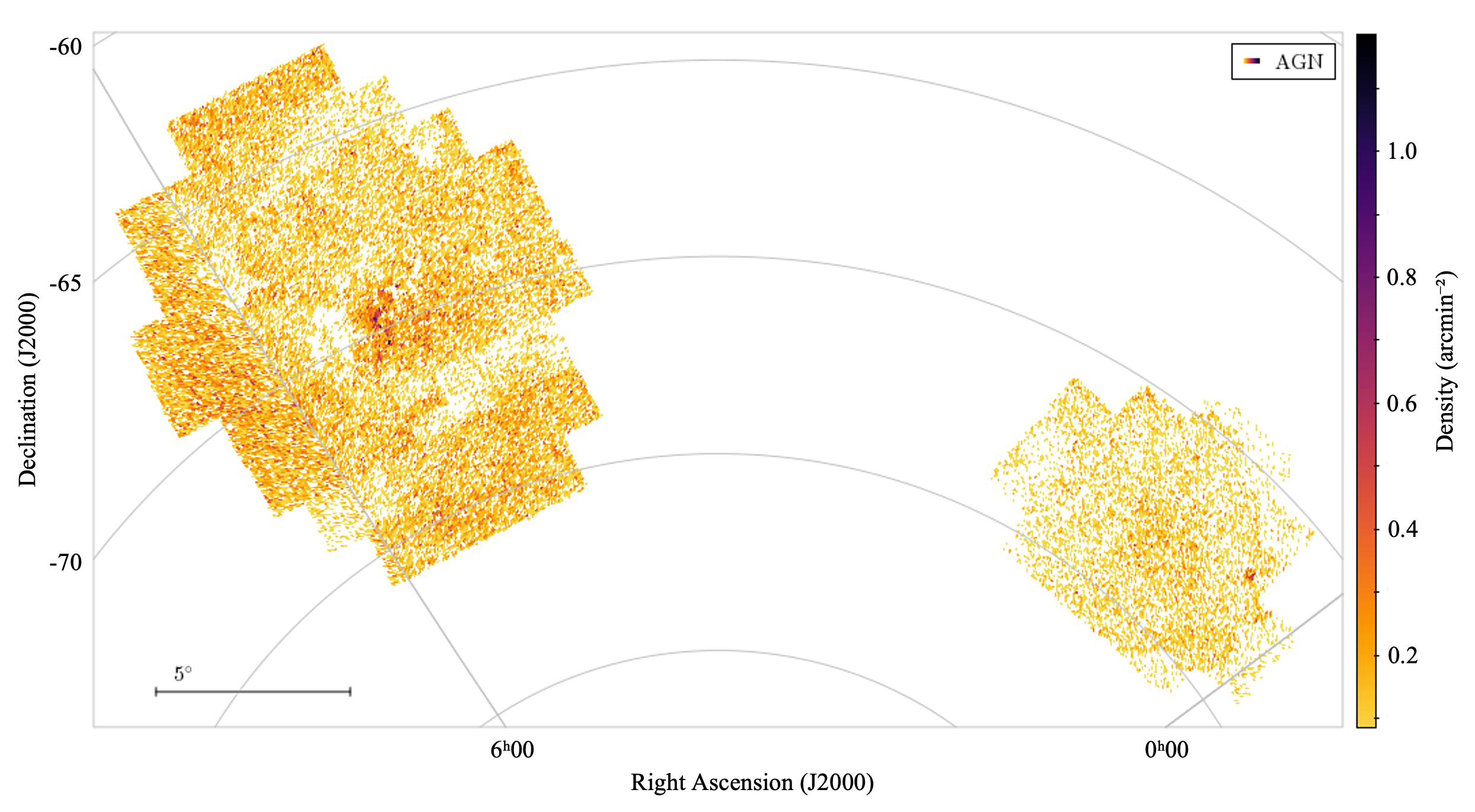}
 \end{tabular}
    \caption{The sky density of AGN with Class probabilities $>$ 80\% for the LMC (left) and SMC (right).}
    \label{Dist_agn80}
\end{figure*}

The AGN spatial distribution, as seen in Figure \ref{Dist_agn80}, is mostly homogeneous across the two fields as expected, with some increase in the number of sources where more surveys (which means more features to inform the classification) overlap. What is unexpected is that there does not appear to be an obvious decrease towards the centre of the SMC and LMC where the stellar density is at its highest.

\begin{figure*}
\centering
\begin{tabular}{c}
	\includegraphics[width=1\textwidth, trim=1mm 1mm 1mm 2mm, clip]{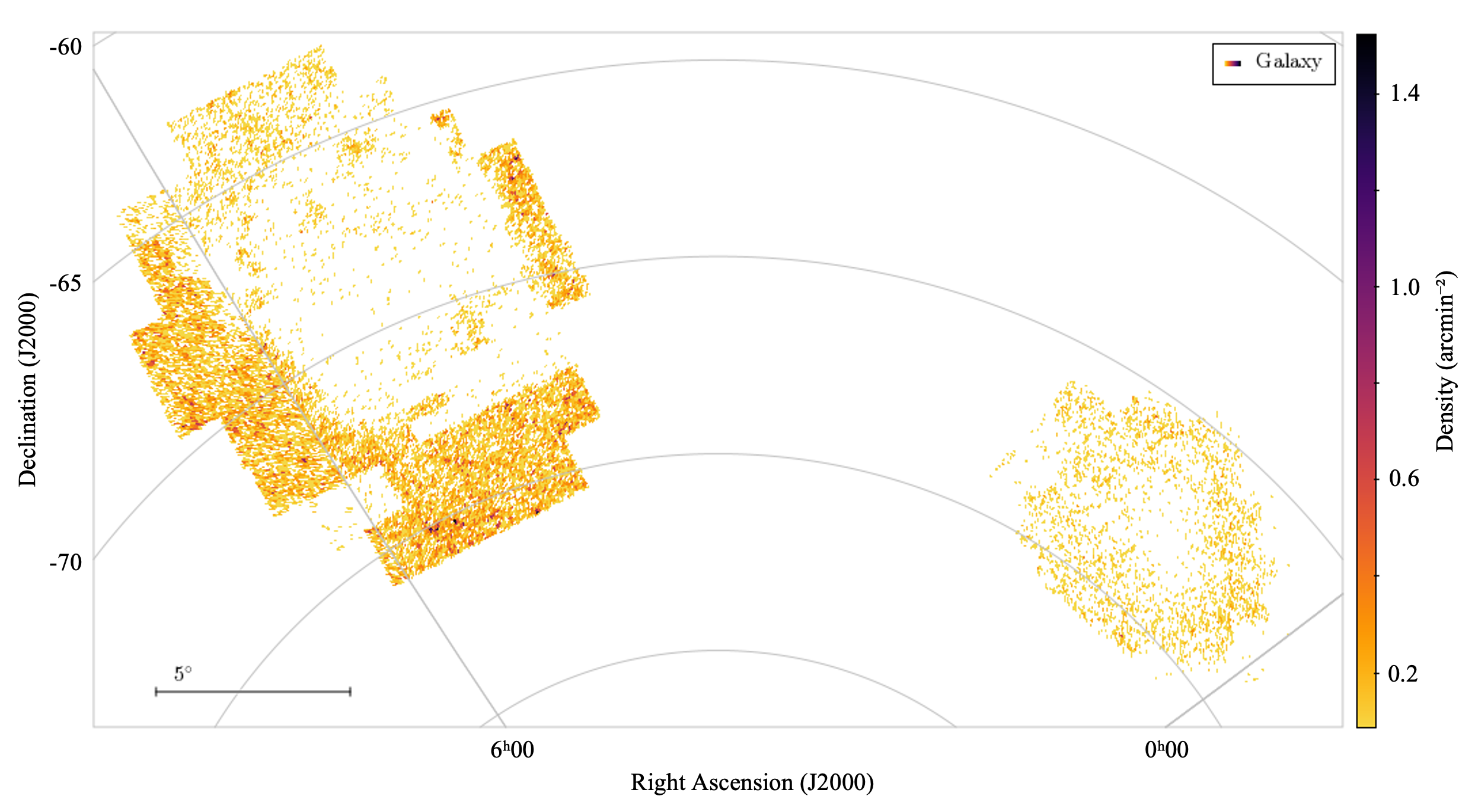}
 \end{tabular}
    \caption{The sky density of galaxies with Class probabilities $>$ 80\% for the LMC (left) and SMC (right).}
    \label{Dist_gal80}
\end{figure*}

The spatial distribution of galaxies, as seen in Figure \ref{Dist_gal80}, is mostly homogeneous away from the centre of the LMC and SMC, and then decreases in number towards the centre of the Clouds, as expected.

\begin{figure*}
\centering
\begin{tabular}{c}
	\includegraphics[width=1\textwidth, trim=1mm 1mm 1mm 2mm, clip]{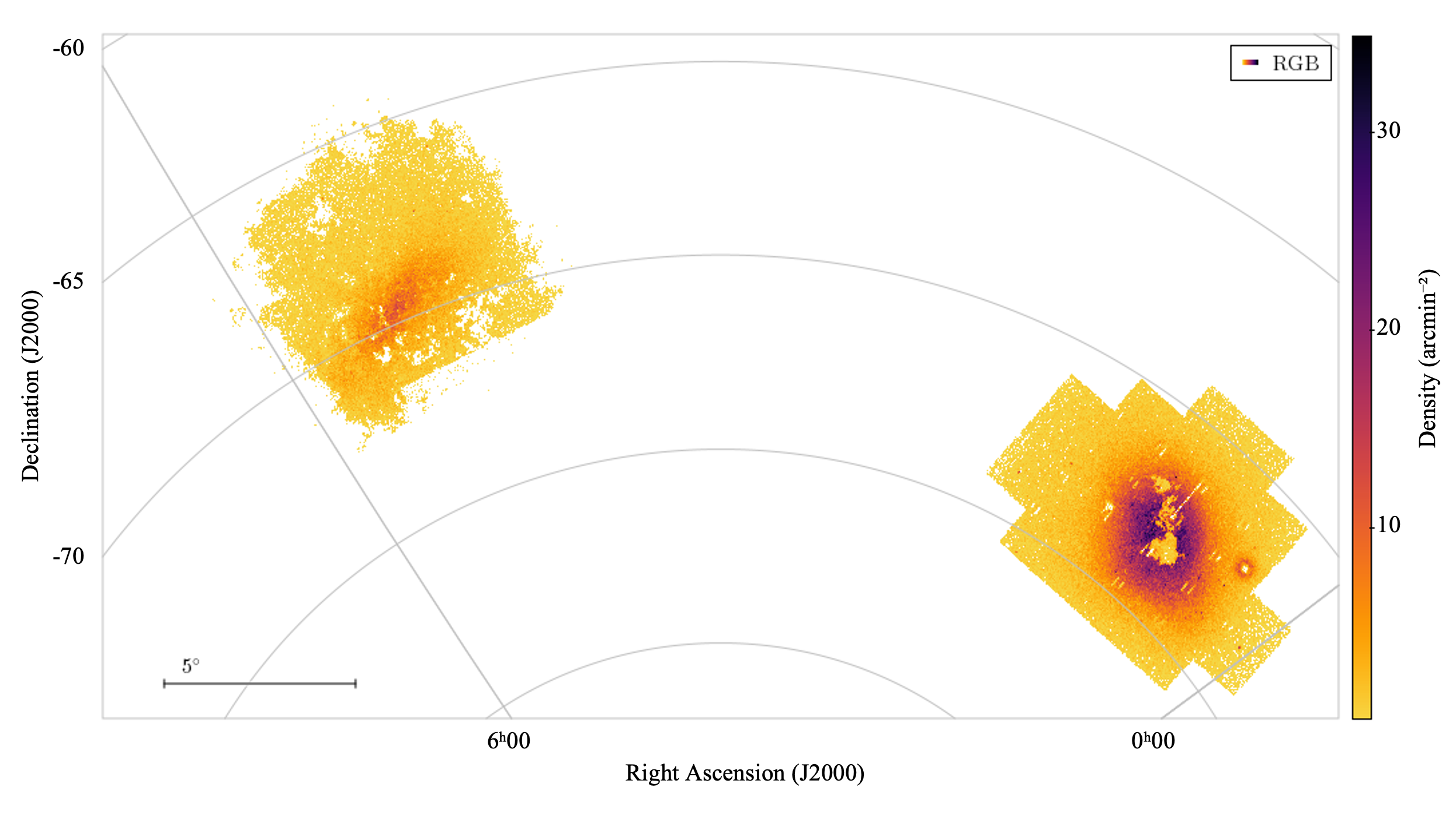}
 \end{tabular}
    \caption{The sky density of RGB with Class probabilities $>$ 80\% for the LMC (left) and SMC (right).}
    \label{Dist_rgb80}
\end{figure*}

The RGB stellar spatial distribution, seen in Figure \ref{Dist_rgb80}, for the SMC is dense in the centre as expected, but is also reaching the edges of the SMC survey footprint with higher than expected numbers. It is possible that some of the redder extragalactic sources are being misclassified as RGBs. For the LMC, the RGBs concentrate mainly in the bar and do not reach the outskirts of the LMC survey field. The shape of the spatial distribution roughly follows the SAGE/HERITAGE survey footprint. It should be noted that there is a difference in the RGB SMC and LMC training set. Unlike the LMC, the SMC has RGBs spectroscopically observed at fainter magnitudes (12.5 mag $<$ $Y$ $<$ 24 mag), whilst LMC has more brighter RGBs (12.5 mag $<$ $Y$ $<$ 16.5 mag).

\begin{figure*}
\centering
\begin{tabular}{c}
	\includegraphics[width=1\textwidth, trim=1mm 1mm 1mm 2mm, clip]{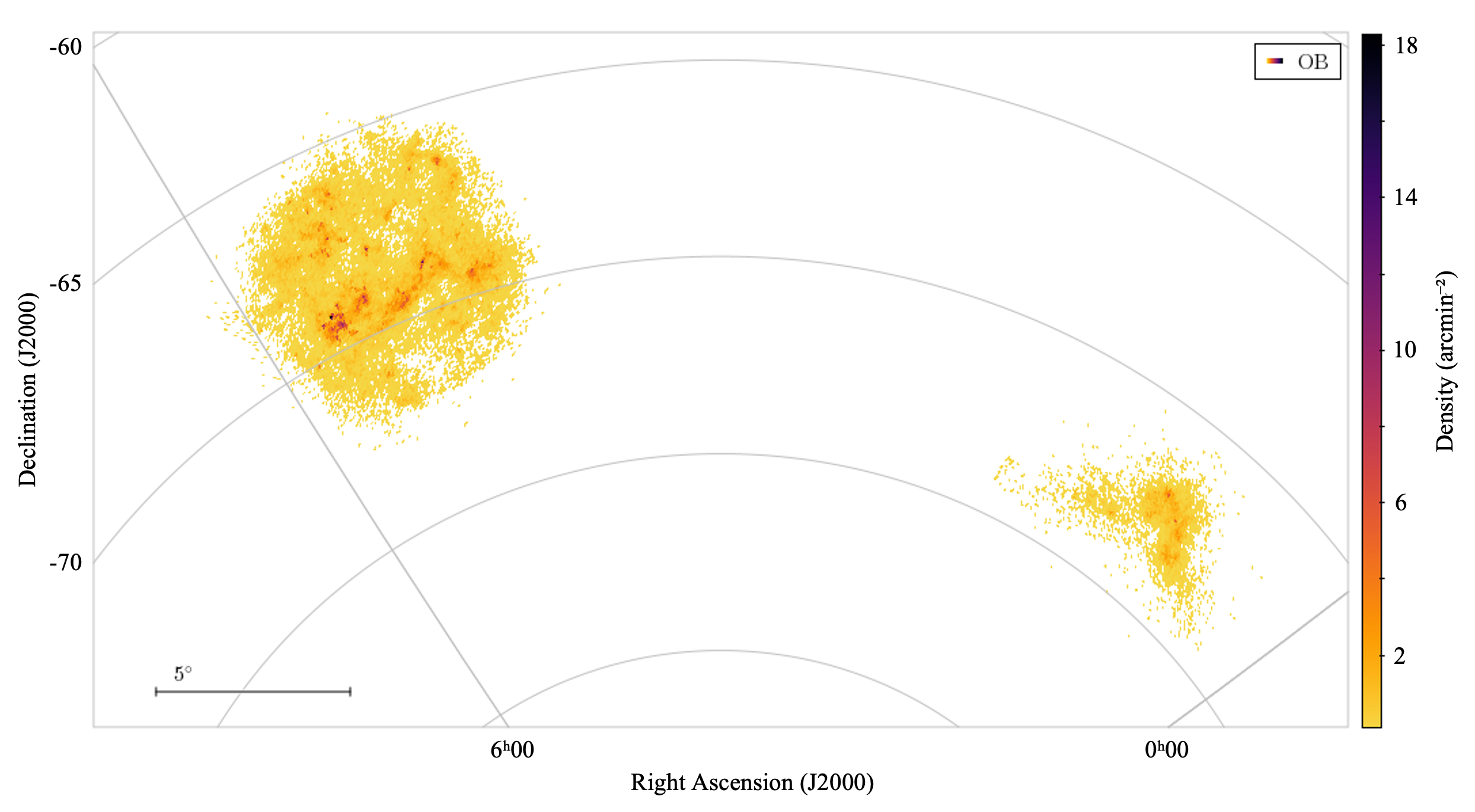}
 \end{tabular}
    \caption{The sky density of OB with Class probabilities $>$ 80\% for the LMC (left) and SMC (right).}
    \label{Dist_ob80}
\end{figure*}

The OB stellar spatial distribution, seen in Figure \ref{Dist_ob80}, concentrates nicely in the centre as expected with fewer sources at the edge of the Magellanic Clouds. The SMC is known to have a bar structure with an extension towards the East, which is what we are seeing here with the OB spatial distribution. Emission-line stars were seen to most likely be classed as OB stars and the slight increase in sources towards the lower left might include emission-line stars (such as Be stars) from the Bridge.

\begin{figure*}
\centering
\begin{tabular}{c}
	\includegraphics[width=1\textwidth, trim=1mm 1mm 1mm 2mm, clip]{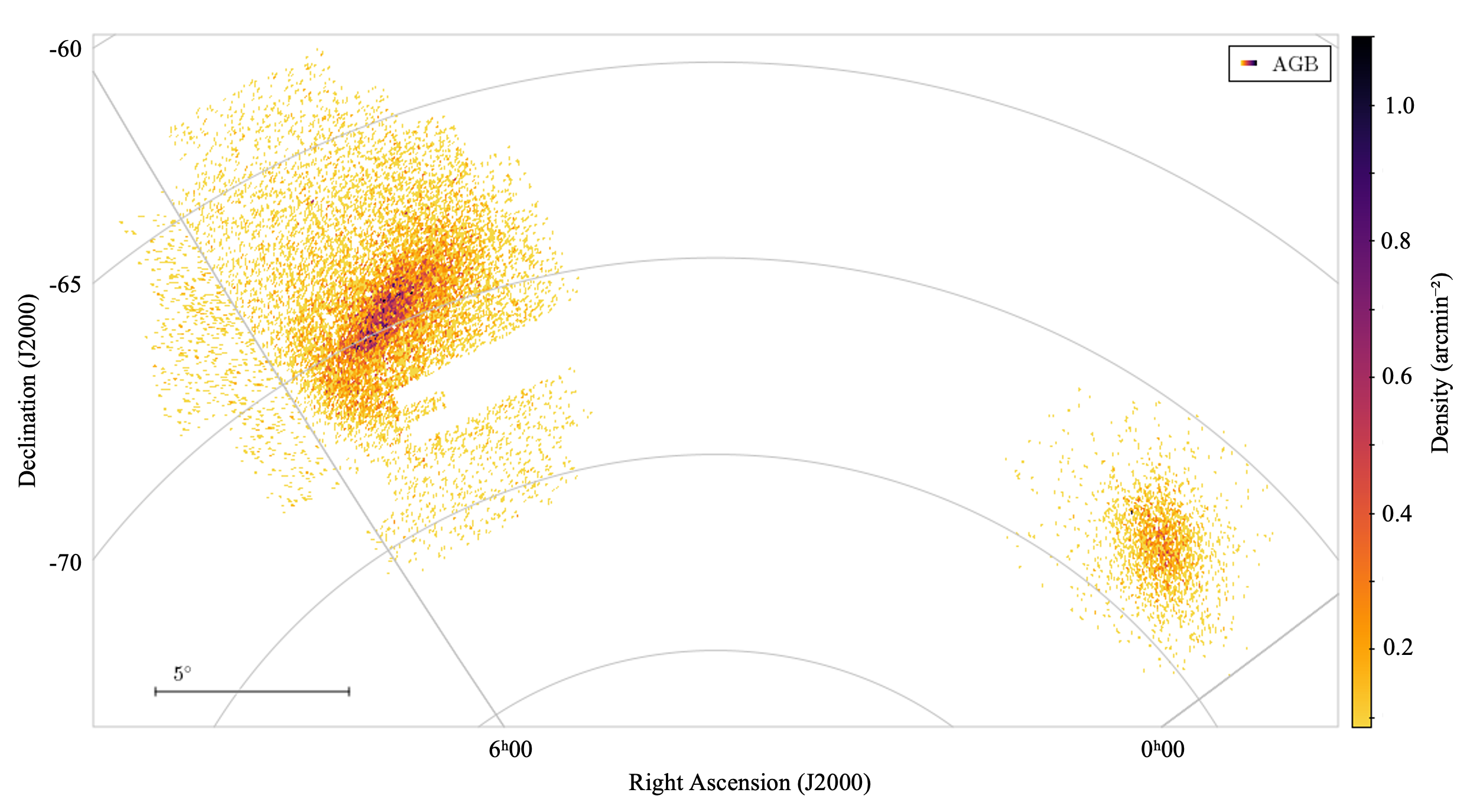}
 \end{tabular}
    \caption{The sky density of AGB with Class probabilities $>$ 80\% for the LMC (left) and SMC (right).}
    \label{Dist_agb80}
\end{figure*}

The AGB stellar spatial distribution, seen in Figure \ref{Dist_agb80}, can be seen to concentrate in the centre of the Magellanic Clouds, as expected. The SMC AGBs roughly concentrate in the shape of the main bar of the SMC and decrease towards the outskirts of the SMC survey footprint. The LMC AGBs concentrate mostly in the bar of the LMC, and reduce in number towards the outskirts of the survey footprint. The LMC AGBs reaching the edges of the survey footprint was expected (Nikolaev \& Weinberg 2000; El Youssoufi et al. 2019) show that the AGBs covering the full area of the VMC survey of the LMC and SMC is expected.

\begin{figure*}
\centering
\begin{tabular}{c}
	\includegraphics[width=1\textwidth, trim=1mm 1mm 1mm 2mm, clip]{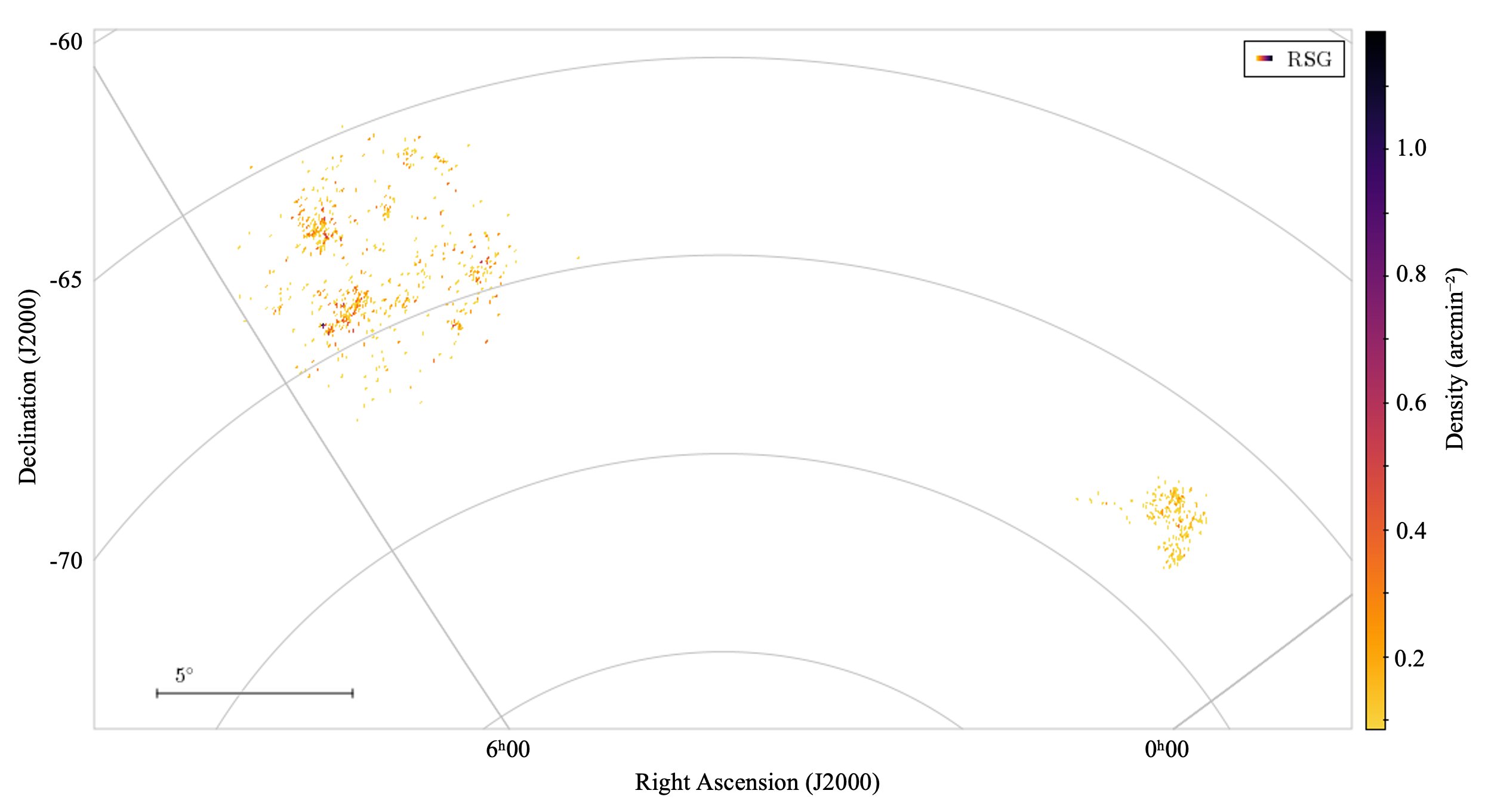}
 \end{tabular}
    \caption{The sky density of RSG with Class probabilities $>$ 80\% for the LMC (left) and SMC (right).}
    \label{Dist_rsg80}
\end{figure*}

\begin{figure*}
\centering
\begin{tabular}{c}
	\includegraphics[width=1\textwidth, trim=1mm 1mm 1mm 2mm, clip]{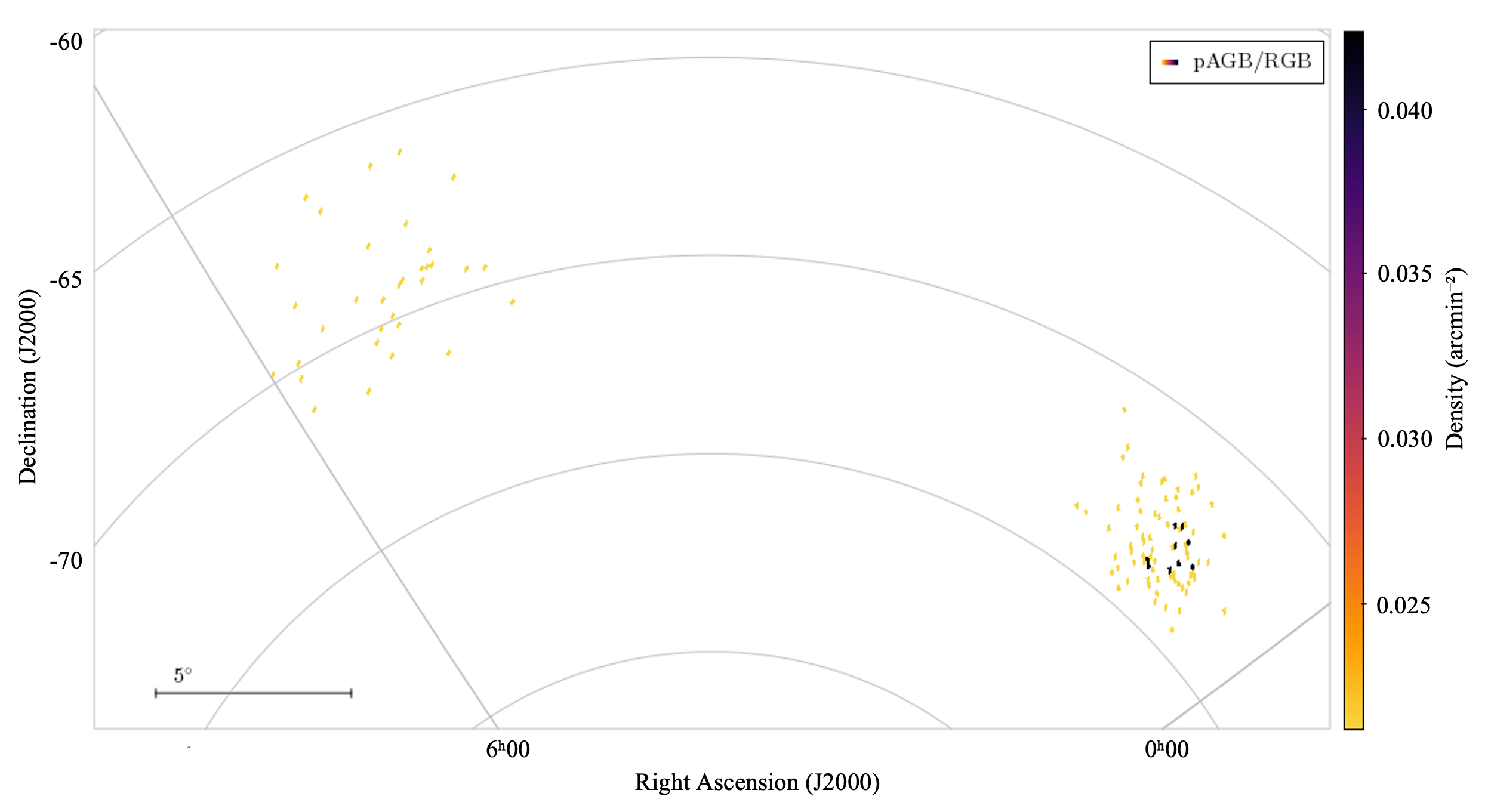}
 \end{tabular}
    \caption{The sky density of post-AGB/RGB with Class probabilities $>$ 80\% for the LMC (left) and SMC (right).}
    \label{Dist_pagb80}
\end{figure*}

\begin{figure*}
\centering
\begin{tabular}{c}
	\includegraphics[width=1\textwidth, trim=1mm 1mm 1mm 2mm, clip]{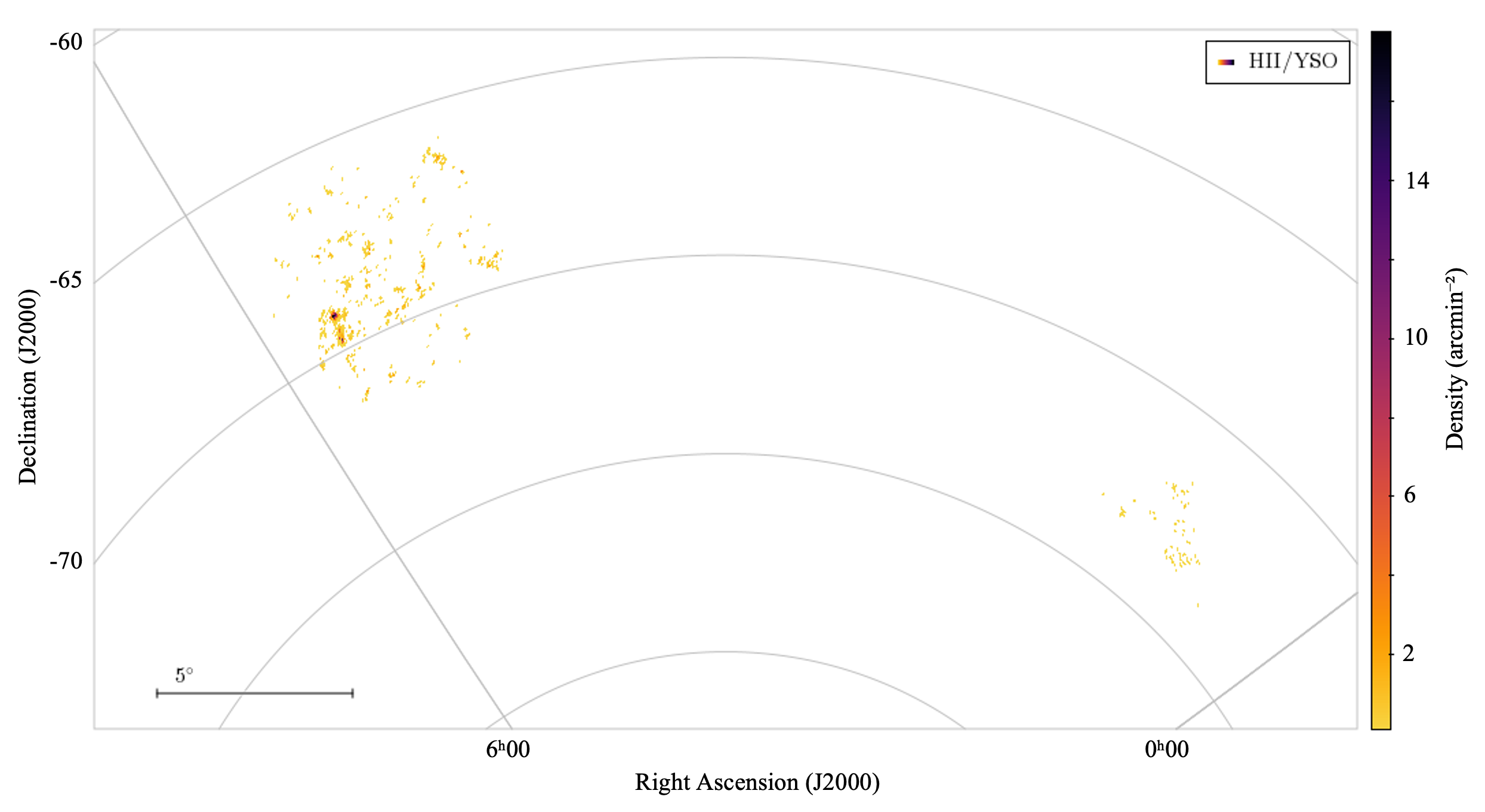}
 \end{tabular}
    \caption{The sky density of H\textsc{ii}/YSO with Class probabilities $>$ 80\% for the LMC (left) and SMC (right).}
    \label{Dist_yso80}
\end{figure*}

\begin{figure*}
\centering
\begin{tabular}{c}
	\includegraphics[width=1\textwidth, trim=1mm 1mm 1mm 2mm, clip]{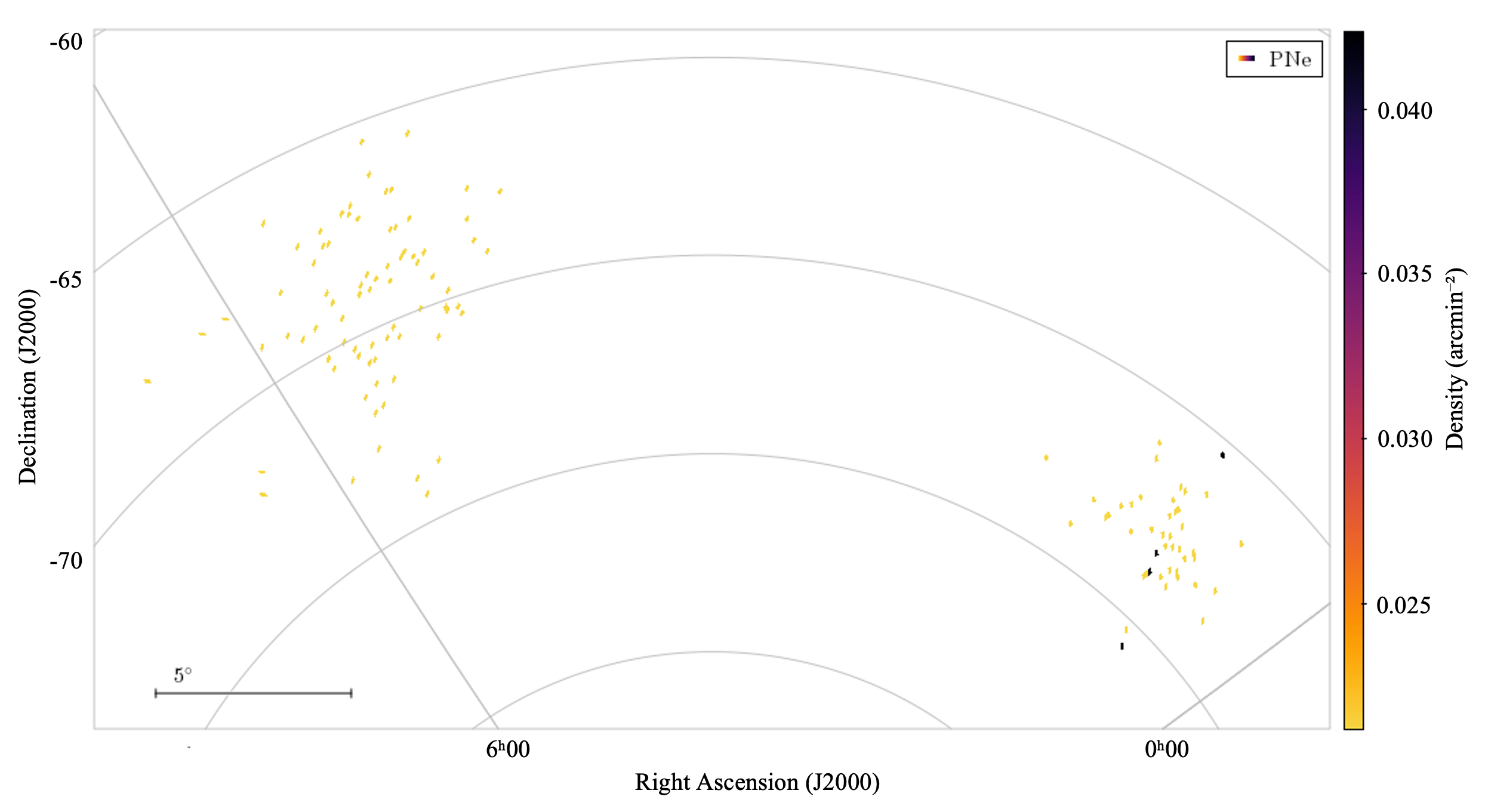}
 \end{tabular}
    \caption{The sky density of PNe with Class probabilities $>$ 80\% for the LMC (left) and SMC (right).}
    \label{Dist_pne80}
\end{figure*}

The RSG, post-AGB/RGB, PNe and H\textsc{ii}/YSOs sources, seen in Figures \ref{Dist_rsg80}, \ref{Dist_pagb80}, \ref{Dist_pne80} and \ref{Dist_yso80}, respectively, also concentrate in the centre of the Clouds as expected. For the LMC the H\textsc{ii}/YSOs have a tendency to clump in areas of high far-IR flux, and therefore areas of emission from star-formation, which is as expected. The RSGs appear concentrated in an extended region centred around 30 Doradus (e.g. Schneider et al. 2018), a region to the north of it, and a region in the south-west. These locations correspond well with the positions of OH/IR stars (the most extreme phases of mass loss of RSGs and massive AGB stars).

\begin{figure*}
\centering
\begin{tabular}{c}
	\includegraphics[width=1\textwidth, trim=1mm 1mm 1mm 2mm, clip]{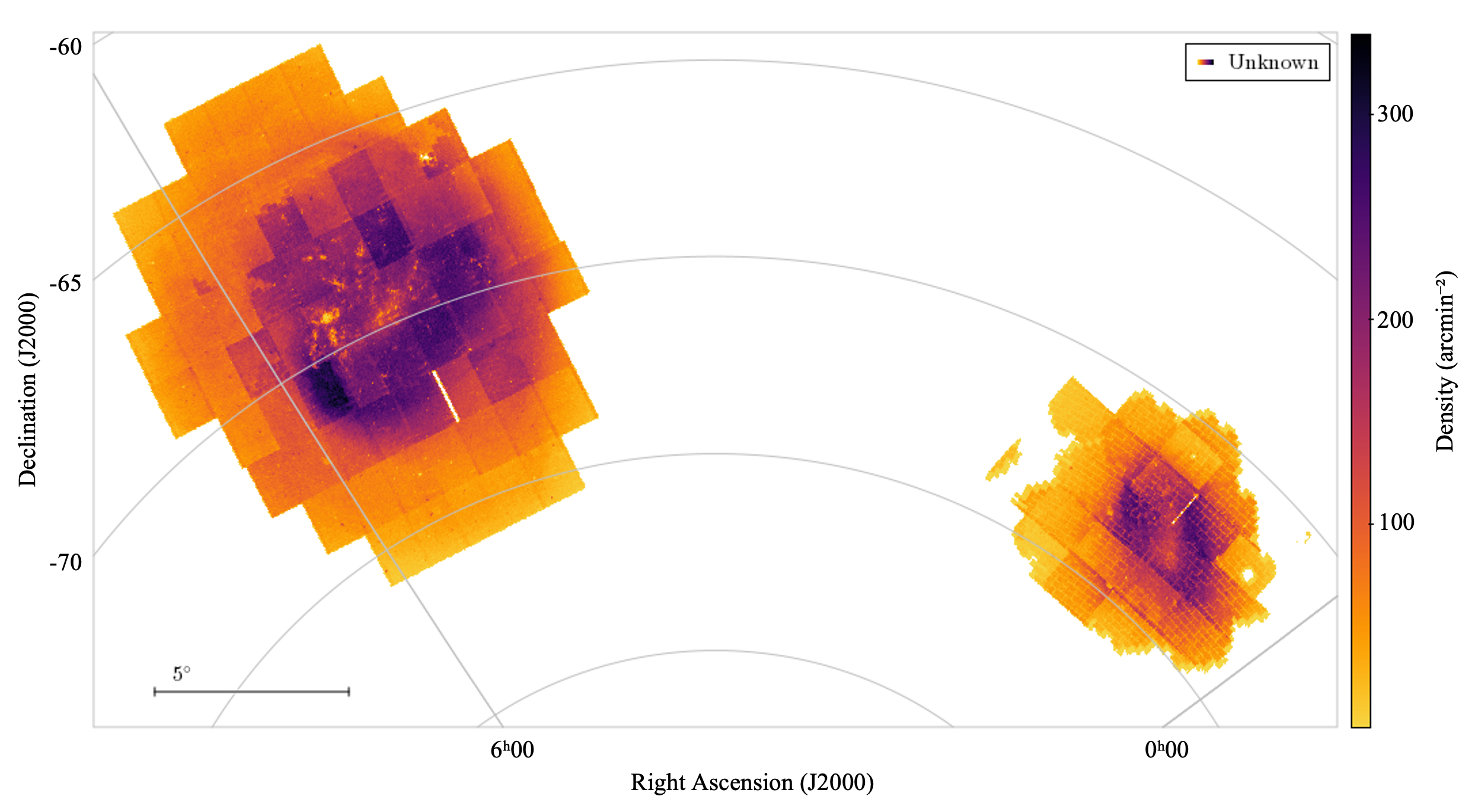}
 \end{tabular}
    \caption{The sky density of Unknown with Class probabilities $>$ 80\% for the LMC (left) and SMC (right).}
    \label{Dist_unk80}
\end{figure*}

The spatial distribution of the Unknowns, as seen in Figure \ref{Dist_unk80}, and cover the majority if the LMC and SMC fields with the highest densities at the centre of the Clouds. The Unknowns represent the majority of sources and are mostly made up of the fainter sources for which there is little to no spectroscopy available to add to the training set, so this spatial distribution is expected. It can be seen that the areas of fewer sources in the centre of the LMC roughly match the location of sources classified as H\textsc{ii}/YSOs in Figure \ref{Dist_yso80}. Furthermore, across both galaxies the pattern of the VMC tiles can be seen. This is most likely because the sources were not filtered for those with large uncertainties and/or flags.

\newpage
\section{Training vs Output}\label{trainvsout}

This section shows the class distributions of the training set in colour--colour and colour--magnitude space compare to those of the PRF classed sources.

\begin{figure*}
\centering
\begin{tabular}{c}
	\includegraphics[width=1\textwidth]{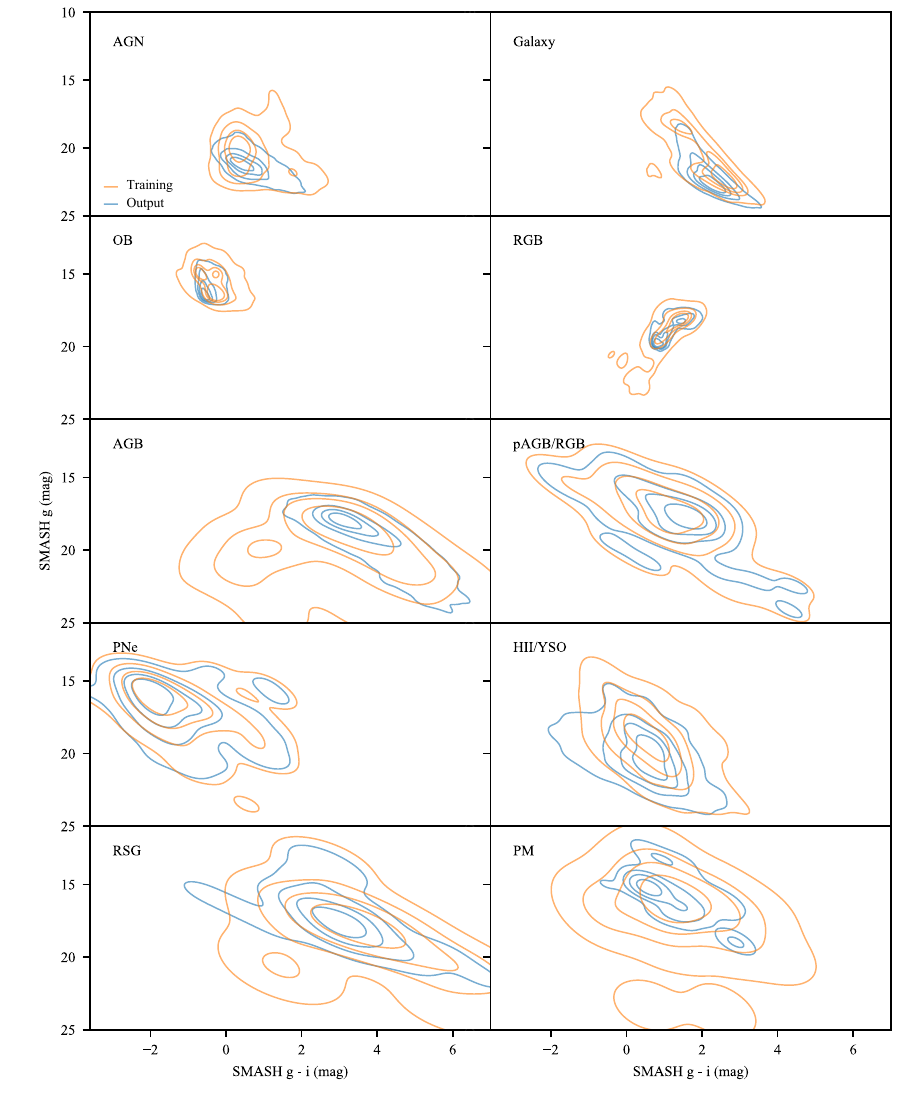}
 \end{tabular}
    \caption{SMASH colour--magnitude diagrams of the sources separated by class for the training set (orange) and the classified output (blue) for sources with probability $>$ 80\% in the SMC and LMC fields. The contours represent a probability distribution in intervals of 0.2.  }
    \label{SMASH_Train}
\end{figure*}

In Figure \ref{SMASH_Train}, showcasing an optical colour--magnitude diagram, we see that for all classes the PRF classed sources concentrate mostly where the training set concentrates. 

\begin{figure*}
\centering
\begin{tabular}{c}
	\includegraphics[width=1\textwidth]{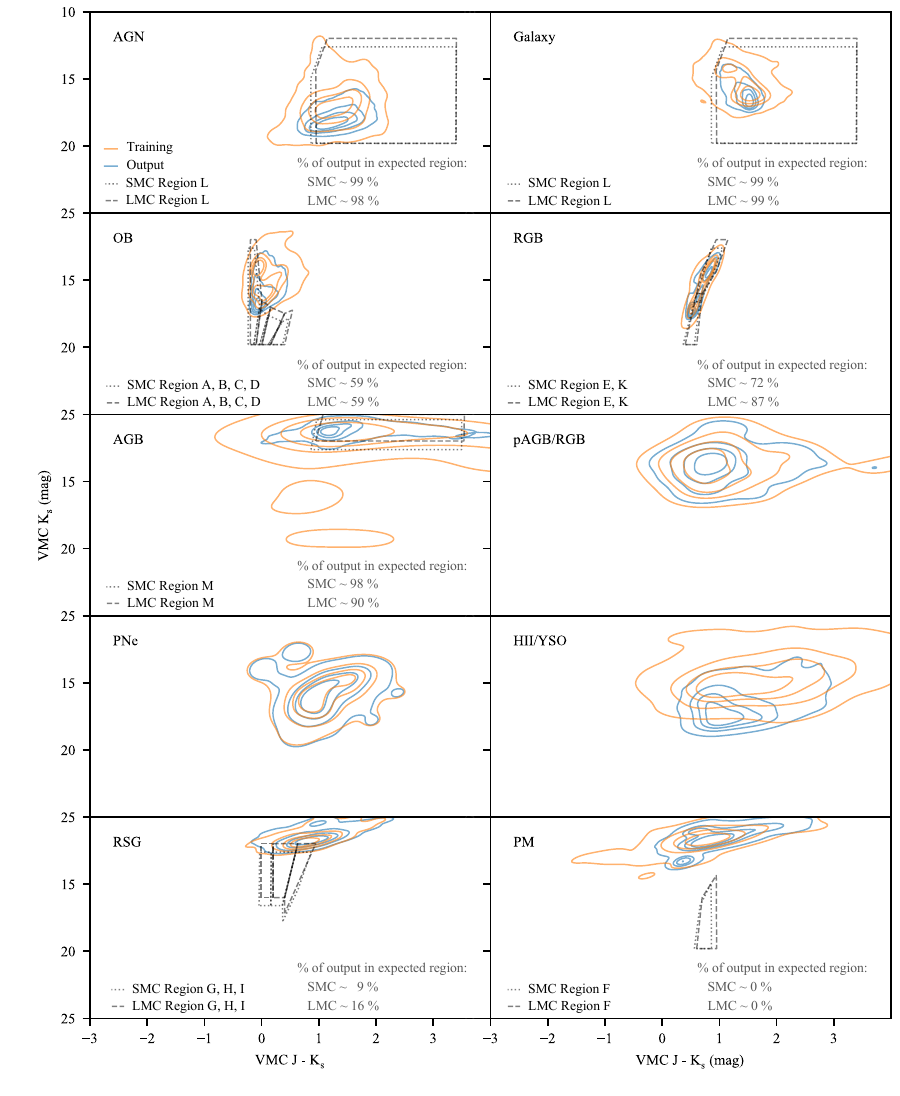}
 \end{tabular}
    \caption{VISTA colour--magnitude diagrams of the sources separated by class for the training set (orange) and the classified output (blue) for sources with probability $>$ 80\% in the SMC and LMC fields. The contours represent a probability distribution in intervals of 0.2. The dotted and dashed lines represent the expected regions calculated for different classes by El Youssoufi et al. (2019) for the SMC and LMC, respectively. }
    \label{VMC_Train}
\end{figure*}

The VISTA near-IR colour--magnitude diagram can be seen in Figure \ref{VMC_Train}. Here we see that for all classes the PRF classed sources concentrate mostly where the training set concentrates. We also plot on top the expected regions calculated for different classes by El Youssoufi et al. (2019).

The training set and output for OB stars mainly lie in region A, but also spread out from the expected main-sequence regions. This could be because some of these OB stars exhibit emission-lines, which can make their appearance more diverse.

For the RGBs, the output for the LMC is in region K but not region E, whereas the SMC RGBs are in both the expected regions, but with more sources in region E. This could be because the SMC training set has more examples of fainter RGBs to train upon, and region E is fainter in K$_s$ than region K. 

For the foreground stars (PM) we see that the expected region for these stars, is fainter than the sources in the training set and output. This implies that the fainter examples of foreground stars are being completely missed by the classifier. 

\begin{figure*}
\centering
\begin{tabular}{c}
	\includegraphics[width=1\textwidth]{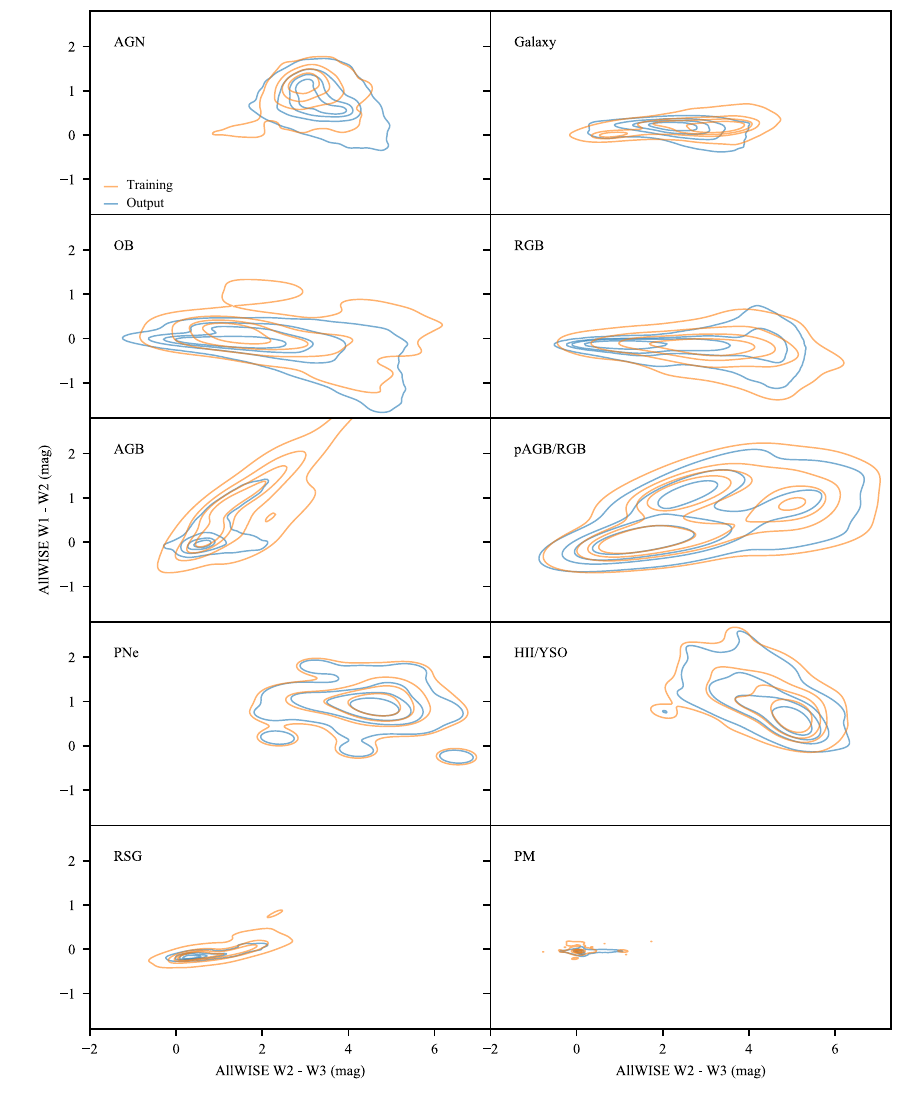}
 \end{tabular}
    \caption{AllWISE colour--colour diagrams of the sources separated by class for the training set (orange) and the classified output (blue) for sources with probability $>$ 80\% in the SMC and LMC fields. The contours represent a probability distribution in intervals of 0.2.  }
    \label{WISE_Train}
\end{figure*}

In Figure \ref{WISE_Train} we see the AllWISE mid-IR colour--colour diagram. Here we see that for all classes the PRF classed sources concentrate mostly where the training set concentrates when compared to Nikutta et al. 2014. 

\section{Unknown VMC CMD population selection}\label{UNKselCMD}

\begin{figure*}
\centering
\begin{tabular}{c}
	\includegraphics[width=\textwidth]{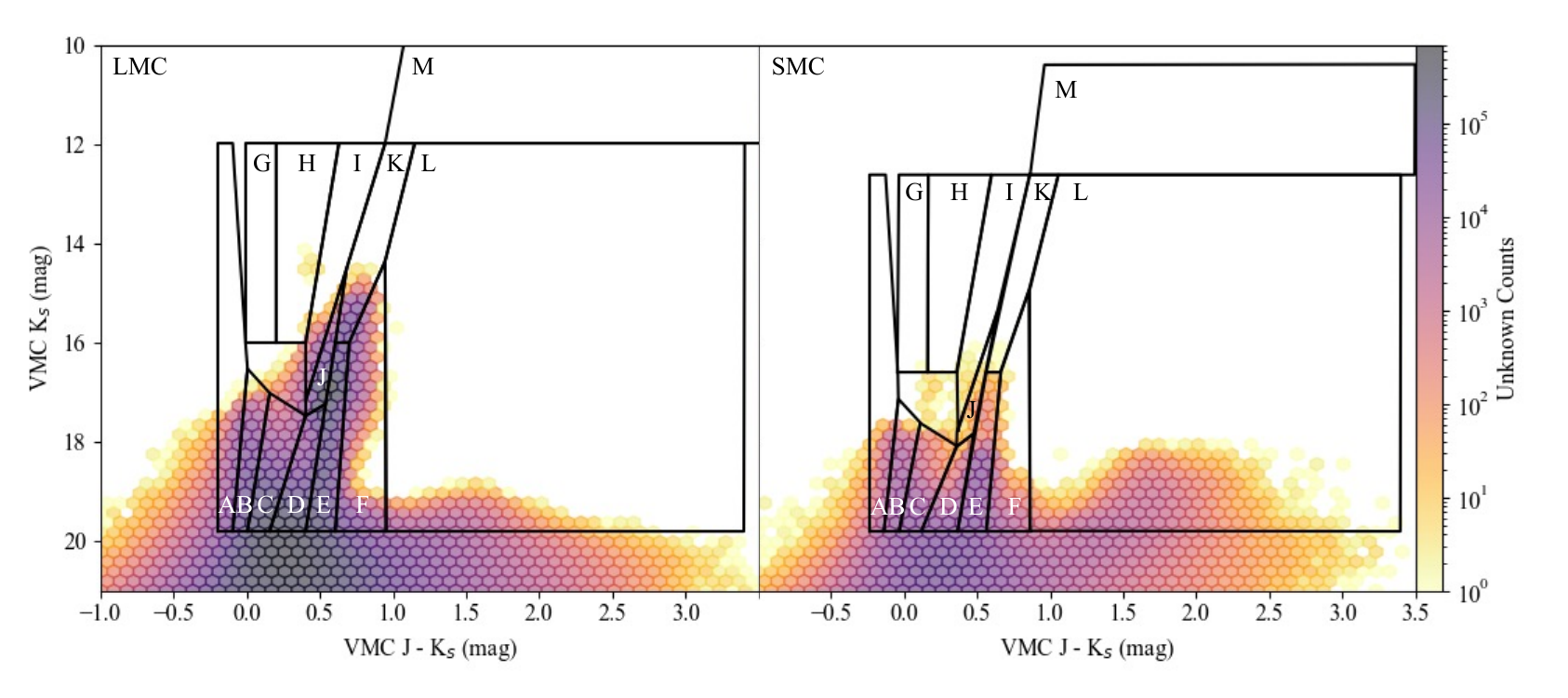}
 \end{tabular}
    \caption{VISTA colour--magnitude diagrams of the sources in the Unknown class with P$_{\rm class}$ $>$ 80\%. Overplotted are the selection criteria from El Youssoufi et al. (2019) that selects different stellar populations.}
    \label{MC_unk}
\end{figure*}

This section shows the distribution of the Unknown class with P$_{\rm class}$ $>$ 80\% in VISTA near-IR colour--magnitude space and how this compares to the stellar population selection criteria (as described in Section 5.8) from El Youssoufi et al. (2019). This can be seen in Figure \ref{MC_unk}.

\bsp	
\label{lastpage}
\end{document}